\documentclass[12pt]{iopart}
\usepackage{iopams}
\expandafter\let\csname equation*\endcsname\relax
\expandafter\let\csname endequation*\endcsname\relax
\usepackage{amsmath}

\usepackage{graphicx,hyperref}
\usepackage{braket}
\usepackage{xcolor,colortbl}
\definecolor{Green}{rgb}{0.23,0.7,0.44}
\definecolor{Yellow}{rgb}{0.94,0.9,0.55}

\definecolor{battleshipgrey}{rgb}{0.52, 0.52, 0.51}
\definecolor{cadet}{rgb}{0.33, 0.41, 0.47}
\definecolor{charcoal}{rgb}{0.21, 0.27, 0.31}
\definecolor{linkblue}{rgb}{0.08,0.3,0.9}

\hypersetup{
	colorlinks   = true, 
	urlcolor     = battleshipgrey, 
	linkcolor    = cadet, 
	citecolor   = charcoal 
}

\newcolumntype{g}{>{\columncolor{Green}}c}
\newcolumntype{b}{>{\columncolor{white}}c}

\renewcommand{\Re}{\mathrm{Re}}

\begin{document}
	
	\title[Ultrafast holographic photoelectron imaging]{It is all about phases: ultrafast holographic photoelectron imaging}	
	
	\author{C. Figueira de Morisson Faria and 
	A. S. Maxwell
}%
	\address{Department of Physics \& Astronomy, University College London\\ Gower Street, London WC1E 6BT, United Kingdom}
	\ead{c.faria@ucl.ac.uk, andrew.maxwell@ucl.ac.uk}

	\date{\today}
	\begin{abstract}
		Photoelectron holography constitutes a powerful tool for the ultrafast imaging of matter, as it combines high electron currents with subfemtosecond resolution, and gives  information about transition amplitudes and phase shifts. Similarly to light holography,  it uses the phase difference between the probe and the reference waves associated with qualitatively different ionization events for the reconstruction of the target and for ascertaining any changes that may occur. These are major advantages over other attosecond imaging techniques, which require elaborate interferometric schemes in order to extract phase differences. For that reason, ultrafast photoelectron holography has experienced a huge growth in activity, which has led to a vast, but fragmented landscape. The present review is an organizational effort towards unifying this landscape. This includes a historic account in which a connection with laser-induced electron diffraction (LIED) is established, a summary of the main holographic structures encountered and their underlying physical mechanisms, a broad discussion of the theoretical methods employed, and of the key challenges and future possibilities. We delve deeper in our own work, and place a strong emphasis on quantum interference, and on the residual Coulomb potential.
	\end{abstract}
\maketitle
\section{Overview}
Attosecond ($10^{-18}s$) science deals with some of the shortest time scales in nature, and has emerged from the study of the interaction of matter with strong laser fields, typically of intensities around $I=10^{13}\mathrm{W/cm}^2$ or higher. The key idea behind it is to probe and ultimately steer  electron dynamics in real time (for reviews see, e.g., \cite{Lein2007,Krausz2009,Salieres2012R}). Among other applications, electrons create or destroy chemical bonds, carry energy in biomolecules, and information in human-made devices, so that the above-mentioned control has the potential to revolutionize many areas of knowledge. For that reason, strong-field phenomena such as high-order harmonic generation (HHG) or above-threshold ionization (ATI) have established themselves as powerful imaging tools.

This is made possible because their underlying physical mechanisms, laser-induced recollision or recombination \cite{Corkum1993}, take place within a fraction of a field cycle.  
In ATI, an electron is released close to the peak of the field and, if it reaches the detector without further interaction, it is called a direct ATI electron. If it is driven back by the field and rescatters with the parent ion, it is known as a rescattered electron  \cite{MilosReviewATI}. For HHG, instead of a laser-induced collision, recombination with a bound state of its parent ion will occur, and the kinetic energy acquired in the continuum will be released as high-frequency radiation \cite{Lewenstein1994}. The time for which the electron will return will be close to a field crossing. Hence, the driving field will dictate the time window for which ionization and recollision occur. This has led to a wide range of applications such as attosecond pulses \cite{Agostini2004,Calegari2016}, high-order harmonic spectroscopy \cite{Marangos2016} and photoelectron holography \cite{HuismansScience2011}.

The information about the target obtained upon recollision is imprinted in the high-harmonic or photoelectron spectra. Particularly in the case of 
photoelectrons, light-induced electron diffraction (LIED) has called a great deal of attention since it has been first proposed \cite{Zuo1996}, due to the high electron recollision currents of around $10^{11}\mathrm{A/cm}^2$ incident on the target \cite{Niikura2002}. This is several orders of magnitude higher than what is available in conventional time-resolved electron microscopy \cite{Ihee2001}.
The physical picture outlined above also makes it possible to associate specific transition amplitudes with electron trajectories. 
Since there may be many pathways for an electron to reach the detector with the same final energy, the corresponding transition amplitudes will interfere. Due to allowing an intuitive physical interpretation, this trajectory-based picture has been incorporated in a myriad of theoretical approaches that allow for quantum interference. Examples are the Strong-Field Approximation (SFA), the Eikonal Volkov Approximation (EVA), Analytical R-Matrix theory (ARM), the Quantum Trajectory Monte Carlo method (QTMC), the Semiclassical Two-Step model (SCTS), the Coulomb Volkov Approximation (CVA), Quantitative Rescattering Theory (QRS), the Coulomb-Corrected Strong-Field Approximation (CCSFA), and the Coulomb Quantum-orbit Strong-Field Approximation (CQSFA). For a review of many of these methods see, e.g., \cite{Popruzhenko2014a} and Sec.~\ref{sec:beyondsfa}.

Throughout the years, ultrafast imaging has moved from purely structural questions such as the reconstruction of molecular orbitals \cite{Itatani2004}, to dynamic effects such as electron migration in larger molecules \cite{Calegari2014,Lepine2014,Kuleff2016,Calegari2016}. In all these studies, quantum interference plays a huge role. A vital and highly non-trivial issue has been how to retrieve relative phases from the high-order harmonic or photoelectron signal. This is exemplified by the fact that it took six years between the seminal paper reconstructing the highest molecular orbital (HOMO) of $\mathrm{N}_2$ using HHG, and the experimental retrieval of the molecular phase shifts using the Reconstruction of Attosecond Burst By Interferences of Two-photons Transitions (RABBIT) technique \cite{Haessler2010}. 
In fact, in \cite{Itatani2004} it was necessary to postulate such phase shifts in order to achieve a successful bound-state reconstruction. 

The wish to record the \textit{magnitude} and the \textit{phase} of photoelectron scattering amplitudes has led to the inception of ultrafast photoelectron holography. In conventional holography \cite{Gabor}, interference patterns between a reference and a probe wave are recorded and the phase differences are used to construct the hologram.  Similarly, in ATI one may employ the quantum interference between different types of electron trajectories to obtain this information. This has been first proposed in \cite{Spanner2004}.
Originally, the definition ``holographic structures'' referred to those 
generated by 
the interference of direct trajectories, for which there is no interaction with the core subsequent to ionization, i.e., ``the reference'', with those recolliding elastically with the core, i.e., ``the probe'' \cite{Bian2011}. As the field progressed this idea has been relaxed and generalized by some research groups, which refer to holographic patterns also as resulting from  the interference between trajectories undergoing different types of rescattering. 

Within this generalized framework, several holographic patterns have been predicted theoretically, some of which are easily identified in experiments, and some of which are fairly obscure.
Well known examples 
are the spider-like structure caused by the interference of two types of forward-scattered trajectories \cite{HuismansScience2011,Hickstein2012}, and fork-  \cite{Moeller2014} or fishbone-like \cite{Bian2012,Haertelt2016,Li2015} structures resulting from backscattered trajectories interfering with forward-scattered orbits. The fishbone holographic pattern is particularly sensitive to the target structure and has been employed in \cite{Haertelt2016} to probe diatomic molecules. 
A largely unexplored and to a certain extent contentious issue is how the residual long-range potential influences the contributing trajectories and the holographic patterns. While earlier models incorporate the binding potential only upon rescattering and employ Coulomb-free orbits during the propagation in the continuum \cite{Spanner2004,HuismansScience2011,Bian2011,Bian2012}, in the past few years there have been detailed studies of Coulomb-distorted orbits \cite{Lai2017,Maxwell2017,Maxwell2017a,Maxwell2018,Shilovski2018,Maxwell2018b}, whose topology is significantly altered by the residual potential \cite{Yan2010,Yan2012}.  A recent example  is the fan-shaped structure that forms near the ionization threshold. It is a known fact that this pattern is caused by the joint influence of the external field and the long range of Coulomb potential, but the precise mechanism behind it has raised considerable debate. For instance, in \cite{Rudenko2004,Arbo2006,Chen2006,Arbo2008,Arbo2008b} it has been attributed to resonances with specific bound states of the Coulomb potential with a well-defined orbital angular momentum. In \cite{Yan2010}, it has been related to the interference of several types of trajectories. In \cite{Lai2017,Maxwell2017} we have shown that this structure is a type of holographic pattern resulting from the interference of direct and forward-deflected trajectories.  The residual Coulomb potential causes an angle-dependent, radial distortion in a pattern known as the ``temporal double slit''. If Coulomb distortions are not incorporated, the temporal double slit corresponds to the interference between two reference, non scattering events and it is thus not classified as holographic \cite{Lindner2005,Arbo2006}. 

There are currently three main research trends related to photoelectron holography: more complex targets, longer wavelengths, in the mid- or far-IR regime, and tailored fields. These trends are not mutually exclusive and may be combined in subfemtosecond imaging and control of electron dynamics.  Longer wavelengths $\lambda$ lead to larger ponderomotive energies $U_p \sim I\lambda^2 $, and thus provide the electron with larger kinetic energies upon return \cite{Wolter2015}. This implies that the holographic patterns will be clearer and closer to those predicted by simplified models, as Coulomb distortions become less prominent \cite{Marchenko2011}. Furthermore, large ponderomotive energies may be achieved with lower driving-field intensities, which means that ionization of the target may be substantially reduced. This makes the mid-IR regime more favourable for probing complex systems, as for larger molecules the ionization potentials are typically lower. Longer wavelengths have been successfully used in LIED of organic molecules, in order to retrieve bond lengths and track bond dynamics in real time \cite{Wolter2016}.  They have also been used in key publications on photoelectron holography, such as those in which spider-like patterns were first identified \cite{HuismansScience2011,Marchenko2011,Hickstein2012}. Targets such as molecules or multielectron atoms allow for assessing/probing dynamic changes such as those involving the coupling of electronic and nuclear degrees of freedom, core polarization, resonances, electron-electron correlation or charge migration. 
Finally, external driving fields can be tailored with the purpose of manipulating specific types of trajectories, and thus expose holographic patterns which are obfuscated by more prominent structures. Examples are orthogonally polarized \cite{Zhang2014,Xie2017} or bicircular \cite{Milos2016,Hoang2017,Eckart2018,Milos2018} fields.

However, this hugely popular research field has evolved in a highly fragmented way. This lack of consensus includes the main holographic patterns encountered, the theoretical approaches developed and the current trends. The aim of the present review is to compare, unify and organize the vast research landscape around photoelectron holography, from its early days to the current trends, with special emphasis on our own work on the subject. This includes a summary of the main interference patterns identified in experiments (Sec.~\ref{sec:patterns}), and the theoretical models used to model photoelectron diffraction and holographic structures (Sec.~\ref{sec:theory}). In  Sec.~\ref{sec:CQSFA}, we delve deeper into the Coulomb Quantum-orbit Strong-Field Approximation, which is the method derived and employed in our publications. Subsequently, in Sec.~\ref{sec:results}, we discuss the key results obtained with the CQSFA. This includes the types of orbits encountered (Sec.~\ref{sec:orbits}), single-orbit photoelectron momentum distributions (Sec.~\ref{sec:subbarrrier}) and several types of holographic structures (Sec.~ \ref{sec:resholographic}). Finally, in Sec.~\ref{sec:trends} we conclude the review with a summary of the existing trends. We use atomic units throughout.

\section{Holographic patterns and types of rescattering}
\label{sec:patterns}
In order to understand photoelectron holography, one must first discuss laser-induced photoelectron diffraction (LIED).  In LIED, photoelectrons emitted in above-threshold ionization (ATI) are employed to image specific targets. As an imaging tool, LIED exhibits three key advantages: (i) It allows for high electron currents, which are orders of magnitude higher than those in standard photoelectron microscopy; (ii) The electron emission is highly coherent, as it is controlled by the external laser field; (iii) It allows for resolving dynamic changes in the target with subfemtosecond resolution.
Since its early days in the late 1990s \cite{Zuo1996}, LIED has led to a wide range of applications such as the tracking of the coupling between electronic and nuclear degrees of freedom, giant resonances, multi-orbital effects \cite{Krecinic2018} and bond dynamics in real time \cite{Wolter2016}. 
One should note, however, that the concept of LIED is broader than that of photoelectron holography, as it does not necessarily involve rescattered trajectories. In fact, even direct ATI trajectories, in which the electron reaches the detector  without rescattering, carry information about the electron's parent ion. Rescattered trajectories are however expected to be far more sensitive with regard to the target, so that they have been widely studied in the context of LIED and photoelectron holography.  In this section, we will commence by discussing both and will subsequently focus on specific holographic patterns that have been predicted or measured. Examples of laser-induced electron diffraction patterns that are not holographic structures are ATI rings, the \textit{spatial} and the \textit{temporal} double slit.
\subsection{ATI rings}

Ionization events associated to ATI happen at specific times, which repeat with the periodicity of the laser field. Thus, the transition amplitudes related to events separated by an integer number of cycles will interfere. The periodicity of the field will lead to discrete peaks of energy
\begin{equation}
\frac{\mathbf{p}_f^2}{2}=n\omega-U_p-I_p,
\label{eq:atirings1}
\end{equation}
where $\mathbf{p}_f$ is the electron momentum at the detector, $\omega$ the field frequency, $U_p$ is the ponderomotive energy, $n$ is an integer and $I_p$ the ionization potential. The remaining frequencies will be averaged out.  In photoelectron momentum distributions, this inter-cycle interference will manifest itself as concentric rings centered around the origin of the plane spanned by the electron momentum components parallel and perpendicular to the driving-field polarization. In the limit of an infinitely long pulse $N_c \rightarrow \infty$, intercycle interference is described by a Dirac-delta comb \cite{Becker2002Review}, i.e., the peaks given by (\ref{eq:atirings1}) are infinitely sharp. In \cite{Maxwell2017} we show that, for a  monochromatic field with a finite number $N_c$ of cycles, the probability modulation associated with inter-cycle interference reads as 
\begin{equation}
\Omega_{N_c}(\mathbf{p}_f)=\frac{\cos\left[\frac{2\pi i N_c}{\omega}\left(I_p+U_p+\frac{1}{2}\mathbf{p}_f^2\right)\right]-1}{\cos\left[\frac{2\pi i }{\omega}\left(I_p+U_p+\frac{1}{2}\mathbf{p}_f^2\right)\right]-1} \label{eq:nintercycle}
\end{equation}
and that the ATI rings remain invariant if the Coulomb potential is incorporated.

\subsection{Temporal and spatial double slits}
\label{subsec:doubleslits}
Temporal and spatial double slits are interference patterns that were first defined for direct ATI, in a theoretical framework for which the residual Coulomb potential is neglected in the electron propagation. A well-known approach that allows for tunneling and quantum interference and falls within this category is the strong-field approximation (SFA). The SFA neglects the electric field when the electron is bound and the binding potential when the electron is in the continuum. Formally, the SFA may be viewed as a Born series with a field-dressed  basis, for which there is a clear-cut definition of whether rescattering events are present or absent. \textit{Direct} ATI electrons correspond to the zeroth order term of such a series, and do not undergo any act of rescattering.
The energy of the direct ATI electrons extends up to $2U_p$. This corresponds to the maximal kinetic energy that a classical electron may acquire in the absence of the Coulomb potential. \textit{Rescattered} ATI electrons correspond to the events determined by the first-order term of the abovementioned series, for which a single act of rescattering is present.

A \textit{spatial} double-slit type of interference is present for direct ATI in aligned diatomic molecules. There will be electron emission at different centers in the molecule, so that the associated transition amplitudes will carry different phases. This will leave quantum-interference  imprints in the ATI spectra and photoelectron angular distributions, which will provide structural information about the molecule, such as 
the internuclear distance \cite{Hetzheim2007}. This idea has been initially proposed in \cite{Muth2000}, and subsequently explored by several research groups both theoretically \cite{Jaron2004,Milos2006,Hetzheim2007,Lai2013,Suarez2016,Suarez2018} and experimentally \cite{Ansari2008,Zhang2018,Kunitski2019}. Since the mid 2000s, it has also been generalized to account for more complex types of molecules, with many scattering centers, multi-electron dynamics, coupling of different degrees of freedom, and different orbital shapes (for a review see, e.g., \cite{Lein2007}). Spatial double slits are also present for high-order harmonic generation \cite{Augstein2012}.

The \textit{temporal} double slit involves the interference of the two different types of trajectories that occur in direct ATI, i.e.,  the long and the short orbits. An electron along the short orbit will be released in the continuum towards the detector, while the electronic wave packet following the long orbit will start on the opposite side of the ion, return and eventually reach the detector \cite{Lindner2005,Arbo2006,Arbo2012}. The electric field at both instants will have opposite signs and the same amplitude.   Diffraction patterns coming from ATI direct trajectories also carry information about the geometry of the bound state from which the electron has been released, and can be manipulated using tailored fields.   This is exemplified in our previous work for both the temporal and spatial double slits in elliptically polarized fields \cite{Lai2013}. It is important to stress that, for the long direct ATI orbit, return is allowed but not rescattering. This is a particularity of the methods used in these studies, which neglect the residual binding potential in the continuum. If the residual Coulomb potential is incorporated, the distinction between direct and rescattered electrons become blurred.   More details will be provided in Sec. \ref{sec:orbits}.

The role of rescattering in laser-induced electron diffraction in diatomics has been first proposed in \cite{Lein2002}. Therein, the emphasis was on how recollision at spatially separated centers would affect the above-threshold ionization photoelectron angular distributions. Nonetheless, key ideas of how to construct simplified classical models in order to compute phase shifts are already to be found. Similar ideas have been used in \cite{Bian2011} in order to construct holographic patterns.

\subsection{Rescattered trajectories and photoelectron holography}

Photoelectron holography uses the phase difference between a recolliding (``the probe'') and a direct (``the reference'') ATI electronic wave packet to map a specific target, as briefly mentioned in Ref.~\cite{Spanner2004}. These wavepackets may be associated with different types of trajectories, employing the recollision physical picture. This seminal paper touches upon the possibility of photoelectron holography as an imaging tool and hints at it being able to resolve amplitudes and phases. Its main focus is however on images obtained by electron-self diffraction of atoms and diatomic molecules, and on how to maximize their resolution. A detailed account of the possible distortions that may be caused by the external driving field is provided, together with prescriptions for avoiding them. 
Nonetheless, there is an extensive discussion of the types of interference that may occur, and on rescattered trajectories, which are classified as forward-scattered and backscattered according to their scattering angle.  Backscattered trajectories may lead to very high photoelectron energies and have been widely studied since the 1990s, in the context of the ATI plateau\footnote{The plateau is a well-known structure with ATI peaks of comparable intensities, which may extend up to the energy of $10U_p$. For a review see \cite{Becker2018}, and for seminal plateau measurements see \cite{Paulus1994}.}. Forward-scattered trajectories, in contrast, typically yield a much lower photoelectron energy range, which is comparable to that of the direct ATI trajectories. They have only called the attention of the strong-field community much more recently. This sudden shift of focus may be attributed to two main reasons:
\begin{itemize}
	\item \textit{ The existence of a myriad of low-energy structures, which have been identified in experiments and theoretical studies.} Within the past decade, there have been several measurements of  near-threshold enhancements in ATI spectra for long-wavelength driving fields. An example of such enhancements is the so-called Low-Energy Structure (LES) reported in \cite{Blaga2009} and attributed to the interplay between the binding potential and the external field.  Other examples are the very low-energy structure (VLES)\cite{Quan2009} and zero energy structure (ZES)\footnote{Sometimes called the near-zero energy structure (NZES)\cite{Pisanty2016a}} \cite{Wolter2014}. Although classical in nature, these features require a strong interaction with the core, which may be viewed as scattering.  Indeed, in \cite{Rost2012} it has been shown that the energy bunching of neighboring low-energy trajectories that turn around the core lead to a series of peaks whose existence neither depends on the range nor on the shape of the potential \cite{Rost2012,Lemell2012}. The binding potential however serves to focus these trajectories and determines the strength and absolute position of these peaks \cite{Lemell2012}. Support for this interpretation has been provided in subsequent work, in which a detailed investigation of the LES has been performed using classical trajectories, analytical models and a systematic mapping of the initial conditions on the plane of the final momentum \cite{Kelvich2016}.  Therein, it has been shown that the binding potential considerably influences the transverse and longitudinal momentum components, and the role of Coulomb focusing and Coulomb defocusing is investigated. The residual Coulomb potential leads to looplike orbits that give rise to a caustic, and  a bunching similar to that described in \cite{Rost2012} occurs. The changes in the electron transverse momenta caused by the potential also play an important role. The existence of caustics, orbits whose transverse momenta are modified by the potential, and their relation to the LES have also been discussed                                                                                                                                                                                                                                                                                                                                                                                                                                                                                                                                                                                                                                                                                                                                                                                                                                                                                                                                                                                                                                                                                                                                                                                                                                                                                                                                                                                                                                                                                                                                                                                                                                                                                                                                                                                                                                                                                                                                                                                                                                                                                                                                                                                                                                                                                                                                                 in \cite{Yan2010,Danek2018}. Within the context of the analytical R-Matrix theory (ARM), the LES and the near-zero energy structure are associated with steep changes in the imaginary part of the action, which may be related to the presence of cusps \cite{Pisanty2016,Pisanty2016a}.
	Multiple forward-scattering events in the LES have been addressed in \cite{Liu2010}. For seminal work on Coulomb focusing see, e.g., \cite{Brabec1996}.
	
	\item \textit{Recent SFA results for Coulomb-type potentials show that the contributions from the rescattered trajectories are not obfuscated by those of the direct trajectories.}  It was widely believed that, for photoelectrons whose energies extend up to the direct-ATI cutoff energy $2U_p$, the direct electron contributions would prevail and thus obfuscate those from scattered trajectories. A largely overlooked issue is however the  large scattering cross section of the Coulomb potential, which enhances the contributions from rescattered electrons by several orders of magnitude. Extensive studies of such types of rescattering have been done in the context of photoelectron holography. Features such as a three-pronged fork-type structure and rhombi that occur for scattering angles almost orthogonal to the polarization axis have been associated with such trajectories \cite{Moeller2014,Becker2015}.
\end{itemize}

\begin{figure*}
	\centering
	\includegraphics[width=0.75\linewidth]{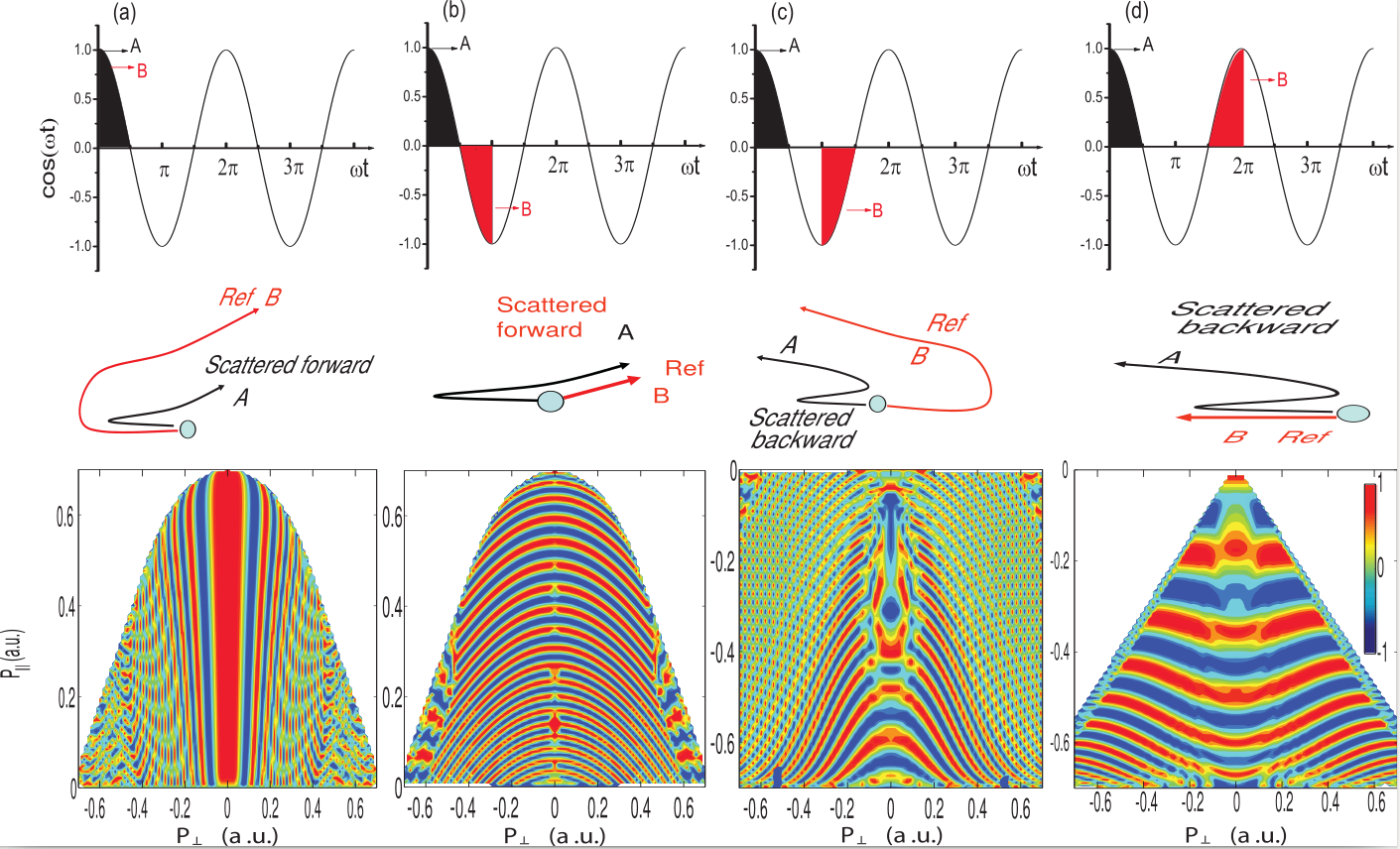}
	\caption{Fig.~3 of \cite{Bian2011} showing the various types of holographic patterns obtained from different types of interfering orbits. Here the notations A and B refer to the probe and reference orbits, respectively.}
	\label{fig:seminal}
\end{figure*}

The interference between direct 
and rescattered 
trajectories and its relation to photoelectron holography was first explored in detail in \cite{Bian2011}. Therein, four types of holographic structures, provided for clarity in Fig.~\ref{fig:seminal}, have been obtained assuming several types of intra-cycle interference. The phase shifts between the probe and the reference depend on the instant of ionization and recollision.  Explicitly, holographic patterns are attributed to the interference between (i) long direct orbits and forward scattered trajectories; (ii) short direct orbits and forward scattered trajectories; (iii) long direct orbits and backscattered trajectories; (iv)  short direct orbits and backscattered trajectories. Therein, it is argued that only pattern (i) would be easily identified in experiments, as it is located along the polarization axis. 
One should note, however, that the patterns in \cite{Bian2011}  markedly differ from those in experiments and in \textit{ab-initio} computations. An exception is perhaps pattern (i), which resembles the spider-like structure identified in \cite{HuismansScience2011}, and is attributed to the same type of interference (see discussion below). Nonetheless, the model in \cite{Bian2011} does not reproduce the converging fringes near the origin. 

This follows from the fact that, in \cite{Bian2011},  simplifying assumptions have been performed. First, the influence of the residual Coulomb potential has been neglected, which means that the real topology of the orbits and the Coulomb phases are not taken into consideration. Second, the distance from the origin at which the electron is born and the ionization rates are  kept fixed. This means that contributions from ionization times close to a field crossing are overestimated. Recent work \cite{Shilovski2018} has shown that, if the residual Coulomb potential is taken into consideration, one must considerably change the time intervals used in order to reproduce the features reported in \cite{Bian2011}.
\subsection{Most common holographic structures} 

\subsubsection{The spider.}

The so-called ``spider'' is possibly the best known  ultrafast holographic structure. It consists of broad fringes nearly parallel to the polarization  axis extending up to very high photoelectron energies. It has been first measured in ATI from a metastable state of Xenon in fields of intensities of the order of $10^{11}\mathrm{W/cm}^2$ and wavelengths around $7\mu m$ \cite{HuismansScience2011,Bian2011} (for even lower frequencies see \cite{Huismans2012}). Shortly thereafter, this structure has been identified experimentally for ATI in rare gases starting from the ground state, with near- and mid-IR driving fields \cite{Marchenko2011,Hickstein2012} of much higher intensities ($\sim 10^{13}\mathrm{W/cm}^2$) and has been reported by several groups in atoms \cite{Moeller2014} and molecules \cite{Marchenko2011,Meckel2014}. Traces of the spider can also be seen in \cite{Bergues2011,Weger2013}.
\begin{figure}
	\centering
	\includegraphics[width=0.5\linewidth]{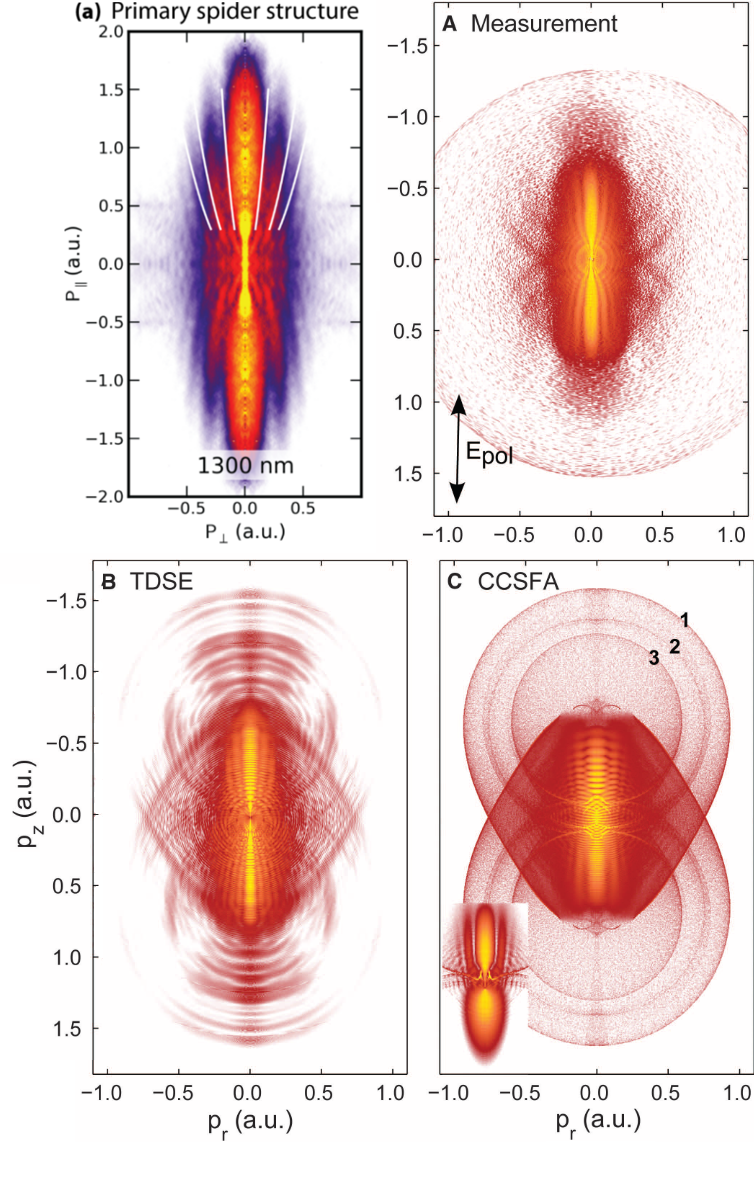}
	\caption{Photoelectron angular distributions showing the the spider-like structure. The upper left panel was taken from the experimental results in \cite{Hickstein2012}, while the remaining panels are from the seminal paper \cite{HuismansScience2011}. The latter three panels (A-C) show the experimental results and
		distributions from computations performed with \textit{ab-initio} methods and the Coulomb-corrected strong-field approximation (CCSFA).}
	\label{fig:firstspider}
\end{figure}

In \cite{HuismansScience2011}, the phase differences observed have also been associated with the interference of two types of forward-scattered trajectories starting from the same half cycle of the driving field. A simple model of this structure that considers the interference of a plane wave with a Coulomb scattering wave has been proposed in \cite{Hickstein2012}. It reproduced not only the spider, but also inner spider-like structures that occur near the ionization threshold and are associated with multiply recolliding orbits. Some features observed in this structure, such as the fringes parallel to the driving-field polarization, as well as the contributing orbits, were predicted qualitatively and compared with the experiments \cite{Bian2011}. This early model was however too simple to account for the diverging patterns close to the threshold and the finer details of the spider-like pattern. Trajectory based models that allow for the residual Coulomb potential and carry quantum phases, such as the Coulomb Corrected SFA (CCSFA) and its variations \cite{Popruzhenko2008a,Yan2010,Yan2012}, the Coulomb Quantum-orbit Strong-Field Approximation (CQSFA) \cite{Lai2015a,Maxwell2017,Maxwell2017a,Maxwell2018} and the Quantum Trajectory Monte Carlo method (QTMC) \cite{Li2014} reproduce the spider in striking detail, in agreement with experiments and \textit{ab-initio} methods.  An alternative explanation for the spider in terms of glory rescattering has been proposed in \cite{Xia2018}.
The type of interference that leads to the spider has also been associated with side lobes encountered in angle-resolved ATI photoelectron distributions \cite{HuismansScience2011}. Studies of how the spider behaves with regard to the field parameters have been reported in \cite{Huismans2012}.

It has been initially argued that the spider-like structure is not very sensitive to the target due to being observed for many different species, and stemming from the interference of two distinct types of forward scattered trajectories, whose interaction with the core is brief.  Further studies however reveal that the spider-like fringes exhibit an offset for aligned molecules \cite{Meckel2014,He2018}, and carry information about nodal planes  and rotational degrees of freedom \cite{Walt2017}. An \textit{ab-initio} computation of photoelectron momentum distributions for diatomic molecules has in fact found that the spider and other structures are very sensitive with regard to the molecular orientation, nodal planes and the coupling of different continua \cite{Fernandez2009}.  For traces of the spider in the context of a multielectron computation for $CO_2$ see \cite{Majety2017}. Recently, it has been proposed that phase differences in the spider-like pattern may be used to resolve   electron motion in molecules by preparing the system in a non-stationary coherent superposition of states \cite{He2018}.

\subsubsection{The fishbone structure.}
\begin{figure}
	\centering
	\includegraphics[width=1.0\linewidth]{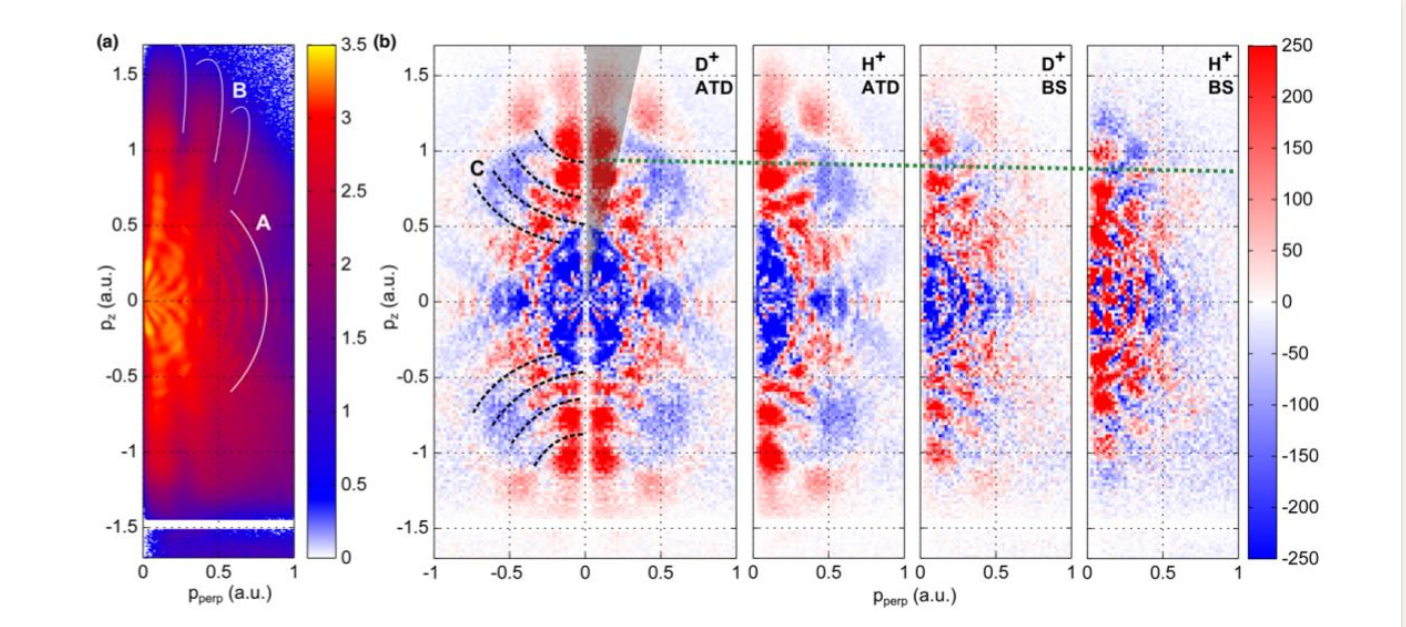}
	\caption{Measured photoelectron momentum distribution for aligned $D_2$ showing the ATI rings and the spider [panel (a)], together with differential momentum distributions showing the fishbone structure [panel (b)] for $H_2$ and $D_2$. The coordinates $p_z$ and $p_{\mathrm{perp}}$ give the momentum components parallel and perpendicular to the laser-field polarization. From  \cite{Haertelt2016}.  }
	\label{fig:fishbone}
\end{figure}

A less prominent holographic pattern is a fishbone structure that forms for moderate to high photoelectron energies, with fringes nearly perpendicular to the polarization axis. This structure stems from the interference of direct and backscattered trajectories. Since backscattered trajectories spend in principle a longer time near the core region, it is expected that they will be more sensitive to the target structure than  the forward-scattered trajectories that lead to the spider \cite{Bian2012,Haertelt2016,Li2015}.  Typically, however, the fishbone structure is obfuscated by the spider-like fringes, so that the latter must artificially be removed. A simple SFA computation proposes several possible types of interfering trajectories for this structure \cite{Li2015}. Therein, however, several disagreements with \textit{ab-initio} methods are pointed out and related to the absence of the residual Coulomb potential in the model. Recently, we have shown that the presence of the Coulomb potential will lead to spiral-like structures that are also usually obfuscated by the spider, but may be identified for large scattering angles \cite{Maxwell2017a,Maxwell2018}.  Within the SFA framework, it has also been proposed that structures involving backscattered trajectories could be retrieved in a realistic setting by using orthogonally polarized fields \cite{Li2016}. For an early experimental example of the fishbone structure see Fig.~\ref{fig:fishbone}.
\subsubsection{The fan.}
\label{sec:fan}

\begin{figure}
	\centering
	\includegraphics[width=0.5\linewidth]{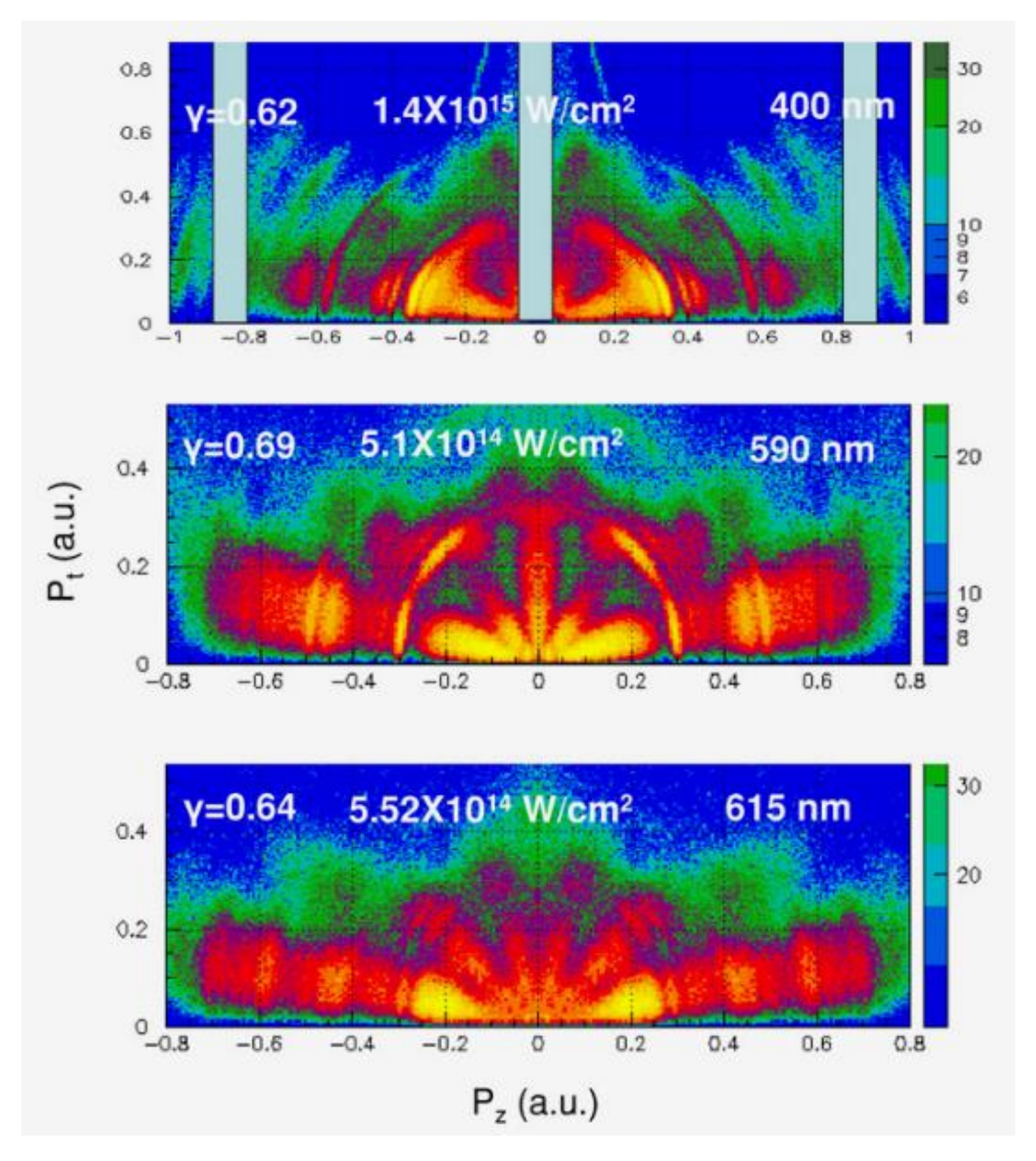}
	\caption{Photoelectron angular distributions showing the near-threshold, fan-like structure measured in the seminal publication \cite{Maharjan2006}. The momentum components $P_z$ and $P_t$ are oriented parallel and perpendicular to the driving-field polarization. }
	\label{fig:fanexp}
\end{figure}
The ``fan'' is a widely known pattern consisting of radial interference fringes, which forms near the ionization threshold \cite{Rudenko2004,Maharjan2006,Gopal2009}. This pattern does not occur for short-range potentials, so that it requires the interplay between the laser field and the Coulomb potential to be able to form. In fact, it is absent both in SFA \cite{Chen2006,Arbo2008} computations and in the photodetachment of negative ions \cite{Korneev2012,Shearer2013,Law2016}, which show fringes nearly perpendicular to the driving-field polarization in this region. In the literature, the fan-shaped structure has been interpreted as a resonant process involving intermediate bound states with a specific angular momentum, both in the seminal experimental work \cite{Rudenko2004,Maharjan2006} and in subsequent theoretical papers \cite{Arbo2006,Chen2006,Arbo2008,Arbo2008b}. Specifically, in \cite{Rudenko2004} the fan's degradation for few-cycle pulses suggested a resonance-like character, and in 
\cite{Arbo2006} the fan has been related to Ramsauer-Townsend fringes, by comparing it to \textit{ab-initio} methods and with CMTC computations. Therein, laser-dressed hyperbolic orbits with neighboring angular momenta have been identified as those responsible for the fan. One should note, however, that classical-trajectory methods do not allow for quantum interference. This means that only indirect statements about how the fan forms can be inferred from this method. 

In \cite{Arbo2008b,Chen2006}, the number of fringes in the fan have been related to a specific angular momentum, which gives the minimal number of photons necessary for the electron to reach the continuum, and empirical rules for predicting the number of fringes have been provided. These rules, however, work well only in the multiphoton regime\footnote{Roughly speaking, the multiphotoen regine is characterized by a Keldysh parameter $\gamma=\sqrt{I_p/(2U_p)}>1$, where $I_p$ is the target's ionization potential. In this regime, tunneling is not the prevalent ionization mechanism, but, rather, the electron is freed by multiphoton absorption. In the multiphoton regime, resonances are extremely important. For a a more rigorous definition see \cite{Popruzhenko2014a}.}. This, together with the fact that he fan-shaped structure is also present for a wide range of species, much lower frequencies and higher intensities \cite{Maharjan2006}, indicates that resonances cannot be the sole mechanism behind it. 

It is worth mentioning that the fan-shaped structure has also been identified using orbit-based methods that allow for quantum interference, such as the Coulomb-Volkov Approximation (CVA)\cite{Arbo2008}, the Coulomb-corrected strong-field approximation (CCSFA) \cite{Yan2010,Yan2012}, the Quantum Trajectory Monte Carlo (QTMC) method  \cite{Li2014} the semiclassical two-step model (SCTS) for strong-field ionization \cite{Shvetsov2016}, and the Coulomb Quantum-orbit Strong-Field Approximation (CQSFA) \cite{Lai2017,Maxwell2017,Maxwell2017a,Maxwell2018,Maxwell2018b}. However, if such methods employ Coulomb-distorted trajectories and solve the direct problem (see discussion in Sec.~\ref{sec:theory}), they require a huge number of contributed orbits to obtain converged photoelectron momentum distributions. This makes it difficult to ascertain how the fan forms. The CQSFA is an exception as it only needs a few contributed trajectories for each value of the final momentum. This means that they can be switched on and off at will. 
For that reason, in our previous publications \cite{Lai2017,Maxwell2017,Maxwell2017a,Maxwell2018} we have focused on the question of what types of orbits are responsible for specific patterns, including the fan.  We have found that the fan  may be viewed as a holographic pattern resulting from the interference of direct with forward deflected trajectories. The latter undergo an angular-dependent distortion due to the presence of the Coulomb potential, which is maximal for momenta parallel to the laser-field polarization, and cancels out for momenta perpendicular to it. These distortions take place in a fraction of the laser field, which are much shorter than typical timescales for which resonances occur. For short-range potentials or the SFA, these distortions are absent. In this case, one may speak of return, but not of deflection, and the temporal double slit is recovered. The CVA obtains the fan by considering Coulomb distortions in the final continuum state, but keeps the same orbits as the SFA. Thus, it does not assess how the Coulomb potential modifies the orbits \cite{Arbo2008}. 
One should also note that the right number of fringes has only been obtained with the CQSFA and the SCTC methods. This is made possible due to an extra phase, which was overlooked in the remaining orbit-based methods mentioned in this section.  More details about this phase will be provided in Sec.~\ref{sec:CQSFA} and can be found in \cite{Shvetsov2016,Lai2017,Maxwell2017}. The CQSFA, in conjunction with analytic approximations, is also an excellent means to disentangle sub-barrier and continuum distortions and the associated phase differences \cite{Maxwell2017a}.
Finally, near-threshold fan-shaped fringes have also been observed in molecules \cite{Veltheim2013,Mi2017}. 
Interesting features encountered therein are a marked enhancement in the fan-like fringes, which has been related to electronic-nuclear coupling in $H_2$ \cite{Mi2017}, and a suppression associated to the population of Rydberg states in dimers \cite{Veltheim2013}.

\subsubsection{Fork and Rhombi-like structures.}
\begin{figure}
	\centering
	\includegraphics[width=\linewidth]{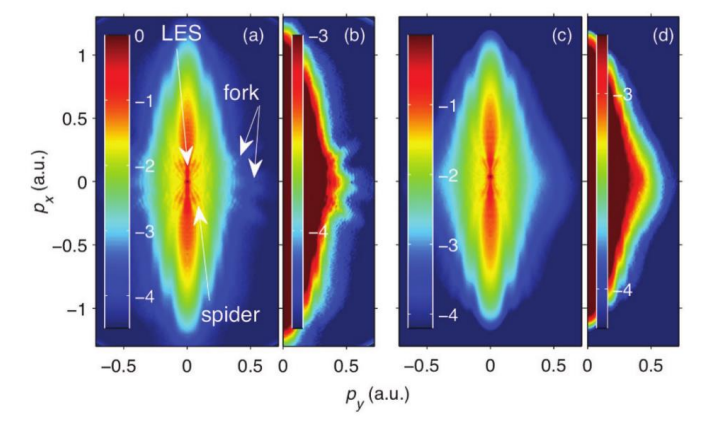}
	\caption{Photoelectron angular distribution showing the fork-like structure measured in \cite{Moeller2014}, together with traces of the spider and of the low-energy structure. The momentum components $p_x$ and $p_y$ are oriented parallel and perpendicular to the driving-field polarization.   }
	\label{fig:fork}
\end{figure}

In \cite{Moeller2014}, a three-pronged fork-like structure was observed experimentally for photoelectron momentum distributions using xenon in long, mid-IR driving pulses (see Fig.~\ref{fig:fork} for a depiction of this structure). This structure is nearly orthogonal to the laser polarization axis and is very sensitive to the pulse length and frequency. In fact, it is markedly enhanced with increasing wavelength and disappears for few-cycle pulses. Physically, it has been explained in terms of low-energy forward scattered trajectories. An SFA computation has shown that this structure is universal and does not necessarily require Coulomb-type potentials to form. However, the divergent Coulomb scattering cross section makes it visible by allowing it to rise above the contributions of the direct electrons.  

A key difference between the fork and the previous holographic patterns is that its existence is of a classical nature, similar to that leading to the low-energy structures (LES). The fork is not caused by quantum-interference effects, but, rather, stems from the kinematical constraints imposed upon the electron's returning energies. This is particularly important for orbits whose excursion times are much longer than a field cycle. This explains their absence for few-cycle pulses. Specifically, SFA studies provide a detailed analysis of the role of forward- and backscattered trajectories in the energy ranges for which the fork appears, and show that it is closely related to the LES \cite{Becker2015}. This raises the question of whether the presence of the Coulomb potential is a necessary condition for the LES to arise. Therein, rhombi-like structures of similar origin have also been identified.  To the present date, there is no study of the influence of the residual Coulomb potential on the fork- or rhombi-shaped structure. Because, however, the fork occurs for momenta nearly perpendicular to the field-polarization axis, the phase distortions introduced by the Coulomb potential when the electron is in the continuum are expected to cancel out \cite{Lai2017}.  This very likely explains the excellent agreement of the SFA with existing experiments in which the fork is observed.

\section{Theoretical methods}
\label{sec:theory}
In order to compute ATI spectra or electron momentum distributions, one must determine the evolution of an electronic wave packet from a bound state to a continuum state associated to a final momentum $\mathbf{p}_f$ when the electron reaches the detector. This evolution is given by the time-dependent Schr\"odinger equation, which can either be solved numerically, or approximately. For one-electron atoms, the \textit{ab-initio} solution of the TDSE is meanwhile  straightforward, and there are well-established Schr\"odinger solvers available to the strong-field community \cite{qprop,Mosert2016,Patchkovskii2016}. For multielectron systems, however, \textit{ab-initio} solutions are a formidable task. Key issues are how to model electron-electron correlation, the core dynamics, and the coupling of different degrees of freedom. A key difficulty of a full \textit{ab-initio} solution is the so called ``exponential wall'', i.e., the fact that the numerical effort increases exponentially with regard to the degrees of freedom. This implies that a series of approximations must be made in order to render the problem tractable. Examples are the use of the time-dependent density functional theory \cite{Runge1984,Burke2005}, trajectory-based grids \cite{Shalashilin2000,Shalashilin2008}, the time-dependent restricted-active-space self-consistent-field theory with space partition (TD-RASSCF-SP) \cite{Miyagi2017}, and the R Matrix with time dependence \cite{Hassouneh2015,Clarke2018}. For discussions of classical, semiclassical, quantum and \textit{ab-initio} methods for correlated two-electron systems in the context of laser-induced nonsequential double ionization (NSDI) see the review \cite{Faria2011}.
Here, we will consider single-electron systems and, unless otherwise stated, use atomic units throughout.  

Our starting point will be the Hamiltonian $H(t)=H_a+H_I(t)$, of an electron under the influence of an external laser field and a binding potential. The field-free atomic Hamiltonian is given by 
\begin{equation}
H_a=\frac{\hat{\mathbf{p}}^{2}}{2}+V(\hat{\mathbf{r}})
\label{eq:H}
\end{equation}
and $H_I(t)$ gives the interaction with the external field. In Eq.~(\ref{eq:H}), $\hat{\mathbf{r}}$ and $\hat{\mathbf{p}}$ denote the position and momentum operators, respectively. The binding potential  $V(\hat{\mathbf{r}})$ is chosen to be of Coulomb type, i.e.,  
\begin{equation}
V(\hat{\mathbf{r}})=-\frac{C}{\sqrt{\hat{\mathbf{r}}\cdot
		\hat{\mathbf{r}}}},\label{eq:potential}
\end{equation} 
where $0\leq C\leq 1$ is an effective coupling. The interaction Hamiltonian in the length and velocity gauge read
\begin{equation}
H_I(t)=-\hat{\mathbf{r}}\cdot \mathbf{E}(t)
\end{equation}
and
\begin{equation}
H_I(t)=\hat{\mathbf{p}}\cdot \mathbf{A}(t)+\mathbf{A}^2/2,
\end{equation}
respectively, where $\mathbf{E}(t)=-d\mathbf{A}(t)/dt $
is the electric field of the external laser field and $\mathbf{A}(t)$ the corresponding vector potential.  The evolution of the system is described by the time-dependent Schr\"odinger equation
\begin{equation}
i\partial_t|\psi(t)\rangle=H(t)|\psi(t)\rangle \,,\label{eq:Schroedinger}
\end{equation}
which is either solved fully numerically in a specific basis set or computed semi-analytically using a series of approximations. Some of them will be discussed below, but our emphasis will be the strong field approximation and beyond, more specifically the Coulomb Quantum-orbit Strong-Field Approximation (CQSFA). 
It is instructive to write the time-dependent Schr\"odinger equation (\ref{eq:Schroedinger}) in integral form using time evolution operators. This leads to \begin{equation}
U(t,t_0)=U_a(t,t_0)-i\int^t_{t_0}U(t,t^{\prime})H_I(t^{\prime})U_a(t^{\prime},t_0)dt^{\prime}\,
,\label{eq:Dyson}
\end{equation}
where the time evolution operators $U_a(t,t_0)=\exp[iH_a (t-t_0)]$ and 
\begin{equation}
U(t,t_0)=\mathcal{T}\exp \bigg [i \int^t_{t_0}H(t^{\prime})dt^{\prime} \bigg],
\label{eq:fullU}
\end{equation}
where $\mathcal{T}$ denotes time-ordering, are associated with the field-free and full Hamiltonians, respectively, evolving from an initial time $t_0$ to a final time $t$.  Above-threshold ionization requires computing the transition amplitude $\left\langle\psi_{\textbf{p}_f}(t) |U(t,t_0) |\psi_{0} \right\rangle $ from a bound state $\left\vert \psi _{0}\right\rangle $ to a final continuum state $ |\psi_{\textbf{p}_f}(t)\rangle$ with momentum $\mathbf{p}_f$, which may be written as 
\begin{equation}
M(\mathbf{p}_f)=-i \lim_{t\rightarrow \infty} \int_{-\infty }^{t }d
t^{\prime}\left\langle \psi_{\textbf{p}_f}(t)
|U(t,t^{\prime})H_I(t^{\prime})| \psi _0(t^{\prime})\right\rangle \,
,\label{eq:transitionamp}
\end{equation}
with $\left | \psi _0(t^{\prime})\right\rangle=\exp[iI_pt']\left\vert \psi _{0}\right\rangle $, where $I_p$ is the ionization potential. One should note that Eq.~(\ref{eq:transitionamp}) is formally exact, and it is used to construct many of the approaches presented below.  Perturbation theory with regard to the laser field is obtained by iterating Eq.~(\ref{eq:Dyson}). This specific series works well for weak fields, but breaks down in the parameter range of interest. Below we will briefly discuss the theoretical orbit-based methods whose starting point is Eq.~(\ref{eq:transitionamp}) approximating the time-evolution operator (\ref{eq:Dyson}). We will examine the SFA and CQSFA in more depth, as these are the approaches that lead to our main results and to our publications that are revised in this work.

\subsection{Strong-field approximation}

The strong-field approximation, also known as the Keldysh-Faisal-Reiss (KFR) \cite{Faisal1973,Reiss1980,Keldysh1965} theory, is the most widespread semi-analytical approach employed in the modelling of strong-field ionization. In its standard form, it consists in replacing the full time evolution operator (\ref{eq:fullU}) by the Gordon-Volkov \cite{Gordon1926,Wolkow1935} time-evolution operator $U^{(GV)}$ related to the Hamiltonian 
\begin{equation}
H^{(GV)}(t)=\frac{\hat{\mathbf{p}}^{2}}{2}+H_I(t)
\label{eq:VolkovH}
\end{equation}
of a free particle in the presence of the laser field.  This brings a key advantage as the time dependent Schr\"odinger equation using the Hamiltonian (\ref{eq:VolkovH}) can be solved analytically.  This procedure will lead to the SFA transition amplitude 
for the ATI direct electrons.

The SFA may also be generalized to account for rescattering by proceeding as follows. One commences by noticing that the full time-evolution operator may also be written as 
\begin{equation}
U(t,t_0)=U^{(GV)}(t,t_0)-i\int^t_{t_0}U(t,t^{\prime})VU^{(GV)}(t^{\prime},t_0)dt^{\prime}\,
,\label{eq:Dyson2}
\end{equation}
in terms of the Gordon-Volkov time evolution operator. If one now iterates Eq.~(\ref{eq:Dyson2}) to first order in $V$ and inserts the resulting expression in Eq.~(\ref{eq:Dyson}), the zeroth and the first-order term in the series will give the direct and the rescattered electrons when the resulting time evolution operator is employed in (\ref{eq:transitionamp}). It is commonly accepted that the ``rescattered electrons'' contain one act of rescattering. One may also write the direct ATI transition amplitudes in terms of the binding potential $V$. Explicitly, this yields
\begin{equation}
M_{d}(\mathbf{p})=-i \int_{-\infty}^{\infty}dt' \langle \mathbf{p}+\mathbf{A}(t')|V|\psi_{0}\rangle e^{i S_{d}(\mathbf{p},t,t')},
\label{eq:ATIdirect}
\end{equation}
where
\begin{equation}  
S_d(\mathbf{p},t,t')=-\lim\limits_{t\rightarrow \infty} \frac{1}{2} \int^{t}_{t'}[\mathbf{p}+\mathbf{A}(\tau)]^2 d\tau + I_pt'\label{eq:Sdir}
\end{equation}
is the action describing a process in which an electron is freed in the continuum by tunnel ionization and reaches the detector without further interaction.  For a thorough explanation and detailed derivations we refer to \cite{Fring1996,Becker1997,Lohr1997,Ivanov2005}. 
One may also combine the direct and rescattered ATI transition amplitudes in order to obtain the expression
\begin{equation}
M_{r}(\mathbf{p})=- \int^{\infty}_{-\infty} dt \int^{t}_{-\infty} dt'\int d^3k\exp[iS_{r}(\mathbf{p},\mathbf{k},t,t')]V_{\mathbf{k}0}V_{\mathbf{pk}},
\label{eq:rescSFA}
\end{equation}
which allows for up to one act of rescattering and contains, apart from the ionization time $t'$ and the final time $t$, the intermediate momentum $\mathbf{k}$ as integration variables. Thereby, the action reads
\begin{align}
&S_{r}(\mathbf{p},\mathbf{k},t,t')=\notag\\&-\frac{1}{2} \int^{\infty}_{t}[\mathbf{p}+\mathbf{A}(\tau)]^2 d\tau -\frac{1}{2} \int^{t}_{t'}[\mathbf{k}+\mathbf{A}(\tau)]^2 d\tau + I_pt'.
\label{eq:Sresc}
\end{align}
In the SFA, the influence of the core is incorporated in the ionization prefactor
\begin{equation}
V_{\mathbf{k}0}=\left\langle \mathbf{k}+\mathbf{A}(t)\right|V \left| \psi_0 \right \rangle
\end{equation}
and in the rescattering prefactor 
\begin{equation}
V_{\mathbf{pk}}=\left\langle \mathbf{p}+\mathbf{A}(t)\right|V \left| \mathbf{k}+\mathbf{A}(t) \right \rangle.
\end{equation}
Although Eq. (\ref{eq:rescSFA}) also contains direct electrons, it is more convenient to employ the transition amplitude (\ref{eq:ATIdirect}) in this case. Calculating the direct electrons from Eq.~(\ref{eq:rescSFA}) would require the computation of limits and is thus more cumbersome. This is referred to by some groups as ISFA (improved SFA).

A very intuitive description can be obtained if Eqs.~(\ref{eq:ATIdirect}) and (\ref{eq:rescSFA}) are solved using saddle-point methods. This requires seeking the values of the ionization times $t'$, the rescattering times $t$ and of the intermediate momentum $\mathbf{k}$ so that the actions (\ref{eq:Sdir}) and (\ref{eq:Sresc}) are stationary, i.e., their derivatives with regard to such variables must vanish. The saddle-point equation obtained from the direct action $S_d(\mathbf{p},t')$ reads
\begin{equation}
[\mathbf{p}+\mathbf{A}(t')]^2 = - 2I_pt',
\label{eq:tunnelsfa}
\end{equation}
which gives the kinetic energy conservation at the ionization time. Eq.~(\ref{eq:tunnelsfa}) has no real solutions, as a direct consequence of tunnel ionization having no classical counterpart. If the condition $\partial S_{r}(\mathbf{p},\mathbf{k},t,t')/\partial t'=0$ is imposed upon (\ref{eq:Sresc}), a formally identical equation will follow, with the final momentum $\mathbf{p}$ being replaced by the intermediate momentum $\mathbf{k}$. This gives
\begin{equation}
\label{eq:tunnelsfaresc}
[\mathbf{k}+\mathbf{A}(t')]^2 = - 2I_pt'.
\end{equation}
The condition $\partial S_{r}(\mathbf{p},\mathbf{k},t,t')/\partial t=0$  leads to the conservation of energy
\begin{equation}
[\mathbf{p}+\mathbf{A}(t)]^2=[\mathbf{k}+\mathbf{A}(t)]^2
\label{eq:saddresc}
\end{equation}
upon recollision, and $\partial S_{r}(\mathbf{p},\mathbf{k},t,t')/\partial \mathbf{k}=\mathbf{0}$ restricts the electron's intermediate momentum  $\mathbf{k}$, such that it must return to the site of its release, i.e., the origin. Explicitly, 
\begin{equation}
\mathbf{k}=-\frac{1}{t-t'}\int_{t}^{t'}\mathbf{A}(\tau)d\tau.
\label{eq:saddk}
\end{equation}
Physically, the approximations introduced above have the following consequences:
\begin{itemize}
	\item \textit{The influence of the laser field is neglected when the electron is bound.} This means that the SFA does not take into consideration processes involving bound states only, such as excitation, the core dynamics, or field-induced distortions such as Stark shifts. It is also quite common to neglect bound-state depletion. One may however modify the SFA in order to include such effects to a certain extent. Bound-state polarization has for instance been introduced in \cite{Spiewanowksi2014}. Furthermore, electron-electron correlation and excitation have been incorporated in our previous work \cite{Faria2004} and \cite{Shaaran2010}, respectively. For a thorough review on these issues see \cite{Symphony}.
	\item \textit{The influence of the residual binding potential is neglected when the electron is in the continuum. } This implies that the field-dressed momentum is conserved, and that the influence of the core is incorporated via prefactors at a single point, namely the origin. Relaxing this approximation leads to several Coulomb-distorted approaches, whose overview will be provided here.
	\item \textit{Formally, the SFA may be viewed as the Born series with a modified, field-dressed basis.} This allows a clear distinction between \textit{direct} and \textit{rescattered} events.  This becomes even clearer if one looks at the saddle-point equations stated above and the corresponding actions. Eqs. (\ref{eq:Sdir}) and (\ref{eq:tunnelsfa}), in principle, allow return, but not rescattering, while Eqs. (\ref{eq:Sresc}), (\ref{eq:tunnelsfaresc}), (\ref{eq:saddresc}) and (\ref{eq:saddk}) allow a rescattering event to occur. 
	\item \textit{If the steepest descent method is used}, one gains an intuitive orbit-based interpretation but \textit{loses} the spatial and momentum widths at the instant of ionization and recombination that would stem from the uncertainty relation. Relaxing these approximations has shown some softening on phase jumps in molecular HHG \cite{Chirila2009} and quantitative differences in the HHG spectra \cite{Perez-Hernandez2009}.
\end{itemize}

One should note, however, that there exists a more general formulation of the SFA which takes into consideration exact scattering waves instead of Volkov states. This formulation has been developed in \cite{Lewenstein1995} and is reviewed in \cite{Symphony}.

\subsection{Beyond the strong-field approximation}
\label{sec:beyondsfa}

\begin{figure*}
	\centering
	\includegraphics[width=0.95\linewidth]{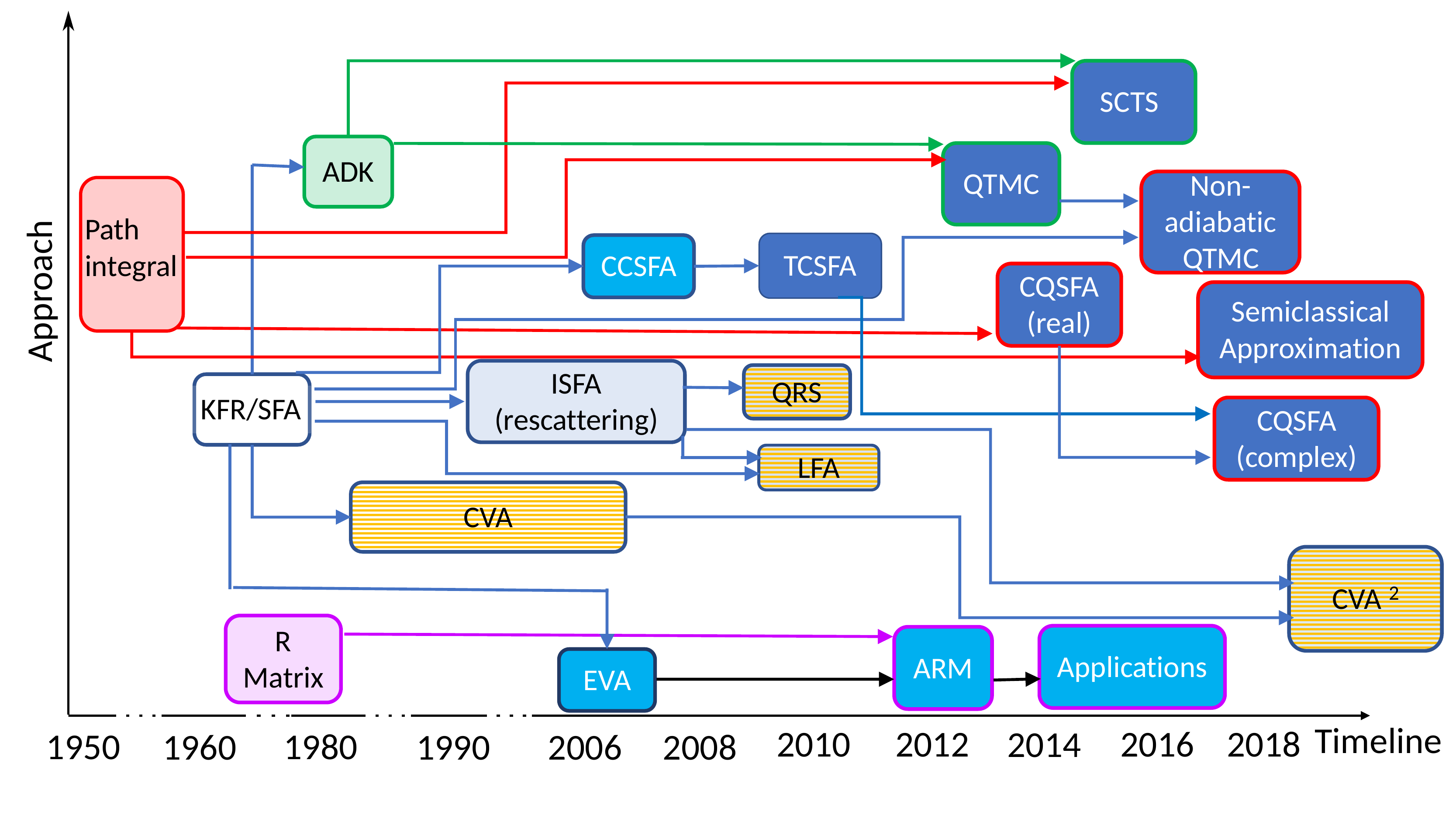}
	\caption{Timeline with a schematic ``family tree'' of the orbit-based approaches discussed in this review. The blue boxes signal changes in the orbits from the plain, direct SFA, by including rescattering (pale blue), incorporating the orbits approximately or perturbatively (sky blue), treating the Coulomb potential and laser field on equal footing (dark blue). The orange color indicates changes in the scattering states, which can either be performed via factorization (CVA, CVA$^2$, QRS), or by including higher-order terms in the field-dressed Born series (LFA). White and pale blue stripes indicate that the direct or rescattered SFA orbits are included, respectively. The outlines of the boxes and arrows indicate to which ``family'' a specific approach belongs. Dark blue outlines refer to approaches which used KFR/SFA methods as a starting point, red outlines refer to those derived from path integral methods using time slicing techniques, and green outlines indicate those for which, in addition to path integral methods, the quasi-static Ammosov-Delone Krainov (ADK) rate has been used to weight the initial distribution of orbits. For the Eikonal Volkov Approximation (EVA), we used a black outline as it is strongly based on the WKB approximation, although it is indirecty related to the KFR theory. The magenta boxes and contours indicate that a partition in space and boundary conditions have been employed, in the spirit of the R-Matrix theory. The approach in \cite{Milosevic2017} is called "Semiclassical Approximation" as it is constructed as the semiclassical limit of the path-integral transition amplitude, but no specific name so far has been given the author. Most approaches in the dark blue boxes use real trajectories in the continuum, with the exception of the complex CQSFA. This is explicitly indicated by the words ``real'' or ``complex''. }
	\label{fig:timeline}
\end{figure*}
In this section, we summarize several orbit-based theoretical methods and models that incorporate Coulomb effects in the continuum.  Some of them, such as the Coulomb Volkov approximation (Sec.~\ref{sec:CVA}) and the Quantitative Rescattering Theory (Sec.~\ref{sec:QRS}), rely on factorization and modified continuum states, but employs orbits from the SFA, either direct or rescattered. The low-frequency approximation (Sec.~\ref{sec:lfa}) is developed along similar lines, but employs a higher-order Born expansion.
Other methods either include Coulomb distortions in the orbits approximately or perturbatively. For instance, the Coulomb-corrected Strong-Field Approximation (CCSFA) (Sec.~\ref{sec:CCSFA}) expand the orbits perturbatively around the SFA, while the Eikonal Volkov Approximation (EVA) (Sec.~\ref{sec:eva}), employ a field-dressed Wentzel-Krammers-Brillouin (WKB) and further approximations in the scattering angles. This makes it possible to compute the orbits recursively starting from the SFA, and makes the EVA suitable for the outer region in the Analytical R-Matrix (ARM) theory (Sec.~\ref{sec:arm}). One may also treat the Coulomb potential and the laser field on equal footing by solving classical equations of motion in which both are present in order to describe the continuum propagation, but incorporating these equations in the action, so that quantum interference is included. This is the procedure taken in the Trajectory-Based Coulomb SFA (TCSFA) (Sec.~\ref{sec:TCSFA}), which is similar to the CCSFA, but non-perturbative with regard to the residual potential, and in path-integral strong-field approaches such as the Quantum Trajectory Monte Carlo (QTMC) method (Sec.~\ref{sec:QTMC}), the semiclassical two-step model (SCTS), the semiclassical approximation for strong-field processes, and the Coulomb-Quantum Orbit Strong-Field Approximation (CQSFA). The latter is the method employed by us and will be discussed in more detail in Sec.~\ref{sec:CQSFA}. In order to facilitate the discussion, a brief timeline with these methods is shown in Fig.~\ref{fig:timeline}. 

Further approximations and differences in implementation exist in all the above-mentioned methods, the most important of which are: 
\begin{itemize}
	\item\textit{Real orbits in the continuum.} All methods with full Coulomb distortion in the orbits  (see dark blue boxes in Fig.~\ref{fig:timeline}), except a very recent version of the CQSFA, employ real orbits in order to describe the electron's continuum propagation. This is due to the fact that complex orbits lead to branch cuts, which are tractable if the orbits are Coulomb free or if approximate methods are employed around the SFA, but are much more difficult to tackle otherwise. 
	\item\textit{ Adiabatic tunneling rates.} Because sub-barrier dynamics are not easy to address if Coulomb-corrected or Coulomb-distorted orbits are taken (sky blue and dark blue boxes in Fig.~\ref{fig:timeline}, respectively), some methods, such as the QTMC and the SCTS employ adiabatic, cycle-averaged tunneling rates to weight the initial orbit distributions (see dark blue boxes in Fig.~\ref{fig:timeline} outlined in green).  Methods that go beyond that must make several approximations in order to deal with this part of the problem, which includes cusps and singularities. A proper treatment of sub-barrier dynamics also requires complex orbits, and the above-mentioned tunneling rates provide a way to avoid them. 
	\item \textit{Direct vs inverse problem.} The full presence of the Coulomb potential in the continuum significantly alters the topology of the orbits. This makes it difficult to guess their shapes, relevance or main features. For that reason, methods such as the TCSFA, QTMC and SCTS solve the \textit{direct} problem, i.e., given the initial momentum distribution, they employ a shooting method and a large number of contributed orbits (typically $10^8-10^9$). Once these orbits have been propagated in time, they are binned according to the resulting final momentum. An advantage of this type of implementation is that it is versatile, as it can be easily adapted to different targets and driving-field shapes. Furthermore, the initial momentum distribution is also known. Two major disadvantages are however the presence of cusps and caustics if sub-barrier corrections are included, and the huge number of contributed orbits. Another possibility is to solve the \textit{inverse} problem, i.e., given a final momentum, find the corresponding initial momentum assuming a specific type of orbit and sub-barrier corrections. This approach is typically used if the Coulomb potential is only incorporated approximately around the SFA (see sky blue boxes in Fig.~\ref{fig:timeline}), and, for fully Coulomb-distorted orbits, in the CQSFA. Advantages of solving the inverse problem are: (i) a much smaller number of contributed orbits, typically four for fixed final momentum components; (ii) ``cleaner'' interference patterns and a better control over cusps and caustics, which facilitates the inclusion of sub-barrier corrections and complex trajectories. A drawbacks is the loss of versatility, i.e., prior knowledge of the approximate behavior of the orbits and precise initial guess conditions are required. This is because the inverse problem is harder to solve.
\end{itemize}
One should bear in mind that this is only a short summary.
A discussion of these methods, focused on their advantages, shortcomings, approximations, implementation and on their success in reproducing specific holographic structures, is provided below. For a summary of the the features included in many of these models see Table \ref{table:comparison} presented subsequent to the discussion.

\subsubsection{Coulomb-Volkov approximation (CVA)}
\label{sec:CVA}	

In order to account for Coulomb effects in the continuum one approach one may replace the Gordon-Volkov states with so called Coulomb-Volkov states, which approximately account for both the laser field and the Coulomb potential. To make the problem tractable it is assumed that the laser distorts the Coulomb continuum adiabatically. Then, the resultant Coulomb-Volkov wavefunction is the product of a Coulomb continuum scattering state and a Gordon-Volkov state. Explicitly, the final state of the system reads
\begin{equation}
\braket{\mathbf{r}|\psi^{(CV)}_{\mathbf{p}}(t)}
=\psi^{(CV)}_{\mathbf{p}}(\mathbf{r},t)=\psi^{(C)}_{\mathbf{p}}(\mathbf{r})\psi^{(GV)}_{\mathbf{p}}(\mathbf{r},t),
\label{eq:CVA}
\end{equation} 
where $\psi^{(C)}_{\mathbf{p}}$ is the field-free Coulomb scattering wave for the (final) continuum state and 
\begin{align}
\braket{\mathbf{r}|\psi^{(GV)}_{\mathbf{p}}(t)}&=\psi^{(GV)}_{\mathbf{p}}(\mathbf{r},t)\notag
\\&=\frac{\exp[i [\mathbf{p}+\mathbf{A}(t)]\cdot \mathbf{r}]}{(2\pi)^{3/2}}\exp{iS_d(t,t')},
\label{eq:Volkovwf}
\end{align}
is the Gordon-Volkov wavefunction. In (\ref{eq:Volkovwf}),  $S_d(t,t')$ is given by the SFA action for direct ATI [Eq.~(\ref{eq:Sdir})]. This ansatz is then used in the direct ATI transition amplitude. The above-stated equations show that, in the CVA, the propagation from the instant of ionization to the instant of detection is Coulomb-free, as it is dictated by the Volkov propagator. 
A summary of the history and main results of the CVA can be found in the review \cite{Popruzhenko2014a}.
This method was originally proposed in a strong field context in Refs.~\cite{Jain1978,Cavaliere1980} and since then there has been a wealth of publications further developing it \cite{Dorr1987,Shakeshaft1987,Basile1988,Reiss1994,Jaron1999,Groeger2000,Macri2003,Rodriguez2004,Arbo2008,Arbo2010,Arbo2012}.

\begin{figure}
	\centering
	\includegraphics[width=\linewidth]{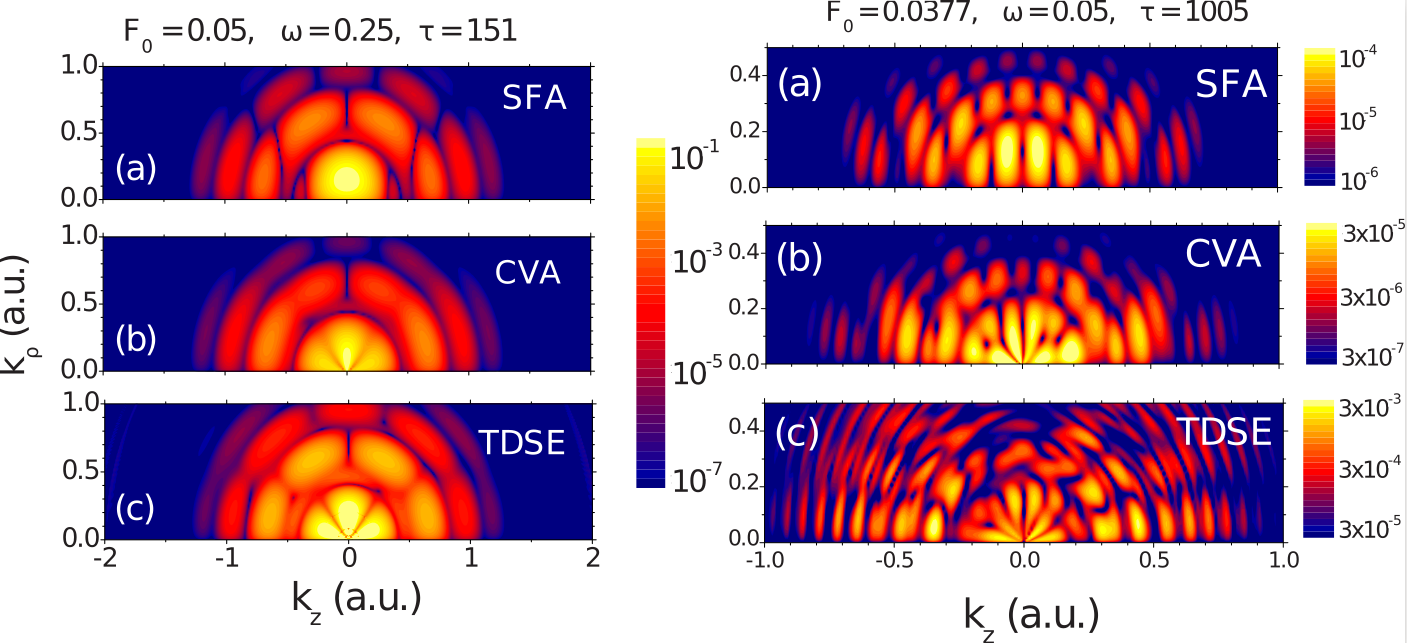}
	\caption{Photoelectron momentum distributions plotted in a logarithmic scale computed using the SFA, CVA, and an \textit{ab-initio} TDSE calculation, [panels (a), (b) and (c), respectively]. The left and the right columns are Figs.~1 and 3 in Ref.~\cite{Arbo2008}, respectively. The  field strength $F_0$, its frequency $\omega$ and the pulse length $\tau$ are given at the the top of both columns.}
	\label{fig:CVA}
\end{figure}

One of the early successes of the CVA was that it was found to break a four-fold symmetry in the momentum distributions for photoelectrons ionized via elliptically polarized fields \cite{Basile1988,Jaron1999} that was exhibited by SFA-like approaches but was not observed in experiments \cite{Bashkansky1988}. In the case of relatively high frequency and low intensity linearly polarised laser fields it was found \cite{Arbo2008} that the CVA had markedly better agreement with \textit{ab-initio} solutions than the SFA (left panels in Fig.~\ref{fig:CVA}) for low photoelectron momenta in the multiphoton regime. 

On the other hand, this agreement worsens for higher photoelectron momenta or lower-frequency fields (right panels in Fig.~\ref{fig:CVA}). In fact, because the continuum propagation uses Volkov states, the CVA is only suitable for a parameter range in which significant Coulomb distortion of the continuum propagation is not required. Thus, it is not surprising that the CVA reproduces the fan in the multiphoton regime, as the Coulomb factor in Eq.~(\ref{eq:CVA}) can describe a resonant behavior appropriately. However, the CVA does not satisfactorily reproduce interference effects in the low-frequency regime, as in this case the Coulomb potential would exert a strong influence on the electron continuum propagation. In the right panels of Fig.~\ref{fig:CVA}, one clearly sees that, in this latter caase, the CVA neither reproduces the correct number of fringes in the near-threshold fan, nor the spider-like structure, which is almost absent. 	

Recently however, an alternative version of the CVA \cite{Lopez2019}, known as the second order CVA or CVA$^2$ has been used to account for rescattering by using the Born expansion, introduced in the above section, replacing the scattering states in the SFA with Coulomb-Volkov states. This enables the CVA to account for hard-rescattering in the same way as the SFA and extends its description to the high energy photoelectron momentum distributions. Care should be taken with this method, as Coulomb-Volkov states are not solutions of the Schr\"odinger equation and their use will introduce errors in a non-standard way. 
Additionally, there is some inconsistency using Coulomb-Volkov states in direct terms if a Born series about the potential has been applied. The direct terms will contain only a Gordon-Volkov time evolution operator so should not contain any Coulomb effects, only the rescattered term will contain the full time evolution operator that will warrant this. 	
Hence, it is not clear what effects are neglected and some parts may be ``doubly''/ over accounted for. However, as the original CVA did not reproduce effects due to rescattering electrons there is an empirical reason to construct such a model and ``doubly'' counted effects will most likely be small.

In its original form the CVA was not suited for use in electron holography, adding no real improvement over the SFA for the typical parameter range. Furthermore, the lack of quantum trajectories/orbits made it difficult to interpret any interference holographically. In its current form, the CVA has been used to study some holographic effects \cite{Lopez2019,Borbely2019} in some cases exploiting the new-found ability to account for rescattering. However, the CVA still tends to closely mimic the SFA for the low frequency fields used in holography (see Fig.~5 in Ref.~\cite{Borbely2019}). This prevents a good description of effects that deviate non-perturbatively from the `traditional' SFA such as the LES\footnote{In fact, it has been shown \cite{Becker2014a} that, by considering previously neglected forward scattered trajectories, the LES can be reproduced within the SFA. Thus, if this were applied to the CVA, presumably it would be able to reproduce the LES as well.} \cite{Lemell2013}, which the CVA does not correctly reproduce. Thus, the CVA can be used to interpret and understand holographic interference, but mainly by exploiting and improving the mechanisms that already qualitatively exist in the SFA.

\subsubsection{Quantitative rescattering theory (QRS).}
\label{sec:QRS}

\begin{figure}
	\centering
	\includegraphics[width=\linewidth]{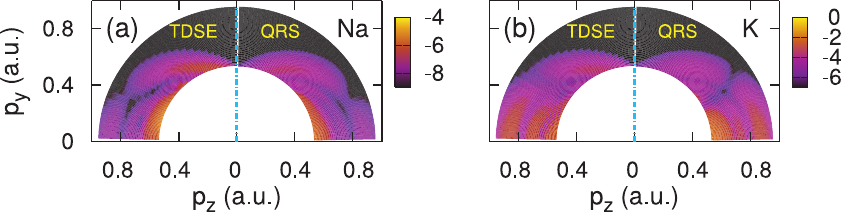}
	\caption{Momentum dependent ATI photoelectron distributions of Na (a) and K (b) atoms computed using the TDSE and the QRS model for a field of peak intensity 1.0 $\times$ 10$^{12}$ Wcm$^{-2}$ and wavelength  3.2 $\mu$m over five optical cycles. The plots use a logarithmic scale. Fig~2 of \cite{Chen2009}.  }
	\label{fig:QRS}
\end{figure}

The QRS, also referred to as scattering-wave SFA (SW-SFA) in earlier papers   \cite{Morishita2008,Le2008a,Le2008,Chen2009,Micheau2009,Le2009,Le2009a,Lin2018}, uses a particular factorization of the momentum dependent transition amplitude. This can be employed to describe processes in which electron recollision or recombination occurs, such as HHG, high-order above-threshold ionization (HATI) and NSDI. For more information on the QRS method and its current status see the review \cite{Lin2018} and the tutorial \cite{Le2016}.  With this factorization the momentum dependent probability distribution is then written as
\begin{equation}
M(\mathbf{p})=W\left(p^{(r)}\right)\sigma\left(p^{(r)},\theta\right),
\end{equation}
where $\sigma(p^{(r)},\theta)$ is the differential elastic scattering cross section between free electrons with momentum  $\mathbf{p}^{(r)}$  at the instant of rescattering, and scattering angle $\theta$ and the ion, while $W(p^{(r)})$ describes the flux of a returning wavepacket. It is shown that the flux term can be treated accurately by the SFA, while the dipole term uses a Coulomb scattering wave to account for effect of the Coulomb attraction of the ion. The final momentum $\mathbf{p}$ is given in terms of the momentum of the returning wavepacket by 
\begin{equation}
\mathbf{p}=\mathbf{p}^{(r)}-\mathbf{A}\left(t^{(r)}\right),
\label{eq:causticmap}
\end{equation}
where $t^{(r)}$ denotes the rescattering time.
This factorization was tested both theoretically \cite{Morishita2008,Le2008a,Micheau2009} and experimentally \cite{Minemoto2008} and found to work well in key cases for atoms \cite{Morishita2008,Le2008,Micheau2009} and molecules \cite{Le2008a,Le2009}, in comparison with \textit{ab-intio} solutions. For an example of a photoelectron spectra for ATI calculated with the QRS see Fig.~\ref{fig:QRS}, where it is compared with \textit{ab-initio} methods. Following this success it was used to explain the species dependence of the high-energy plateau in HHG \cite{Morishita2008,Le2008,Le2008a} and HATI \cite{Chen2009} and was applied to HHG of small \cite{Le2009} and even polyatomic \cite{Wong2013,Le2013,Le2013a} molecules  where \textit{ab-intio} solutions would be too intensive. Despite this, it has been stated that \cite{Lin2018} using the QRS to model HHG for complex polyatomic molecules is troublesome as it is a single active electron model that does not account for the electron or nuclear correlations. Recently, in Ref.~\cite{Krecinic2018}, it was argued that for $CF_3I$ there are contributions from many orbitals for the photoelectron distributions. Thus, single active electron models, where only the highest-occupied molecular orbitals are considered (such as the QRS) will not be successful.

The QRS method works well for the range of intensities of interest for photoelectron holography. However, in contrast to CVA it does not describe the direct electrons in ATI and thus is not suited to study most holographic interference patterns as it will not account for the reference signal. Despite employing a trajectory-based approach for the returning wavepacket this description uses SFA orbits. Thus the Coulomb potential is neglected in the continuum propagation besides hard scattering events. Then, although this method can certainly be used for imaging purposes such in HHG spectroscopy (e.g. \cite{Le2009a,Le2013}) or using a LIED technique in ATI \cite{Blaga2009,Hoang2017}, the target information will not be encoded in the trajectories but in the scattering cross-section term. 

\subsubsection{Low-Frequency Approximation (LFA).}	
\label{sec:lfa}

The idea behind the LFA is to factorize the transition amplitude in a similar fashion to the QRS. However, this is  achieved by extending the Born series expansion about the potential to beyond first order. A early form of the LFA was proposed in \cite{Kroll1973} but it was developed and used in a strong field context in the following publications \cite{Cerkic2009a,Milosevic2010LasPhy,Milosevic2010,Fetic2011a,Milosevic2014a,Milosevic2018}. Eqs.~(\ref{eq:ATIdirect}) and (\ref{eq:rescSFA}) give the zeroth and first-order terms in the Born expansion, which are usually associated with direct and rescattered electrons, respectively. The inclusion of the first-order term is sometimes referred to as the improved SFA (ISFA) or SFA2. 
The Born series is obtained by iteratively substituting Eq.~(\ref{eq:Dyson2})\footnote{Note that an alternative form of Eq.~(\ref{eq:Dyson2}) is used for LFA, where the full and Gordon-Volkov time evolution operators are swapped in the integrand.} into Eq.~(\ref{eq:transitionamp}). Both 
equations are formally exact and an approximation is only made when, after iteration, the remaining full time-evolution operators are replaced by their Gordon-Volkov counterparts. 
Repeating this process twice and retaining the full time time evolution operator adds an additional ``correction'' term to the standard rescattering transition amplitude, namely
\begin{align}
\label{eq:LFA}
M^{(2)}_r(\mathbf{p})&=(-i)^2\lim\limits_{t\rightarrow\infty}\int_{-\infty}^{t}\!\! dt'\int d^3\mathbf{k}
\int_{t'}^{t}\! dt''
\Bigg[V_{\mathbf{p}\mathbf{k}}V_{\mathbf{k}0}e^{iS_r(\mathbf{p},\mathbf{k},t,t')} \notag\\
&\qquad-\underbrace{i\int_{t''}^{t}\! dt'''\braket{\psi^{(GV)}_{\mathbf{p}}(t''')|VU(t''',t'')V|\psi^{(GV)}_{\mathbf{k}}(t'')}}_{M^{(c)}(\mathbf{p},\mathbf{k})}\braket{\psi^{(GV)}_{\mathbf{k}}(t')|H_I(t')|\psi_0(t')}
\Bigg],
\end{align}
where the Gordon-Volkov states are defined in Eq.~(\ref{eq:Volkovwf}).
The extra term can be interpreted as describing two rescattering events. Including it makes the total transition amplitude formally exact. Now as in \cite{Cerkic2009a}, one of the seminal papers, the LFA is used to further evaluate $M^{(c)}(\mathbf{p},\mathbf{k})$. The assumption is made that, if the laser intensity is high and its  frequency is low, the laser-dressed momentum will change only a little between rescattering events.  This means that one may use a \textit{stationary} approach in the second-order term in Eq~(\ref{eq:LFA}). Explicitly,  the time evolution operator $U(t''',t'')$ is replaced by the time-independent Green's operator $ G_V(E_{\mathbf{k}+\mathbf{A}(t''')})\delta(t'''-t'')$ calculated at time  $t'''$ for the field-dressed instantaneous energy  
\begin{equation}
E_{\mathbf{k}+\mathbf{A}(t''')}=\frac{1}{2}\left[\mathbf{k}+\mathbf{A}(t''')\right]^2.
\end{equation}

Using this approximation, it is possible to write the whole rescattered contribution, in a similar fashion to the QRS, as a product of a differential cross section of the laser-free continuum electron back to the initial bound state and a returning electron wavepacket part. This procedure changes the scattering states but leads to the same orbits as the improved SFA as given by Eq.~(\ref{eq:Sresc}), and the resulting transition amplitude can be evaluated using the saddle-point or uniform approximation.

In \cite{Fetic2011a}, it was argued that the QRS is equivalent to taking only a single (short) orbit in the above expression. It has even been stated \cite{Frolov2009,Milosevic2014a} that the QRS can be derived entirely from the LFA by making some additional approximations. However, an exact correspondence between these two methods is not completely clear as the cross-section in the QRS is between a continuum-scattering state (accounting for the Coulomb potential) and the ion, which is evaluated exactly using \textit{ab-initio} techniques. However, in the LFA the cross-section is between Gordon-Volkov states and the initial atomic state. A further difference with regard to the QRS is that the transition amplitude in the LFA is in integral form, due to applying the relations for the time-evolution operator. This  integral is then evaluated using the saddle-point approximation.

\begin{figure}
	\centering
	\includegraphics[width=0.75\linewidth]{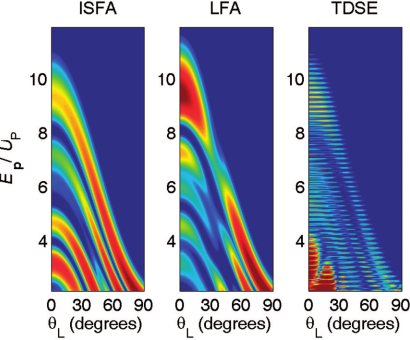}
	\caption{Computations from Fig.~2 of \cite{Fetic2011a} using the ISFA, LFA and TDSE, left, middle and right, respectively, which show the photoelectron detachment rate of $F^{-}$ over the electron emission angle $\theta_L$ and the detached electron energy $E_{\mathbf{p}}/U_{\mathbf{p}}$ in a logarithmic scale. The driving field was taken to be linearly polarized, with a peak intensity of 1.3$\times 10^{13}$W/cm$^2$ and wavelength of 1800 nm.
		The TDSE computation employed a pulse with a cos$^2$ envelope over 15 optical cycles.}
	\label{fig:LFA}
\end{figure}
There has been extensive comparison between the LFA and the ISFA with improvements found both in the high-energy ATI spectra \cite{Cerkic2009a,Milosevic2010LasPhy,Fetic2011a} and in the low-energy part when studying the LES \cite{Milosevic2014a}. Reasonable agreement has been found when compared to experiment for the ATI spectra of atoms \cite{Milosevic2010,Fetic2011a}. A comparison with \textit{ab-initio} methods has been performed for the case of photodetachment of $F^{-}$ \cite{Fetic2011a}, see Fig.~\ref{fig:LFA}, and in the extension of the LFA to HHG \cite{Milosevic2018} with bicircular laser fields. In both cases very good agreement with the TDSE was found. The LFA has also been extended to beyond the dipole approximation for ATI of H$_2^+$, where good agreement was found along the momentum components parallel to the laser field polarization when compared to \textit{ab-initio} methods \cite{Brennecke2018}.

The LFA is a fully fledged quantum-orbit method and unlike the QRS it has been shown to give improvements in the \textit{low} and \textit{high-}energy part of the ATI spectrum. For these reasons, it seems appropriate to apply this method to electron holography. Despite this, the LFA does not satisfactorily reproduce the holographic interference of interest that arises for linearly polarized fields. This may be because only hard-scattering Coulomb effects are considered and Coulomb-distortion of the orbits is not incorporated. These distortions include deflections and soft collisions, which may happen away from the origin. For linearly polarized fields these ``soft'' Coulomb effects have been shown to be crucial for the correct formation of interference structures present in the distributions \cite{Lai2017, Maxwell2017}. The Born series will in theory, if taken to high enough order, describe all Coulomb effects. However, the direct orbits in Born-series approaches neglect Coulomb distortion and deflection. Instead, these effects will be described in terms of the higher order terms that relate to orbits undergoing multiple hard recollisions. This is an inappropriate basis with which to describe ``soft'' Coulomb effects and as such should not be expected to converge quickly for cases where this plays an important role.

\subsubsection{Coulomb-Corrected SFA (CCSFA).}
\label{sec:CCSFA}

The CCSFA \cite{Popruzhenko2008c,Popruzhenko2008d,Popruzhenko2008a,Popruzhenko2009} introduces Coulomb effects perturbatively to the SFA in the action. For a detailed introduction to the CCSFA see the review \cite{Popruzhenko2014a}. The peturbative expansion is implemented  after the application of the saddle-point approximation to Eq.~(\ref{eq:transitionamp}). Both the action and the trajectories can be expanded in the following way
\begin{align}
\mathbf{r}(\mathbf{p}_f,t)&=\mathbf{r}^{(0)}(\mathbf{p}_f,t)+\mathbf{r}^{(1)}(\mathbf{p}_f,t)+\dots\\
\mathbf{p}(\mathbf{p}_f,t)&=\mathbf{p}^{(0)}(\mathbf{p}_f,t)+\mathbf{p}^{(1)}(\mathbf{p}_f,t)+\dots\\
S(\mathbf{p}_f,t)&=S^{(0)}(\mathbf{p}_f,t)+S^{(1)}(\mathbf{p}_f,t)+\dots,
\end{align}
where the zeroth order gives the Coulomb-free trajectories and the direct SFA action [Eq.~(\ref{eq:Sdir})].
The first order correction to the action is then given by
\begin{equation}
S^{(1)}(\mathbf{p}_f,t)=\int_{t_s}^t d\tau \frac{C}{\left|\mathbf{r}^{(0)}(\mathbf{p}_f,\tau) \right|},
\label{eq:CoulombPhase}
\end{equation}
where a hydrogenic Coulomb potential has been used and $C$ is the Coulomb correction factor, set to unity for singly charged ions. The corrections are all dependent on the final momentum, thus the CCSFA uses an inverse solution approach. The final transition amplitude can be written as
\begin{equation}
M(\mathbf{p}_f)=\sum_s \mathcal{C}(t_s)\exp\left(i S(\mathbf{p}_f,t_s) + i F(\mathbf{p}_f,t_s) \right),
\end{equation} where $t_s$ are the corrected times of ionisation given by inserting a corrected initial momentum $\mathbf{p}_0$ into to the solutions of the saddle-point equation (\ref{eq:tunnelsfa}),  $\mathcal{C}(t_s)$ is the SFA prefactor, 
\begin{equation}
F(\mathbf{p}_f,t)=\mathbf{r}\cdot \left[\mathbf{p}_f+\mathbf{A}(\tau)\right]\bigg\vert_{t_s}^{t}\approx F^{(0)}(\mathbf{p}_f,t)\pm C/\sqrt{2 I_p},
\end{equation} 
The Coulomb phase is integrated along the whole temporal contour, even up to the singularity. A regularization procedure has been introduced to remove the singularity, in which the asymptotic form of the initial bound state wavefunction is matched to the action for a matching time $t_*$, where the laser is dominant over the Coulomb force but only a fraction of a laser cycle has taken place since ionization. A detailed formulation of the procedure is given in \cite{Popruzhenko2008c,Popruzhenko2014a}. Bound-state regularization was also applied to other methods that use a Coulomb phase such as the EVA \cite{Smirnova2006} and the CQSFA \cite{Maxwell2018b} discussed below. 

This method is in theory very quick to calculate. However, a particularly debilitating issue that affects it (and all methods that employ a Coulomb phase) is branch cuts, which arise due to the complex trajectories in the square root of Eq.~(\ref{eq:CoulombPhase}). This causes defects in the momentum distributions if crossed in the temporal integration contour. Recently, algorithms that avoid the branch cuts have been introduced \cite{Popruzhenko2014b,Pisanty2016,Maxwell2018b}, these work well for such perturbative methods, enabling \textbf{2D} photoelectron momentum distributions to be produced without defects or additional approximations. (For an example see Fig.~\ref{fig:Coulombfree}).  This will be discussed in more detail later in this review in Secs.~\ref{subsubsec:branchcuts1}, \ref{sec:complextrajs} and \ref{sec:branchcutPADs}.
The CCSFA, together with branch-cut corrections, was successfully used to explain orders of magnitude difference between \textit{ab-initio} methods and the SFA around the direct ATI cutoff, in terms of soft recollisions \cite{Keil2016}. This description required complex trajectories so that the overall probability could be modified during recollision.

The CCSFA provides a consistent and clear extension of the SFA but as a perturbative extension of the direct SFA there are many effects it can not account for. One obvious extension is to solve Newton's equations for the trajectories, including both the laser and Coulomb interaction on an equal footing, instead of using a perturbative expansion. This extension of the CCSFA is referred to as the TCSFA and will be discussed next. Another option could be to, similar to the CVA, include the SFA rescattered electrons. However, using two alternative perturbative series could make the method lose consistency and it may be unclear what effects are included or left out. Recently, the CCSFA has been extended to describe HHG \cite{Popruzhenko2018b}, which shows it is possible to incorporate the returning wavepacket in this method. 

\subsubsection{Trajectory-Based Coulomb SFA (TCSFA)}
\label{sec:TCSFA}

\begin{figure}
	\centering
	\includegraphics[width=0.5\linewidth]{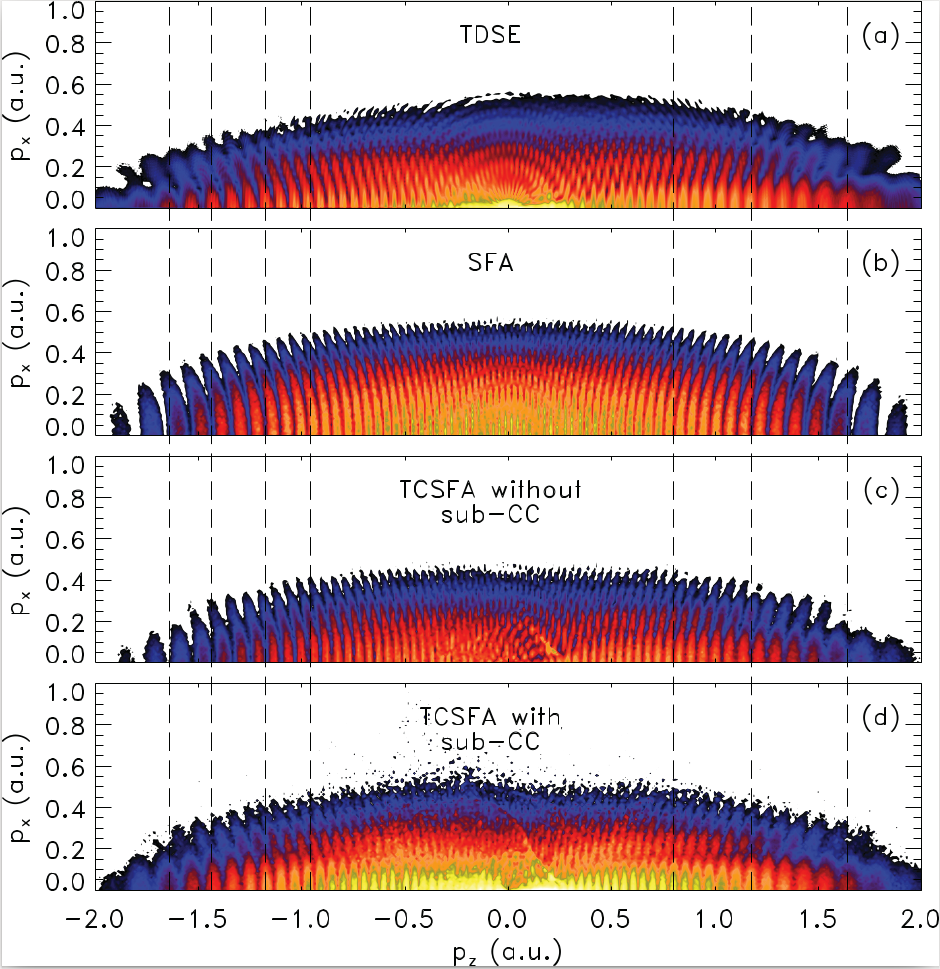}
	\caption{Photoelectron momentum distribution for H(1s), computed using the TDSE, plain SFA and TCSFA without and with sub-CC, [panels (a-d), respectively]. The driving field has a peak intensity of 10$^{14}$ W/cm$^2$ wavelength of 2$\mu$m, and a sin$^2$-envelope over four optical cycles. The plots are presented in a logarithmic scale.  Fig.~2 of \cite{Yan2012}.}
	\label{fig:TCSFA}
\end{figure}

In the TCSFA \cite{Yan2010,Yan2012} the CCSFA is built on by including equations of motion that include the Coulomb potential for continuum trajectories. Thus the trajectories are now described by
\begin{align}
\dot{\mathbf{r}}(\tau)=\mathbf{p}+\mathbf{A}(\tau) && \dot{\mathbf{p}}(\tau)=-\nabla V(\mathbf{r}(\tau)).
\end{align}
Solving these equations of motion in conjunction with the tunnel ionization [Eq.~(\ref{eq:tunnelsfa})] led to a new categorization of quantum orbits \cite{Yan2010} (later also used in the CQSFA) of four possible contributing trajectories for each final momentum point. Methods that solve the full Newton's equations of motion including the Coulomb potential are some times called Coulomb-distorted trajectory approaches. This approach led to unprecedented qualitative agreement with \textit{ab-initio} methods for the direct ATI 2D momentum distributions. In an additional publication the TCSFA was used to show the importance of including the Coulomb phase in the action for the tunnelling part of the trajectory, known as sub-barrier Coulomb-corrections (sub-CC), see Fig.~\ref{fig:TCSFA}. It was shown that as well as influencing the overall tunnelling probability, the interference patterns in momentum distributions were also affected by sub-CC \cite{Yan2012}. Recently, a version of the TCSFA was formulated that goes beyond the dipole approximation \cite{Keil2017}.

In contrast to the CCSFA and CQSFA, the TCSFA solves the direct problem by specifying the initial rather than final momentum in a shooting method. The resulting final momenta are binned  similar to some Monte Carlo approaches such as the QTMC and SCTS methods, see Table \ref{table:comparison}. Because the momentum is not conserved, an extra $\dot{\mathbf{p}}\cdot \mathbf{r}$ term must be incorporated in the action, as it is written in phase space. This term is not included in the TCSFA. It has been conjectured \cite{Maxwell2017} that this results in some discrepancies when compared with \textit{ab-initio} solutions, such as the wrong number of fringes appearing on the inner ATI ring for the fan-like structure \cite{Yan2012}. Branch cuts are an even bigger problem for Coulomb-distorted trajectory models because of the complex square root in equations of motion. There are no full solutions to this branch cut problem, but a partial one has been suggested in \cite{Maxwell2018b}. In the TCSFA (and most other models of this type), branch cuts are avoided by taking real trajectories in the continuum. This is only a partial remedy  because, as shown in Refs.~\cite{Torlina2013,Popruzhenko2014b,Keil2016}, complex trajectories are essential for capturing certain physical effects, such as the effective deceleration of the electron wavepacket due to the Coulomb potential. In principle, the TCSFA could be used for holographic imaging. However, the caustics in the 2D momentum distributions may cause some difficulties. Additionally, the binning of trajectories into final momenta makes it is more difficult to isolate particular interference effects than other methods. On the other hand, the binning approach means that classes of solutions for the trajectories do not have to be known in advance. This makes it much easier to generalize the TCSFA to more complex systems and/or driving fields, for which the classes of solutions for the trajectories may change significantly.

\subsubsection{The Eikonal-Volkov Approximation (EVA).} 
\label{sec:eva}

The EVA \cite{Smirnova2006,Smirnova2007,Smirnova2008} uses a totally different approach to most quantum trajectory methods. The Coulomb field is accounted for by corrections to Gordon-Volkov states using
the eikonal approximation. The wavefunction is written using the WKB (Wentzel-Kramers-Brilliouin) ansatz
\begin{equation}
\psi(\mathbf{r},t)=P(\mathbf{r},t)\exp(i S(\mathbf{r},t)/\hbar).
\label{eq:WKB}
\end{equation}
Here $S(\mathbf{p},t)=S^{GV}_{\mathbf{k}}(\mathbf{r},t)+G_{k}(\mathbf{r},t)$, where the first term is the phase of a Gordon-Volkov and the second is a correction/ distortion due to the Coulomb potential. This is then inserted into the Schr\"odinger equation, expanded about $\hbar$ and terms of the same order are collected. This results in an equation that determines $G_{k}$, which can be solved to integral form 
\begin{equation}
G_{k}(\mathbf{r},t)=-\int_{T}^t d\tau V(\mathbf{r}_0(\tau))+G_{0\mathbf{k}}(\mathbf{r}_0(\tau)),
\label{eq:EVA}
\end{equation}
where $\mathbf{r}_0$ denotes the Coulomb-free electronic coordinate in the laser field. 
Note that, in order to get to Eq.~(\ref{eq:EVA}), the approximation that the change in electron momentum during scattering is small i.e. $|\nabla G_{\mathbf{k}}(\mathbf{r},t)|\ll|\mathbf{k}|$ is used. When an electron travels close to the core, in many cases, the large change in momentum will cause this condition to fail. However, in conjunction with this method the analytical R-Matrix (ARM) method was developed, which enables the use of an alternative approach for electrons that come close to the core. This is discussed in the next section.

Using the EVA one can construct continuum states $\ket{\mathbf{p}^{EVA}_T(t)}$, that are defined by backpropagating the field-free eikonal states from after the laser is off at time $T$ to the desired time $t$. These continuum states in position representation are given by
\begin{align}
\braket{\mathbf{r}|\mathbf{p}^{EVA}_T(t)}=\frac{1}{(2\pi)^{3/2}}&\exp
\left[i\left(\mathbf{p}+\mathbf{A}(t)\right)\cdot \mathbf{r}
-\frac{i}{2}\int_{T}^{t}d\tau(\mathbf{p}+\mathbf{A})^2
\right.\notag\\
& \left.
-i\int_{T}^{t}d\tau V(\mathbf{r}_0(\tau))+iG_{0\mathbf{p}}(\mathbf{r}_0(T))
\right].
\label{eq:pEVA}
\end{align}

\subsubsection{Analytical R-Matrix (ARM) Method}
\label{sec:arm}

The ARM method \cite{Torlina2012,Kaushal2013,Torlina2014} splits the space into an inner and outer region, near and away from the atom/ molecule, respectively. In the inner region, exact \textit{ab-initio} methods are used, while in the outer region the EVA is employed. This has the capacity to give very accurate results, while retaining the concept of quantum trajectories.
The Hamiltonian can be split into two parts that describe the dynamics in  the inner and outer regions, denoted $+$ and $-$, respectively, given by $\hat{\mathcal{H}}^{\pm}=\hat{H}+\hat{L}^{\pm}(a)$, where $\hat{L}^{\pm}$ is the Bloch operator, which preserves the hermiticity of the both the regions and is defined in position notation as
\begin{equation}
\hat{L}^{\pm}=\pm\delta(r-a)\left(\frac{d}{d r}+\frac{1-b}{r}\right),
\end{equation}
where $a$ is the radius of the circular boundary which separates the two region and b is an arbitrary constant. In Ref.~\cite{Torlina2012}  $b$ is ultimately set to $b=C/\sqrt{2I_p}$, which is unity in the case of hydrogen. 
A solution of the TDSE for the inner and outer region may be written  integral form
\begin{equation}
\psi^{\pm}(\mathbf{r},t)=-\int_{-\infty }^t d t' \int d^3 \mathbf{r}' G^{\pm}(\mathbf{r},t,\mathbf{r}',t')L^{\pm} \psi(\mathbf{r}',t'),
\label{eq:ARMIntegral}
\end{equation}
where $G^{\pm}(\mathbf{r},t,\mathbf{r}',t')=\braket{\mathbf{r}|U^{\pm}(t,t')|\mathbf{r}'}$ is the Green's function in the inner or outer region. The initial wavefunction $\psi(\mathbf{r}',t')$ can be taken as the atomic ground state before the laser is turned on. This can be inserted into the outer equation to describe the ionization process
\begin{align}
\psi^{-}_{(1)}(\mathbf{r},t)&=-\int_{-\infty }^t d t' \int d^3 \mathbf{r}' G^{-}(\mathbf{r},t,\mathbf{r}',t')
\notag\\
&\times
L^{-} \psi_0(\mathbf{r}')\exp(iI_p t').
\end{align}
To describe a wavepacket returning to the region of the core $\psi^{-}_{(1)}$ may be substituted into the inner equation to get $\psi^{+}_{(2)}$, then finally the scattered wavepacket is described by inserting $\psi^{+}_{(2)}$ into the outer equation to get $\psi^{-}_{(3)}$. In theory this process could be repeated for further recollision processes but the parameter $a$ is chosen such that return to the inner region results in a ``hard'' recollision and thus the electron wavepacket is unlikely to return. The saddle point approximation is applied to the integrals stemming from Eq.~(\ref{eq:ARMIntegral}), resulting in a complex time of ionization in common with other quantum orbit approaches. For the propagation in the outer region the EVA is exploited to make this region easy to solve. The Green's function for the outer region is replaced with one constructed via the EVA
\begin{align}
G^{-}(\mathbf{r},t,\mathbf{r}',t')&=\braket{\mathbf{r}|U^{EVA}(t,t')|\mathbf{r}'}\\
&=\theta(t-t')\int d\mathbf{p}\braket{\mathbf{r}|\mathbf{p}^{EVA}_T(t)}\braket{\mathbf{p}^{EVA}_T(t')|\mathbf{r}'} ,
\end{align}
where the states $\ket{\mathbf{p}^{EVA}_T(t)}$ are given by Eq.~(\ref{eq:pEVA}).
One of the main benefits of the ARM method is that it incorporates a rigorous approach to ``hard'' scattering, something that is missing from many approaches such as the CCSFA, TCSFA and CQSFA. Additionally, it is easy to extend to many electron systems, shown in \cite{Torlina2012a}, as the R-matrix method is already designed for this.

The combination of the EVA and ARM has given very good results. For example it has been used in the case of nearly circularly polarized fields leading to the interpretation of ionization times in the attoclock\footnote{The attoclock exploits nearly circularly polarized fields to use the offset in photoelectron ejection angle away from the peak of the field to give a measure of time taken for the electron to tunnel.} \cite{Torlina2014,Torlina2015}. This demonstrates that, for hydrogen, there is no tunnelling delay once effects due to the Coulomb potential have been accounted for. A high level of accuracy was achieved for the ARM computations for nearly circularly polarized light, as, in comparison with three different \textit{ab-initio} solutions, deviations did not exceed 2\% \cite{Torlina2015}. Furthermore, it was used to show that including the imaginary part of the complex trajectories improves agreement with \textit{ab-initio} solutions of the TDSE and can be related to wave-packet deceleration by the Coulomb potential \cite{Torlina2013}. In a recent set of three companion papers \cite{Kaushal2018a,Kaushal2018b,Kaushal2018c}, the ARM method is extended to look at the effect of fields with non-vanishing ellipticity upon initial bound electronic states with co- and counter-rotating orbital angular momentum. Thereby, differences in ionization yields, the interplay of the Coulomb potential, its effect on the tunnel exit velocities and coordinate, and the high level of spin polarization possible in the photoelectrons were examined.

Despite the high level of success of the ARM method, it has not been used to calculate 2D photoelectron momentum distributions in order to understand the holographic interference structures present for linearly polarised fields. 
Given that the EVA is ultimately perturbative and assumes small changes in momentum, the strong deflections by the Coulomb potential that are not quite classified as ``hard'' collision by the ARM method may not be fully accounted for. These intermediate interactions occur much more commonly for linear polarized fields. 
Since the EVA employs complex trajectories, it will also suffer from branch cuts. Unlike the TCSFA and CQSFA, the incomplete solution to take the trajectories to be real in the continuum does not seem to have been used, perhaps to preserve consistency of the method. It seems likely that a proper treatment of branch cuts is the main stumbling block for modelling holographic patterns from linear polarized fields for the ARM method.

\subsubsection{Quantum-Trajectory Monte Carlo (QTMC).}
\label{sec:QTMC}

The QTMC \cite{Li2014,Geng2014,Li2014c,Richter2015,Xie2016,Li2016PRA,Liu2016} is an extension upon the classical-trajectory Monte Carlo (CTMC) method \cite{Liu1998}, which uses the quasi-static Ammosov-Delone-Krainov (ADK) tunnelling theory \cite{Ammosov1986} to describe the first step of ionization, while continuum propagation is described by classical trajectories using Newton's equations of motion accounting for both the Coulomb and laser fields. The final momentum distribution is then constructed, via a direct approach, by randomly choosing a spread of initial conditions, in the Monte Carlo fashion. These initial conditions are used to both provide a weight via the ADK tunnelling rate and to compute trajectories in order to calculate the final momenta. Then all the trajectories are binned via their final momentum, such that ADK probabilities are summed if they correspond to trajectories with momenta that lies in a particular bin. 

In the classical case, the probability distribution in the $i-$th bin corresponding to a momentum $\mathbf{p}_{if}$ reads
\begin{align}
\Omega_\epsilon(\mathbf{p}_{if})&=\sum_{\mathbf{p}_{jf} \in \mathbb{S}^{\mathrm{bin}}_i}W(t_0,p_{j0\perp}),
\notag\\
\mathbb{S}^{\mathrm{bin}}_i:&=\left\{\mathbf{p}_f : |p_{fk}-p_{ifk}|<\epsilon \quad \forall k\in\{\parallel,\perp\}\right\},
\end{align}
where $W(t_0,p_{j0\perp})$ is the ADK rate and $\epsilon \ll 1$. Note that computed trajectories are denoted by the index $j$, with final momentum $\mathbf{p}_{jf}$, initial transverse momentum $p_{j0\perp}$ and tunnelling time $t_{0}$. The probability distribution is calculated for a discrete set of momenta $\mathbf{p}_{if}$, each centred on a bin. Square bins are shown here but in theory any shapes could be used.

The CTMC is improved on by the QTMC by including a phase (the action) with each trajectory, allowing for interference effects. The phase is given by the classical action and no additional fluctuation factor/prefactor term is included. Hence, this corresponds to a zeroth-order expansion of the action in the continuum propagator given by Feynman's path integral around classical saddle-points (a.k.a. the classical limit).
\begin{equation}
\braket{\mathbf{p}_f|U(t,t_0)|\mathbf{r}_0}\approx(2\pi)^{-3/2}\exp(iS[\mathbf{r}]-i\mathbf{p}_f\cdot\mathbf{r}_f).
\end{equation}
The probability in the QMTC is given by
\begin{align}
\Omega_\epsilon(\mathbf{p}_{if})&=\left|\sum_{\mathbf{p}_{jf} \in S^{\mathrm{bin}}_i}\sqrt{W(t_{j0},p_{j0\perp})}\exp(i S[\mathbf{r}_j])\right|^2
\notag\\
S[\mathbf{r}_j]&=\int_{t_{j0}}^{\infty}d t
\left(\frac{1}{2}\dot{\mathbf{r}}_j(t)^2+I_p+V(\mathbf{r}_j(t))\right).
\end{align}
Some approximations have been used to arrive at this formula. Firstly, only a real probability is provided by the ADK tunnelling rate formula, so that any phases accumulated in the tunnelling process will be lost. It has been shown in various studies that important phases are picked up in this process (e.g. see \cite{Yan2012,Kaushal2018a,Maxwell2017} for a small sample of such work). Secondly, the Monte Carlo approach only approximately solves the saddle-point equation and there is an error dependent on the bin size, something shared by all direct approaches. Given that the use of the saddle-point approximation is motivated by a fast varying action, the bin size may have to be very small to account for this. Also the initial momenta may vary considerably from the final, thus a large spread of initial momenta should be taken. All this means that many trajectories must be used in order to populate the bins and resolve interference features. Thirdly, as shown in \cite{Shvetsov-Shilovski2016}, there should be a factor of two in front of the potential. In the QTMC, this does not appear because the term $\dot{\mathbf{p}}\cdot \mathbf{r}$ that should appear in a phase-space action has been neglected. The use of the ADK rate does remove the caustics that were present in the TCSFA. This is because the ionization time and initial conditions will all be real. Thus, the problem of branch cuts is avoided.

\begin{figure}
	\centering
\hspace*{1cm}	\includegraphics[width=0.9\linewidth]{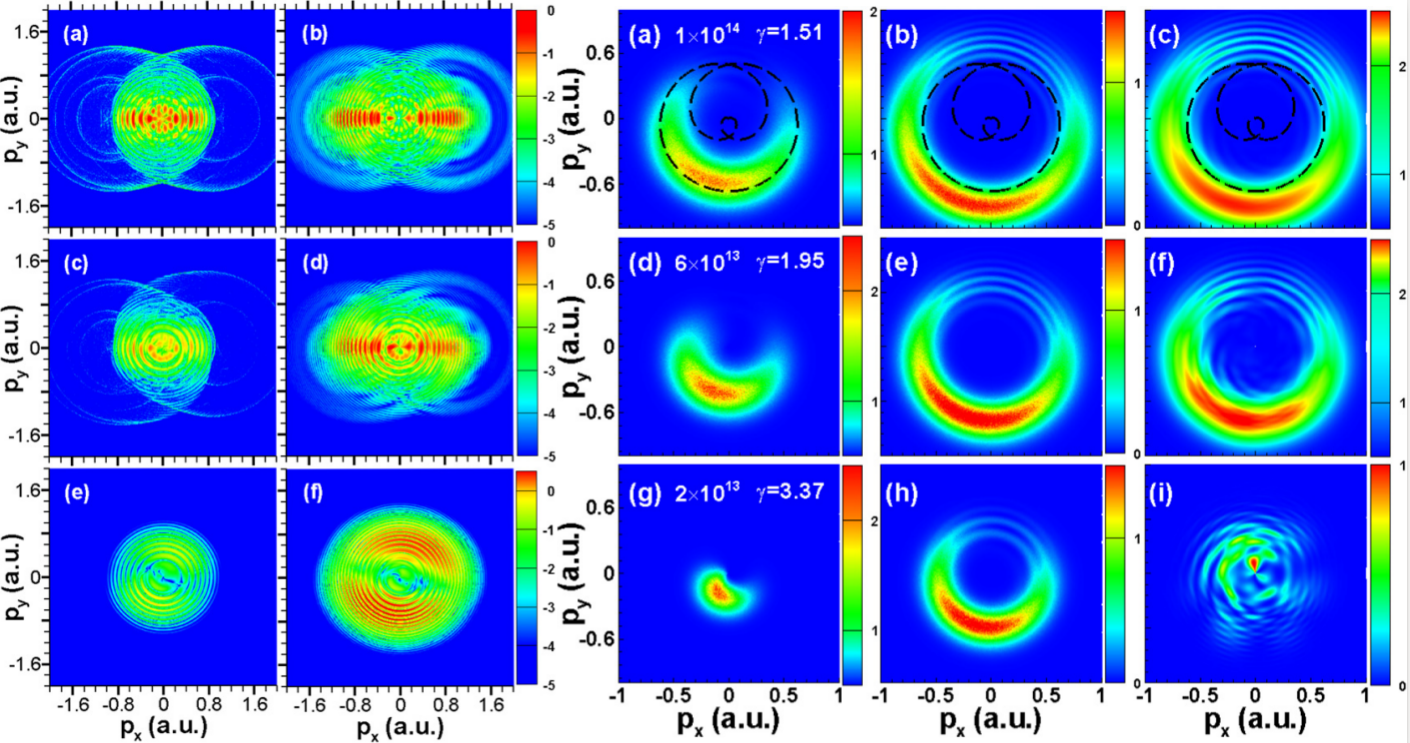}
	\caption{Photoelectron momentum distributions computed for H  in a laser field of wavelength $\lambda=800$ nm over four optical cycles. In the left panels, three different values of ellipticity have been employed: $\epsilon$ = 0 [panels (a) and (b)], $\epsilon$ = 0.15 [panels (c) and (d)] and $\epsilon$ = 0.75 [panels (e) and (f)]. Therein, the  distributions were calculated using the QTMC (left column) and TDSE (second column from left). In the right panels, a circularly polarized field and three different Keldysh parameters were taken: $\gamma$ = 1.51 [panels (a)-(c)], 1.95 [panels (d)-(f)], and 3.37 [panels (g)-(i)]. Therein, the distributions were calculated using the QTMC (the middle column), the CCSFA (the column second from right), and the TDSE (the right column). Throughout, a logarithmic scale has been used. The figure was presented as Figs.~1 and 2 of \cite{Geng2014} (left and right panels, respectively). }
	\label{fig:QTMC}
\end{figure}

The 2D-momentum dependent probability distributions produced by the QTMC have been compared with \textit{ab-initio} methods in \cite{Geng2014} and many qualitative features agree. For an example see Fig~\ref{fig:QTMC}. 
However, in the same study \cite{Geng2014} it was shown that the QTMC does not reproduce many qualitative features in the 2D-momentum dependent probability distributions, if fields with very high ellipticity are used. For instance, the peak of the momentum distribution over the radial momentum component differed by more 50\% over a range of intensities. This is due to the adiabatic/quasi-static approximation used in the ADK tunnelling rate. On the other hand, it was found the CCSFA (which is a non-adiabatic method) was able to give very good agreement with solutions of the TDSE. In response to this, in \cite{Li2016PRA}  an extension of the QTMC was proposed, where the non-adiabatic features of the SFA were introduced to replace the ADK tunnelling rate. In this new version of the theory, the SFA is used to calculate a tunnel-ionization rate, while the the Monte Carlo approach is used for continuum propagation. This makes it more similar to the TCSFA, but the phases and sub-barrier Coulomb-corrections in the ionization step are neglected \cite{Yan2012}. Given that the QTMC now uses a complex time, the tunnel exit is also complex. Thus, its imaginary part must be neglected to circumvent the difficulty of branch cuts.

The QTMC has been used to model the fan-like structure \cite{Li2014c}, the spider-like structure \cite{Geng2014,Richter2015} and distributions that contain features from both of these \cite{Li2014,Xie2016,Li2016PRA}. Because the QTMC is a trajectory-based description with a phase associated to each path, it is well placed to analyze holographic interference. However, given that a tunnelling rate is used, phases that come from the ionization step or the initial bound state of the atom or molecule are ignored. Such phases may be very important to capture some interference features \cite{Yan2012,Kaushal2018a,Maxwell2017}.

\subsubsection{Semiclassical Two-Step Model (SCTS).}
\label{sec:SCTS}

In the SCTS \cite{Shvetsov-Shilovski2016,Shvetsov-Shilovski2018,Shvetsov-Shilovski2019,Lopez2019} a similar approach to the QTMC is used, with a tunnelling rate calculated by ADK theory weighting the initial conditions. 
In order to improve the speed of the calculation, this weight is included via importance sampling, where initial conditions are chosen with a biased random distribution instead of a uniform one. This means that less computational time is wasted calculating trajectories with very low probabilities.
Additionally, the SFA is used to provide an initial condition for the initial momentum coordinate parallel to the laser field. Just as in the QTMC, the trajectories are propagated using Newton's equations of motion, the action is included for each trajectory via the classical limit and they are combined via a binning method is of the final momenta. This gives the final momentum-dependent probability distribution as
\begin{align}
\Omega_\epsilon(\mathbf{p}_{if})&=\left|\sum_{\mathbf{p}_{jf} \in \mathbb{S}^{\mathrm{bin}}_i}\exp(i S[\mathbf{r}_j])\right|^2
\\
S[\mathbf{r}_j]=&-\mathbf{p}_{j0}\cdot\mathbf{r}_{j0}+I_p t_{j0} \notag\\
&+\int_{t_{j0}}^{\infty}d t
\left(\frac{1}{2}\dot{\mathbf{r}}_j(t)^2+2V(\mathbf{r}_j(t))\right),
\end{align}
with tunnelling times $t_{j0}$ and $p_{j0\perp}$, $\mathbf{r}_j$ distributed via $\sqrt{W(t_{j0},p_{j0,\perp})}$ in order to compute the trajectories $\mathbf{r}_j(t)$. In this work \cite{Shvetsov-Shilovski2016} a careful analysis of the action in different representations (e.g. position space and phase-space) was presented, which leads to the inclusion of the $\dot{\mathbf{p}}\cdot \mathbf{r}$ term that was neglected in other approaches such as the TCSFA and QMTC. This is what leads to the factor of two in front of the potential, which also is used in the case of the CQSFA \cite{Maxwell2017}. The SCTS also calculates the asymptotic momentum by modelling the electron as a Kepler hyperbola once the laser field has been turned off at $t_f$. Using this the action from $t_f$ to $\infty$ may be calculated analytically, enabling a more efficient and accurate computation.

\begin{figure}
	\centering
	\includegraphics[width=0.9\linewidth]{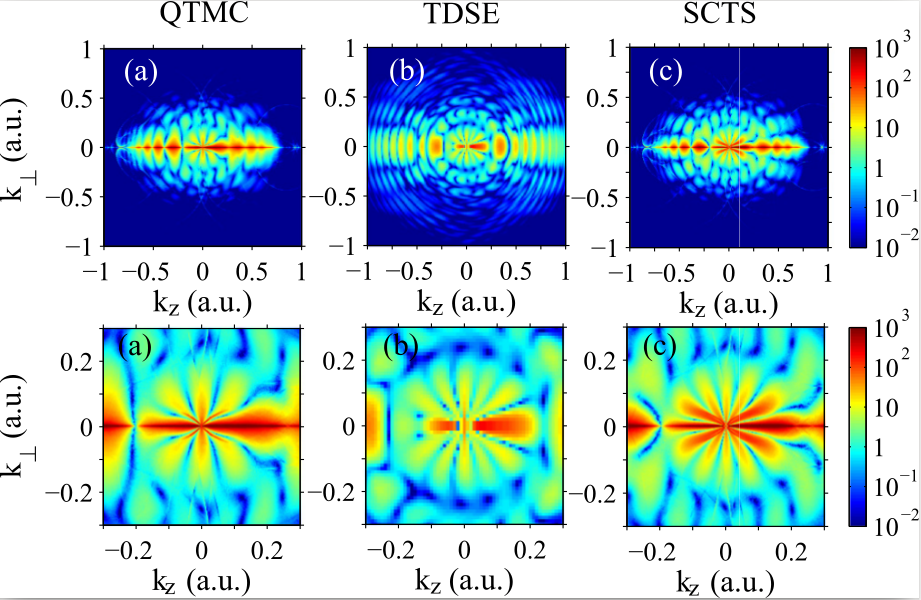}
	\caption{Momentum-dependent probability distributions computed for H in a laser pulse with peak intensity of 0.9 $\times$ 10$^{14}$ W/cm$^2$,  wavelength of 800~nm and duration of eight optical cycles. The distributions were calculated using the QTMC (a), the TDSE (b), and the SCTS (c). The bottom row is a magnification of the top for $|k_i|<0.3$ a.u., $i=\{z,\perp\}$. A logarithmic scale is used throughout. The figure was presented as Figs~1 and 2 in \cite{Shvetsov-Shilovski2016}. }
	\label{fig:SCTS}
\end{figure}

In Refs.~\cite{Shvetsov-Shilovski2016, Shvetsov-Shilovski2019,Lopez2019} a direct comparison of the SCTS, QTMC and \textit{ab-initio} methods is performed, see Fig.~\ref{fig:SCTS}. In the low-energy region it was found the STMC has an improved angular distribution, that is closer to the TDSE results than the QTMC. This was traced back to the improved initial conditions given by the SFA.  The SCTS, like the QTMC, can be applied to understand holographic interference patterns. 
The SCTS was also extended to model ionization of $H_2$, and again discrepancies with regard to the QTMC were identified in the low energy region \cite{Lopez2019}. With this extension, holographic imaging of small molecules using the SCTS is not far off. 

\begin{table*}
	\small
	\begin{tabular}{|g|c|c|c|c|c|c|c|c|c|}
		\hline
			\rowcolor{Yellow}
		Method & Non- & Sub-Barrier & Cont. Coulomb &
		\multicolumn{2}{c|}{Complex Traj.} & Direct/ & Collisions & $\mathbf{\dot{p}}\cdot \mathbf{r}$\\
		\rowcolor{Yellow}
		&Adiabatic&Corrections& Distortions.& Sub. & Cont. &Inverse&&\\
		\hline 
		CQSFA & Yes & Yes & Full & Yes & No& Inverse & S/D& Yes\\ \hline
		CCSFA & Yes & Yes & Approx. & Yes & Yes& Inverse & S/D& N/A\\ \hline
		TCSFA & Yes & Yes & Full & Yes & No& Direct & S/D& No\\ \hline
		EVA & Yes & Yes & Approx. & Yes & Yes& ? & D& N/A\\ \hline
		ARM & Yes & Yes & Approx. & Yes & Yes& ? & H/S/D& N/A\\ \hline
		QTMC & No/Yes & No & Full & No & No& Direct & S/D& No\\ \hline
		SCTS & No & No & Full & No & No& Direct & S/D& Yes\\ \hline
	\end{tabular}
	\caption{Summary of main features in the methods discussed in the present section whose orbits have been modified in order to incorporate the residual Coulomb potentials. The acronyms D, S, and H refer to deflections, soft collisions and hard collisions, respectively. The abbreviations Sub. and Cont. have been used to refer to the sub-barrier/ tunnelling and continuum propagation, respectively.}
	\label{table:comparison}
\end{table*}

\section{The Coulomb Quantum-Orbit Strong-Field Approximation}
\label{sec:CQSFA}
The CQSFA combines the power of the path integral description of the photoelectron with the semi-classical approximation. This enables the use of quantum trajectories that put the Coulomb and laser fields on the same footing, with neither being treated perturbatively. The CQSFA bears some resemblance to the TCSFA method and contains the same type of trajectories but there are a few key differences: 1) The CQSFA solves the inverse problem enabling a description with much fewer trajectories. 2) The TCSFA neglects the phase space action term $\mathbf{\dot{p}}\cdot \mathbf{r}$, included in the CQSFA. 3) The CQSFA considers perturbations in the action around the the saddle point solutions to second order, which leads to a stability prefactor term not included in any of the other models listed above. This term leads to important features that are visible in the final probability distributions, such as the ``holographic'' sidelobes \cite{Maxwell2017}. A new model that also uses a similar path integral approach, which is still currently under development \cite{Milosevic2017}, referred to as the semi-classical approximation, perforns an in-depth derivation of this prefactor.

In order to derive the Coulomb-quantum orbit strong-field approximation, we insert the closure relation $\int d
\mathbf{\tilde{p}}_0 |\mathbf{\tilde{p}}_0\rangle\langle
\mathbf{\tilde{p}}_0 | =1$ in Eq.~(\ref{eq:transitionamp}). This yields
\begin{align}
&M(\mathbf{p}_f)=\notag\\&-i \lim_{t\rightarrow \infty}
\int_{-\infty }^{t }\hspace*{-0.2cm}d t'
\int d \mathbf{\tilde{p}}_0 \left\langle  \mathbf{\tilde{p}}_f(t)
|U(t,t') |\mathbf{\tilde{p}}_0\right \rangle 
\left \langle
\mathbf{\tilde{p}}_0 | H_I(t')| \psi
_0(t')\right\rangle \, ,
\label{eq:Mpp}
\end{align}
where $|\mathbf{\tilde{p}}_f(t)\rangle=|\psi_{\mathbf{p}}(t)
\rangle$. The variables  $\mathbf{\tilde{p}}_0=\mathbf{p}_0+\mathbf{A}(t')$ and $\mathbf{\tilde{p}}_f(t)=\mathbf{p}_f+\mathbf{A}(t)$ 
give the electron's initial and final field-dressed momenta, respectively.  The matrix element $ \left\langle  \mathbf{\tilde{p}}_f(t)
|U(t,t') |\mathbf{\tilde{p}}_0\right \rangle $ is then calculated using time-slicing techniques  \cite{Kleinert2009,Milosevic2013JMP}. This leads to the CQSFA transition amplitude
\begin{align}
M(\mathbf{p}_f)&=-i\lim_{t\rightarrow \infty
}\int_{-\infty}^{t}dt' \int d\mathbf{\tilde{p}}_0
\int_{\mathbf{\tilde{p}}_0}^{\mathbf{\tilde{p}}_f(t)} \mathcal {D}'
\mathbf{\tilde{p}}  \int
\frac{\mathcal {D}\mathbf{r}}{(2\pi)^3}\notag\\
&\hspace{2cm}\times  
e^{i S(\mathbf{\tilde{p}},\mathbf{r},t,t')}
\langle
\mathbf{\tilde{p}}_0 | H_I(t')| \psi _0  \rangle \, ,
\label{eq:CQSFApath}
\end{align}
where $\mathcal{D}'\mathbf{p}$ and $\mathcal{D}\mathbf{r}$ are the integration measures for the path integrals \cite{Kleinert2009,Lai2015a}, and the prime indicates a restriction. The action in Eq.~(\ref{eq:CQSFApath}) is given by
\begin{equation}\label{stilde}
S(\mathbf{\tilde{p}},\mathbf{r},t,t')=I_pt'-\int^{t}_{t'}[
\dot{\mathbf{p}}(\tau)\cdot \mathbf{r}(\tau)
+H(\mathbf{r}(\tau),\mathbf{p}(\tau),\tau]d\tau,
\end{equation}
and the Hamiltonian reads as
\begin{equation}
H(\mathbf{r}(\tau),\mathbf{p}(\tau),\tau)=\frac{1}{2}\left[\mathbf{p}(\tau)+\mathbf{A}(\tau)\right]^2
+V(\mathbf{r}(\tau)),
\label{eq:Hamiltonianpath}
\end{equation}
where both the intermediate momentum $\mathbf{p}$ and coordinate $\mathbf{r}$ have been parameterized in terms of the time $\tau$ and the tilde indicates laser dressing, i.e., $\mathbf{\tilde{p}}=\mathbf{p}(\tau)+\mathbf{A}(\tau)$.

Eq.~(\ref{eq:Hamiltonianpath}) is then solved using saddle-point methods. This means that we must seek variables $\mathbf{p}$, $\mathbf{r}$ and $t'$ such that the action is stationary, i.e., its partial derivatives with regard to these variables vanish. This gives
\begin{equation}
\frac{\left[\mathbf{p}(t')+\mathbf{A}(t')\right]^2}{2}+V(\mathbf{r}(t'))+\dot{\mathbf{p}}(t')\cdot \mathbf{r}(t')=-I_p,
\label{eq:tunncc}
\end{equation}
\begin{equation}
\mathbf{\dot{p}}=-\nabla_rV(\mathbf{r}(\tau))
\label{eq:p-spe}
\end{equation}
and
\begin{equation}
\mathbf{\dot{r}}= \mathbf{p}+\mathbf{A}(\tau). 
\label{eq:r-spe}
\end{equation}
Eq.~(\ref{eq:tunncc}) gives the tunneling probability at $t'$, and Eqs.~(\ref{eq:p-spe}) and (\ref{eq:r-spe}) give the subsequent electron propagation. These equations are quite different from their SFA counterparts in the sense that, according to (\ref{eq:p-spe}) the field-dressed momentum is no longer conserved due to the presence of the Coulomb potential. 

Within the saddle-point approximation, the ATI transition amplitude reads
\begin{equation}
\label{eq:MpPathSaddle}
M(\mathbf{p}_f)\propto-i \lim_{t\rightarrow \infty } \sum_{s}\bigg\{\det \bigg[  \frac{\partial\mathbf{p}_s(t)}{\partial \mathbf{r}_s(t_s)} \bigg] \bigg\}^{-1/2} \hspace*{-0.6cm}
\mathcal{C}(t_s) e^{i
	S(\mathbf{\tilde{p}}_s,\textbf{r}_s,t,t_s)} ,
\end{equation}where $t_s$, $\mathbf{p}_s$ and $\mathbf{r}_s$ are determined by Eqs.~(\ref{eq:tunncc})-(\ref{eq:r-spe}), the term in brackets is associated with the stability of the orbit, and $\mathcal{C}(t_s)$ is given by
\begin{equation}
\label{eq:Prefactor}
\mathcal{C}(t_s)=\sqrt{\frac{2 \pi i}{\partial^{2}	S(\mathbf{\tilde{p}}_s,\textbf{r}_s,t,t_s) / \partial t^{2}_{s}}}\langle \mathbf{p}+\mathbf{A}(t_s)|H_I(t_s)|\Psi_{0}\rangle.
\end{equation}
In practice, we use the stability factor $\partial
\mathbf{p}_s(t)/\partial \mathbf{p}_s(t_s)$ instead of that stated in Eq.~(\ref{eq:MpPathSaddle}), which may be obtained employing a Legendre transformation. As long as the electron starts from the origin, this choice will not affect the action. We normalize  Eq.~(\ref{eq:MpPathSaddle}) so that the SFA transition amplitude is obtained in the limit of vanishing binding potential, and, unless otherwise stated, take the electron to be initially in a $1s$ state \cite{Lai2015a}. We will refer to the product of the stability factor with Eq.~(\ref{eq:Prefactor}) as ``the prefactor''.

One should note that, although the initial momentum states $\left |\mathbf{\tilde{p}}_0\right \rangle $ form a complete basis, by inserting the above closure relation we pay a price. While the CQSFA is very good for reproducing the dynamics of continuum and possibly high-lying bound states, it is not accurate for describing processes involving deeply bound excited states, such as excitation. For that purpose, a closure relation $\int d
\tilde{\phi}_c |\tilde{\phi}_c\rangle\langle
\tilde{\phi}_c|+\sum_n |\psi_n\rangle\langle
\psi_n|=1$ involving the exact continuum states and the system's bound states would be more appropriate. In this case, however, how to apply the time slicing in order to obtain path integrals would be much less straightforward. 
Furthermore, the presence of the Coulomb potential, together with the fact that Eq.~(\ref{eq:tunncc}) has complex solutions, brings a great deal of difficulty. First, since the initial conditions are taken at the origin, a singularity is expected at $V(\mathbf{r}(t'))$. Second, strictly speaking, Eqs.~(\ref{eq:tunncc})-(\ref{eq:r-spe}) should be solved in the complex plane. This means that such equations themselves will contain poles or branch cuts. In order to solve these equations, we have performed a series of approximations, which have been improved from publication to publication. They will be briefly outlined below. For details, we refer to the original papers \cite{Lai2017,Maxwell2017,Maxwell2017a,Maxwell2018,Maxwell2018b}.
\subsubsection{Contours and explicit expressions}
\label{subsubsec:contours}
Unless otherwise stated, we compute the ATI transition amplitude along a two-pronged contour, whose first and second parts are parallel to the imaginary and real-time axis, respectively. This is the most widespread contour used in the implementation of Coulomb-corrected approaches \cite{Popruzhenko2008a,Yan2012,Torlina2012,Torlina2013}, and it is very convenient for subsequent approximations. The contour starts at the complex time  $t'=t'_r+it'_i$. Its first arm, from $t'$ to $t'_r$, may be related to the sub-barrier dynamics, and the second arm of the contour, from the real ionization time $t'_r$  to the final time $t$, can be associated with continuum propagation.  The total action is then given by \begin{equation}
S(\mathbf{\tilde{p}},\mathbf{r},t,t')=S^{\mathrm{tun}}(\mathbf{\tilde{p}},\mathbf{r},t'_r,t')+S^{\mathrm{prop}}(\mathbf{\tilde{p}},\mathbf{r},t,t_r').
\end{equation}

In the first arm of the contour, the electron momentum is assumed to be constant and is kept as $\mathbf{p}_0$.  Physically, this means that the acceleration caused by the Coulomb potential is neglected in the under-the-barrier dynamics.
The action in the first arm of the contour then reads
\begin{align}
S^{\mathrm{tun}}(\mathbf{\tilde{p}},\mathbf{r},t'_r,t')&=I_p(it'_i)-\frac{1}{2}\int_{t'}^{t'_r}\left[ \mathbf{p}_0+\mathbf{A} (\tau)\right]^2d\tau \notag\\ &-\int_{t'}^{t'_r}V(\mathbf{r}_0(\tau))d\tau, \label{eq:stunn}
\end{align}where $\mathbf{r}_0$ is defined by 
\begin{equation}
\mathbf{r}_0(\tau)=\int_{t'}^{\tau}(\mathbf{p}_0+\mathbf{A}(\tau'))d\tau',
\label{eq:tunneltrajectory}
\end{equation} 
which will be referred to as ``the tunnel trajectory''. This type of trajectory has also been defined in many publications, in which the EVA \cite{Smirnova2008} or the CCSFA \cite{Yan2010,Yan2012} have been used.  
For the Coulomb potential, this approximation renders Eq.~(\ref{eq:tunncc}) formally identical to its SFA counterpart (\ref{eq:tunnelsfa}). This is highly advantageous as in some cases, such as monochromatic linearly polarized fields, this can be solved analytically.
In practice, however, the CQSFA solution for the tunnel ionization equation will be different from that obtained with the SFA. It must be matched to the saddle-point equations (\ref{eq:p-spe}) and (\ref{eq:r-spe}) giving the continuum propagation in the \textit{presence} of the Coulomb potential. Thus, we must use the initial rather the final momentum as the independent variable. The action $S^{\mathrm{tun}}(\mathbf{\tilde{p}},\mathbf{r},t'_r,t')$ inside the barrier is then calculated from the origin until the tunnel exit $\mathbf{r}_0(t'_r)$. 

A widely used approximation is to take a real tunnel exit 
\begin{equation}\label{eq:exitreal}
z_0=\mathrm{Re}[r_{0z}(t'_r)].
\end{equation}
Although this will considerably simplify the computations by rendering the trajectories in the continuum real, this makes the problem contour dependent.  The main issue is that, while the actual problem should be contour independent and satisfy the Cauchy Riemann conditions, with this approximation analyticity is lost. Furthermore, from a physical point of view, this  assumption cannot be justified, as the only  variable that must be real is the final momentum at the detector. 
Still, as we will discuss next this will lead to an overall good agreement with \textit{ab-initio} computations and avoids the issue of branching points and branch cuts.
Under  assumption (\ref{eq:exitreal}), the imaginary parts of the variables that will determine the single-orbit ionization probabilities, and probability densities, will be ``picked up'' in the sub-barrier part of the contour. The subsequent propagation will mainly lead to phase shifts in the real part of the action, which will influence the holographic patterns. Inclusion of the imaginary part of the tunnel exit will lead to complex trajectories, which may cause further suppression or enhancement in the ATI transition amplitudes during the continuum propagation. Complex trajectories may also lead to effective, further deceleration or acceleration, with regard to their real counterparts. This latter effect has been highlighted in \cite{Torlina2013}, and is related to the shifting probabilities of the trajectories in the continuum.

Another issue that must be addressed is the fact that there will be a logarithmic divergency at the endpoint $t'$ of the contour, as in this case the tunnel trajectory (\ref{eq:tunneltrajectory}) vanishes. We have applied several strategies in order to deal with this issue, such as introducing a very small imaginary time, close to $t'$ and computing the limit of this extra term tending to zero \cite{Lai2015a,Maxwell2017}, or using the regularization method outlined in \cite{Popruzhenko2014a,Popruzhenko2014b}, in which the contribution to the action coming from the Coulomb potential is matched to the asymptotic value of the bound-state Coulomb wave function. Further restrictions to the contour will be dictated by the condition $\mathbf{r}^2=0$, which is not necessarily at the origin. 
For more detailed discussions we refer to our recent publication \cite{Maxwell2018b}.

In the second part of the contour, the action reads
\begin{align}
S^{\mathrm{prop}}(\mathbf{\tilde{p}},\mathbf{r},t,t'_r)&=I_p(t'_r)-\frac{1}{2}\int_{t'_r}^{t}\left[ \mathbf{p}(\tau)+\mathbf{A} (\tau)\right]^2d\tau \notag
\\& - 2\int_{t'_r}^{t}V(\mathbf{r}(\tau))d\tau, \label{eq:sprop2}
\end{align}
where the factor 2 in front of the Coulomb integral is obtained as, for the specific, Coulomb-type potential used in our publications, $\mathbf{r} \cdot \dot{\mathbf{p}}=V(\mathbf{r}(\tau))$. This extra factor is essential in order to obtain a good agreement with \textit{ab-initio} methods, such as the right number of fringes in the near-threshold fan-shaped structure \cite{Lai2017,Maxwell2017,Shvetsov2016}. 
It is important to bear in mind that concepts such as ``the tunnel exit'' have raised considerable debate.  Indeed, if analyticity is preserved, the problem should be contour independent. This would render the tunnel exit arbitrary and hence of no physical significance  \cite{Popruzhenko2014b}. 
It is convenient to write  the action (\ref{eq:sprop2}) in terms of $\mathbf{p}(\tau)=\mathbf{\mathcal{P}}(\tau)+\mathbf{p}_f$, such that $\mathbf{\mathcal{P}}\rightarrow 0$ when $\tau \rightarrow t$.
This leads to 
\begin{align}
S^{\mathrm{prop}}(\mathbf{\tilde{p}},\mathbf{r},t,t'_r)&=I_p(t'_r)-\frac{1}{2}\int_{t'_r}^{t}\left[ \mathbf{p}_f+\mathbf{A} (\tau)\right]^2d\tau \notag \\ &+S_{\mathcal{P}}(\mathbf{\mathcal{P}},t,t'_r) +S^{\mathrm{prop}}_{C}(\mathbf{r},t,t'_r), \label{eq:sprop3}
\end{align}
where 
\begin{equation}
S_{\mathcal{P}}(\mathbf{\mathcal{P}},t,t'_r)=-\frac{1}{2}\int_{t'_r}^{t}\mathbf{\mathcal{P}}(\tau) \cdot \left[ \mathbf{\mathcal{P}}(\tau) +2 \mathbf{p}_f +2 \mathbf{A}(\tau)
\right]
\label{eq:accphase}
\end{equation}
is the phase difference due to the acceleration caused by the Coulomb potential, and 
\begin{equation}
S^{\mathrm{prop}}_{C}(\mathbf{r},t,t'_r)=\int_{t'_r}^{t}\frac{2C}{\sqrt{\mathbf{r}(\tau)\cdot \mathbf{r}(\tau)}}d\tau
\label{eq:Coulombphase}
\end{equation}
is the Coulomb phase acquired during the electron's propagation in the continuum. These two phases, together with the sub-barrier Coulomb phase 
\begin{equation}
S^{\mathrm{tun}}_{C}(\mathbf{r},t'_r,t')=\int_{t'}^{t'_r}\frac{C}{\sqrt{\mathbf{r}_0(\tau)\cdot \mathbf{r}_0(\tau)}}d\tau,
\label{eq:subCoulombphase}
\end{equation}
vanish if the Coulomb coupling is zero, i.e., in the limit $C\rightarrow 0$. Then $V(\mathbf{r})\rightarrow 0$, $\mathbf{p}_f\rightarrow \mathbf{p}_0\rightarrow\mathbf{p}$ and $\mathbf{\mathcal{P}}\rightarrow 0$, so that the SFA action (\ref{eq:Sdir}) is recovered.

\subsubsection{Analytic approximations}
\label{sec:analyticCQSFA}
For a monochromatic, linearly polarized driving field, Eqs.~(\ref{eq:stunn}) and (\ref{eq:sprop2}) describing the action in the sub-barrier and continuum parts of the contour can be computed analytically using a series of approximations. Thereby, a key issue is how to calculate the phases (\ref{eq:accphase}), (\ref{eq:Coulombphase}) and (\ref{eq:subCoulombphase}). This was the main topic of our previous publication \cite{Maxwell2017a}, to which we refer for details. Below we briefly sketch the main points to be considered. 
\begin{enumerate}
	\item[1)] \textit{Sub-barrier dynamics.} In order to compute the sub-barrier action analytically, we apply the long wavelength approximation to the imaginary part of the ionization time to first order. This leads to convenient approximations for the tunnel trajectory, and to an analytic solution for the sub-barrier Coulomb phase (\ref{eq:subCoulombphase}). It is worth stressing that this is an excellent approximation to the full CQSFA solution. This is justified by the fact that, in the full CQSFA, the momentum along the tunnel trajectory is already approximated by a constant value $\mathbf{p}_0$. Thus, the only difference between both models is the long wavelength approximation applied to $\mathrm{Im}[t']$ and to hyperbolic functions that depend on it.  The resulting transition amplitudes may then be factorized into a Coulomb- and an SFA-like part, which are very useful for interpretational purposes \cite{Maxwell2017,Maxwell2017a}. 
	\item[2)] \textit{Continuum propagation. } To find approximations that render the action $S^{\mathrm{prop}}_{C}(\mathbf{r},t,t'_r)$ analytically solvable is more involved, as one cannot assume the intermediate momentum is fixed. It is however possible to find analytic expressions that make physical sense by (i) using the long-wavelength approximation around relevant times (not necessarily the ionization time); (ii) finding simple approximate functions for the intermediate momentum when dealing with the acceleration integral (\ref{eq:accphase}); (iii) assuming that the field-dressed momentum is constant or piecewise constant when computing the Coulomb phase (\ref{eq:Coulombphase}). Typically, we take the difference $\mathbf{\mathcal{P}}(\tau)$ between a generic momentum $\mathbf{p}_j$ at a  relevant time and the final momenta $\mathbf{p}_f$ to be exponentially decaying and to vanish at the final time $t$. The relevant times for the long-wavelength approximation in (i) and the number of subintervals chosen in (iii) will depend on the orbit considered and will carry different physical meanings. One may construct a single or more subintervals using  the time of ionization, recollision, etc. If the CQSFA orbit is relatively close to its Coulomb free, SFA counterpart, it suffices to assume a single interval from $t'_r$ to $t$ for the continuum propagation, use the final momentum $\mathbf{p}_f$ to compute the Coulomb phase $S^{\mathrm{prop}}_{C}(\mathbf{r},t,t'_r)$ and impose that, in this time interval, the momentum will tend monotonically from $\mathbf{p}_0$ to $\mathbf{p}_f$. 
	\item[3)] \textit{Incorporating soft recollision.} If, for a particular orbit, there is pronounced interaction with the core, apart from the approximations employed in 1) and 2), one must incorporate one act of rescattering in order to obtain a good approximation for the full CQSFA solution. Specifically, the continuum propagation will require two subintervals: (i) from the ionization time $t'_r$ to the time $t_c$ of closest approach to the core, for which the electron coordinate parallel to the core vanishes, and (ii) from $t_c$ to the final time $t$. We also assume that, in the first subinterval, the electron momentum perpendicular to the laser field polarization will remain the same, while the parallel momentum can be computed using 
	\begin{equation}
	\int_{t'_r}^{t_c}(p_{oc\parallel}+A(\tau))d\tau +z_{o0}=0,
	\label{eq:p3c}
	\end{equation}
	where $p_{oc\parallel}$ is the parallel component of the momentum at the instant of recollision, and $z_{o0}$ is the tunnel exit for a generic orbit $o$ that interacts strongly with the core. 
	Upon recollision, the energy of the electron is conserved, the momentum changes instantaneously $\mathbf{p}_{oc}$ to $\mathbf{p}_{o}$  and the rescattering angle does not change until the end of the pulse. These assumptions were ``borrowed'' from rescattered ATI, with the difference that, in our analytic approach, rescattering does not occur at the origin, but at a distance $r_c$ perpendicular to the polarization axis. In the last subinterval, we assume the same monotonic behavior as for the previous orbits for the phase $S_{\mathcal{P}_o}(\mathbf{\mathcal{P}_o},t,t'_r)$ associated with the acceleration caused by the Coulomb potential in the continuum. 
\end{enumerate}
An advantage of these analytical expressions is that they allow a detailed assessment of how the phases (\ref{eq:subCoulombphase}), (\ref{eq:Coulombphase}), and (\ref{eq:accphase}) influence the continuum dynamics and the holographic patterns. A drawback of the present analytic approach is that it is not a stand-alone model and requires the times of closest approach $t_c$ to be obtained from the full CQSFA. 
\subsubsection{Treatment of branch cuts and singularities}
\label{subsubsec:branchcuts1}
The ideal scenario and most rigorous way to proceed would be to relax the assumptions in the previous section and solve the full complex problem. One should note, however, that the Coulomb phases (\ref{eq:Coulombphase}) and (\ref{eq:subCoulombphase}) 
present in CQSFA the transition amplitude contain a square root of the complex variable $u=\mathbf{r}(\tau)\cdot \mathbf{r}(\tau)$. This means that, if saddle-point methods are used,  there will be branch cuts and two Riemann sheets for nonvanishing perpendicular final momenta. If the standard convention is applied, the branch cuts will lie in the negative real half axis of $\mathbf{r}(\tau)\cdot\mathbf{r}(\tau)$, i.e., for
\begin{equation}
\mathrm{Re}\left( \mathbf{r}(\tau)\cdot\mathbf{r}(\tau) \right)<0 \qquad \mathrm{and} \qquad\mathrm{Im}\left( \mathbf{r}(\tau)\cdot\mathbf{r}(\tau) \right)=0.
\label{eq:Branchcut}
\end{equation}
Thereby, an important issue is that only one of the two Riemann sheets leads to physically meaningful results. Hence, it is vital to define a contour so that branch cuts are not crossed and the physical Riemann sheet is employed. This implies that the branch cuts must be mapped into the complex time domain, as $\tau$ is the variable with regard to which $\mathbf{r}\cdot\mathbf{r}$ is parametrized. 
This is a non-trivial task if the residual Coulomb potential is incorporated in the electron's (complex) equations of motion for $\mathbf{r}(\tau)$ and $\mathbf{p}(\tau)$, as they will have branch cuts themselves. A first approximation is to employ Coulomb-free trajectories. Physically, this approximation works if 
the Coulomb acceleration is not substantial, such as for driving fields of large ellipticity, but is inaccurate for linearly polarized fields \cite{Maxwell2018b}. Still, it illustrates the branch cuts' overall behavior even in the latter case.  
\begin{figure*}
	\begin{center}
		\includegraphics[width=\linewidth]{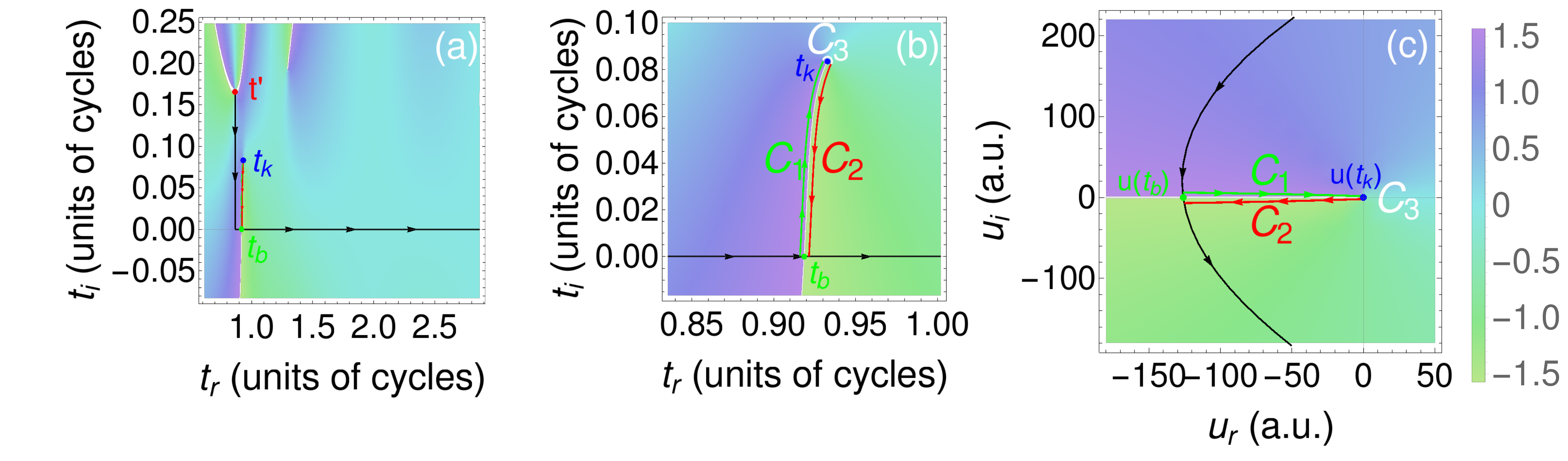}
	\end{center}
	\caption{Contour around a branch cut built for hydrogen in a linearly polarized monochromatic field of intensity $I=2 \times 10^{14}\hspace*{0.1cm}\mathrm{W/cm}^2$ and wavelength $\lambda = 800 \hspace*{0.1cm}\mathrm{nm}$ following the steps outlined in this section, using Coulomb-free trajectories and the momentum components $(p_{f\parallel},p_{f\perp})=(-1.4 \hspace*{0.1cm}\mathrm{a.u.}, 0.7\hspace*{0.1cm}\mathrm{a.u.})$. Panel (a) shows the standard contour in the complex time plane, while panel (b) provides a zoom in of the region around the branch cut. Panel (c) provides the contour and the branch cut in terms of the complex variable $u=\mathbf{r} \cdot \mathbf{r}$. From \cite{Maxwell2018b}.}
	\label{fig:BranchDiagram}
\end{figure*}

A temporal mapping using Coulomb-free trajectories for linearly polarized fields shows that the branch cuts occur in pairs and are separated by gaps \cite{Popruzhenko2014b,Pisanty2016}. 
In \cite{Pisanty2016}, a prescription for building contours that make use of this structure has been proposed. First, one calculates the middle point of each of the above-mentioned gaps. Subsequently, for a specific momentum value, a contour that passes through the middle point of each such gates is sought. The branching point pairs are solutions of $\mathbf{r}(\tau)\cdot\mathbf{r}(\tau)=0$, while the condition $\mathbf{r}(\tau)\cdot\mathbf{v}(\tau)=0$ give the times of closest approach, which are located at the center of the gate, as well as the classical turning points. 

This method has been applied in \cite{Keil2016} to above-threshold ionization in the context of soft recollisions under the Coulomb barrier. Therein, the trajectories were taken to be Coulomb-free.  It was shown to improve the agreement of the SFA and \textit{ab-initio} computations in orders of magnitude, especially near the direct ATI cutoff energy of $2U_p$. This was a counter-intuitive  and rather intriguing result related to  under-the-barrier collisions. 
Nonetheless, its implementation poses a series of difficulties. First, one must find the midpoints in advance, which may not always be possible. For instance, this could prove particularly difficult for Coulomb-distorted trajectories.
Second, to calculate photoelectron momentum distributions using this method, it would be necessary to construct and apply a different contour for each pair of final electron-momentum components.

Therefore, we apply a different method: we keep the standard two-pronged contour discussed in Sec.~\ref{subsubsec:contours}, whose first and second arms are parallel to the imaginary and the real time axis, respectively. After computing the trajectory, we test whether it crosses a branch cut. If it does, we deform the contour in such a way that it goes around the branch cut.  If a specific time  $t_b$ along the second arm of the contour satisfies the branch-cut conditions (\ref{eq:Branchcut}), it may be employed to find the branching point $t_k$ defined  by
\begin{align}
\mathbf{r}(t_k)\cdot\mathbf{r}(t_k) = 0+0 i.
\label{eq:BranchEnd}
\end{align}

The intersection $t_b$ of the branch cut with the real axis and the complex time $t_k$ associated with the branching point can then be used to build a three-pronged contour along which the Coulomb phase is computed. Explicitly,
\begin{equation}
S^{(cut)}_{C}=\int_{c_1}V[\mathbf{r}(\tau)]d \tau+\int_{c_2}V[\mathbf{r}(\tau)]d \tau+\int_{c_3}V[\mathbf{r}(\tau)]d \tau,
\label{SC-0}
\end{equation}
where $c_1$ goes from $t_b$ to $t_k$ following the upper side of the branch cut, $c_2$ returns from $t_k$ to $t_b$ along the other side of the cut and $c_3$ connects them together with a semi-circle around $t_k$. This approach is applicable if the transverse momentum component is non-vanishing. For vanishing transverse momentum the branching points will merge into a pole and the integrals along $c_1$ and $c_2$ become divergent.  An example of this contour is presented in Fig.~\ref{fig:BranchDiagram}. Physically, a branch cut may be associated with a collision (see a detailed discussion in \cite{Maxwell2018b} and a brief explanation in Sec. \ref{sec:complextrajs}). 

In order to make the problem tractable for Coulomb-distorted complex trajectories, additional approximations must be made. First, we assume that their imaginary parts behave as their Coulomb-free counterparts during the continuum propagation. This means that they will be kept constant when the time contour is chosen to be along the real time axis. This approximation may be loosely justified as follows: (i) In the sub-barrier part of the contour, we neglect the Coulomb potential for the CQSFA, thus describing the dynamics in the same way as in the standard, Coulomb-free SFA; (ii) The final momentum $\mathbf{p}_f$ must be real, as it is an observable, which means that the $\lim\limits_{t\rightarrow \infty} \mathrm{Im}[\mathbf{r}(t)] =const.$  If a branch cut is met, we move vertically in the complex time plane by assuming that the influence of the slowly varying Coulomb field on the electron intermediate momentum can be neglected. This is consistent with the physical picture for which collisions happen in much shorter time scales than the driving-field cycle. The approximations described in (ii) simplify the problem by  rendering all momenta real in the second part of the contour.

One should note that the treatment of branch cuts and singularities in Coulomb-distorted strong-field approaches is a contentious issue. For instance in \cite{Cajiao2015,He2018b} it has been suggested that, when the electron is close to the core, such features are absent. Therein, an effective potential has been derived with no singularities or branch cuts, and it has been argued that the classical action employed in most Coulomb-distorted approaches is a limit that is only valid for negligible quantum spreading of the electronic wave packet. A softened effective potential is also obtained in the coupled coherent states method, in which the full Hamiltonian is expanded into a trajectory-guided, coupled coherent state basis \cite{Zagoya2014}. Rigorous ways of treating branch cuts, caustics and singularities are more established in other areas of research (see, e.g., \cite{Koch2018,Koch2018b} for in depth discussions).

\section{Orbits, transition amplitudes and holographic patterns}
\label{sec:results}
In this section, we highlight the main results of our previous publications, but provide a more unified, broader discussion. Such results have been obtained with linearly polarized monochromatic fields of frequency $\omega$ and amplitude $E_0=2\omega \sqrt{U_p}$, where $U_p$ is the ponderomotive energy. The corresponding vector potential reads
\begin{equation}
\mathbf{A}(t)=2\sqrt{U_p} \cos \omega t \hat{e}_z.
\end{equation}
This simplifying assumption has several advantages. First, the ionization times can be written analytically as functions of the momentum components parallel and perpendicular to the driving-field polarization. Second, the periodicity of the field facilitates the interpretation of many holographic patterns and can be used in analytical estimates. 
For this particular case, the CQSFA action obtained combining the sub-barrier and continuum contributions reads
\begin{eqnarray}
S(\mathbf{\tilde{p}}, \mathbf{r},t,t')
& =&\left(I_p + U_p \right) t'
+\frac{1}{2}\mathbf{p}^2_f t'_{r}+\frac{i}{2}\mathbf{p}^2_{0}  t'_{i}
+\frac{U_p}{2\omega}\sin(2\omega t')
\notag\\&&   
+ \frac{2\sqrt{U_p}}{\omega}\left[ p_{ 0\parallel}\sin(\omega t')-(p_{0\parallel}-p_{f\parallel})\sin(\omega t'_{r}) \right] 
-\int^{t'_{r}}_{t'} V(\mathbf{r}_{0}(\tau))\mathrm{d}\tau
\nonumber\\& &
-\frac{1}{2}\int_{t'_{r}}^{t}\mathbf{\mathcal{P}}(\tau)\cdot (\mathbf{\mathcal{P}}(\tau)+2\mathbf{p}_f+2\mathbf{A}(\tau))\mathrm{d}\tau
-2 \int^{t}_{t'_{r}} \hspace*{-1mm}V(\mathbf{r}(\tau))\mathrm{d}\tau, 
\label{eq:CQSFA_mono}
\end{eqnarray}
where $p_{\nu \parallel}$ and  $p_{\nu \perp}$, with $\nu=0,f$, give the momentum components parallel and perpendicular to the laser-field polarization. The explicit expression for the direct ATI transition amplitude within the SFA is given by
\begin{align}
\label{eq:ActionSFAmono}
S_d(\mathbf{p},t')&=\left(\frac{p^2_{\parallel}+p^2_{\perp}}{2}+I_p+U_p\right)t'\notag\\ &+\frac{2 p_{\parallel}\sqrt{U_p}}{\omega}\sin[\omega t']+\frac{U_p}{2 \omega}\sin[2\omega t'],
\end{align}
where the electron momentum remains constant throughout (i.e., $\mathbf{p}_0=\mathbf{p}_f=\mathbf{p}$). 
A direct inspection of the SFA transition amplitude (\ref{eq:ActionSFAmono}) shows that their dominant term is proportional to $t'$. The remaining terms are either bounded, or increase more slowly. This implies that the rough behavior of the action is expected to mirror that of the solutions of the saddle-point equations. 
For the CQSFA transition amplitude (\ref{eq:CQSFA_mono}), this will depend on the orbit in question. If the orbits remain in regions for which the laser field is dominant, i.e., if it is a direct orbit or if the Coulomb potential merely deflects the electron, then the dominant term will be proportional to  $t'$. If the electron crosses a region in which the residual binding potential becomes dominant, then the Coulomb phase may prevail and further investigation is necessary. 

\subsection{Trajectory classification}
\label{sec:orbits}
The solutions of the CQSFA saddle-point equations and of their SFA counterparts can be associated with electron trajectories, whose quantum interference leads to the  ultrafast holographic patterns. For the SFA transition amplitude(\ref{eq:ATIdirect}) leading to direct ATI, the only integration variable is the ionization time $t'$. Within the steepest descent method, $t'$ can be obtained by solving Eq.~(\ref{eq:tunnelsfa}). Specifically for linearly polarized monochromatic fields,  this equation has an analytical solution. Thus,  one may parametrize the ionization times in terms of the electron momentum components parallel and perpendicular to the laser-field polarization. 
The general SFA solution related to an event $e$ within a cycle $c$ is 
\begin{align}
t'_{ec} = \frac{2\pi n}{\omega}\pm\frac{1}{\omega} \arccos\left(\frac{-p_{\parallel}\mp i\sqrt{2 I_\mathrm{p}+p^2_{\perp}}}{2\sqrt{U_\mathrm{p}}}\right),
\label{eq:timesSFA}
\end{align}
where $n$ is an integer.  The times predicted by Eq.~(\ref{eq:timesSFA}) occur in conjugate pairs, and only those in the upper half complex plane, i.e., fulfilling the condition $\mathrm{Im}[t'_{ec}]>0$, have physical significance. 

\begin{figure}
	\centering
	\includegraphics[width=0.9\linewidth]{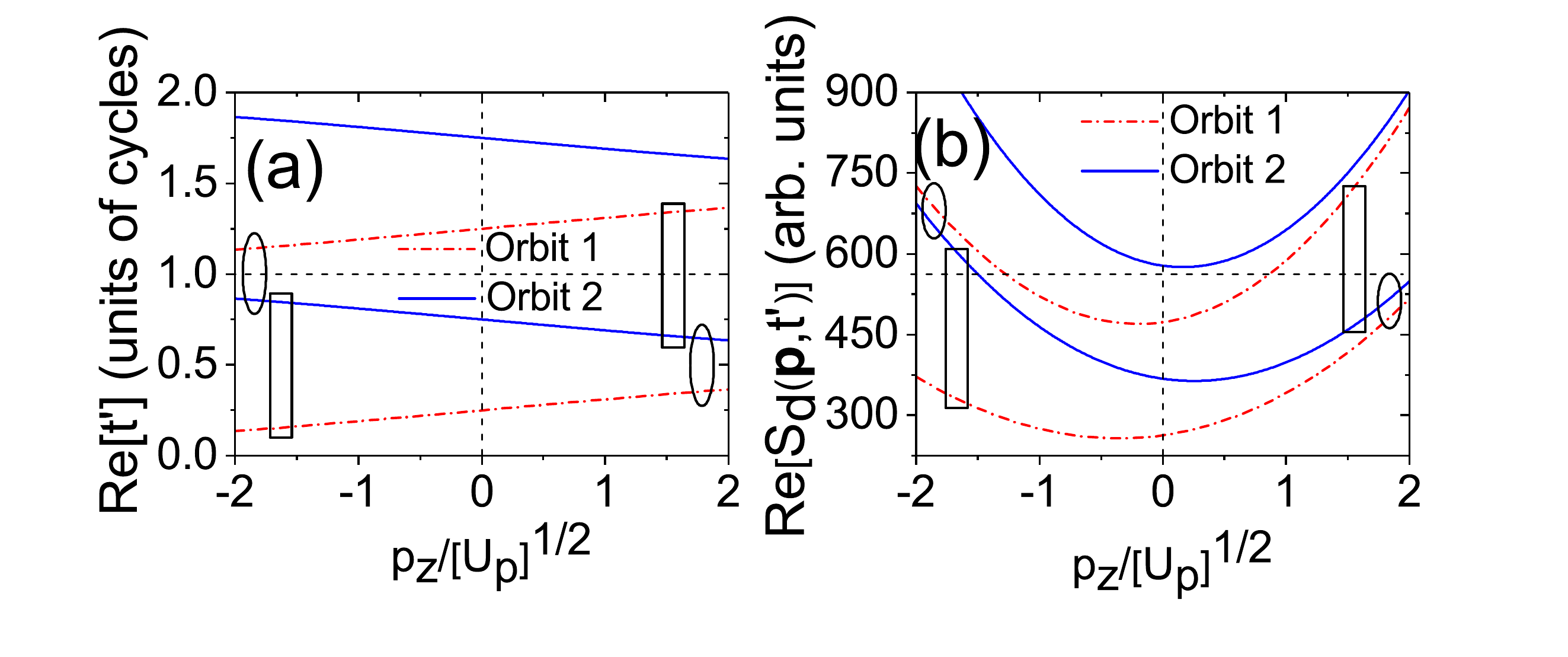}
	\caption{Real parts of the ionization times  and actions as functions of the electron momentum component $p_{\parallel}$ parallel to the driving-field polarization [panels (a) and (b), respectively], computed using the direct SFA for vanishing perpendicular momentum component  $p_{\perp}$.  The circles (squares) indicate interfering orbits for which the difference $\mathrm{Re}][t'_{2c}-t'_{1c}]$ is less than (greater than) half a cycle, which lead to type A (type B) intracycle interference. We have renormalized $p_{\parallel}$ in terms of $\sqrt{U_p}$. From \cite{Maxwell2017}.}\label{fig:MomReTimes}
\end{figure}

Within a single cycle, the SFA exhibits two solutions: the electron may either be released in the direction of the detector, or half a cycle later in the opposite direction. In the latter case, the electron will be freed in the ``wrong'' direction, brought back to its parent ion and then reach the detector. These solutions are known as the short orbit and the long orbit, respectively, and have been briefly mentioned in the context of the temporal double slit (Sec.~\ref{subsec:doubleslits}).
Specifically, we use the solutions
\begin{align}
t'_{1c} &= \frac{2\pi}{\omega}-\frac{1}{\omega} \arccos\left(\frac{-p_{\parallel}+ i\sqrt{2 I_\mathrm{p}+p^2_{\perp}}}{2\sqrt{U_\mathrm{p}}},\right) \label{eq:t1s}\\
t'_{2c}&= \frac{1}{\omega} \arccos\left(\frac{-p_{\parallel}- i\sqrt{2 I_\mathrm{p}+p^2_{\perp}}}{2\sqrt{U_\mathrm{p}}}\right) \label{eq:t2s}
\end{align} 
for each cycle. Solution $t'_{1c}$ ($t'_{2c}$) corresponds to the short (long) orbit for $p_{\parallel}>0$, with the scenario reversing for $p_{\parallel}<0$. Their imaginary parts will be the same, and their real parts will differ. Fig.~\ref{fig:MomReTimes} shows the behavior of the real parts of the two SFA solutions and of the corresponding actions, as functions of the parallel momentum component.  Therein, analytic expressions for the SFA case can also be found.  In general, the expression for the coordinate $\mathbf{r}$ obtained from the SFA is complex. This will also be the case in the CQSFA as the tunnel exit $\mathbf{r}(t'_r)$ is not real if no further approximations are used. 

Once the Coulomb potential has been included, the number and the topology of the orbits will change. The central Coulomb force will act upon the transverse momentum component of the electron and thus considerably alter the dynamics.  We will first analyze these changes for real trajectories by neglecting the imaginary part of the tunnel exit. This is the approach employed in most Coulomb-distorted strong-field approaches, and significantly simplifies the computations. Subsequently,  we will discuss the complex-trajectory case.
\subsubsection{Real trajectories}
\label{sec:realtrajsCQSFA}
\begin{figure}
	\centering
	\includegraphics[width=0.75\linewidth]{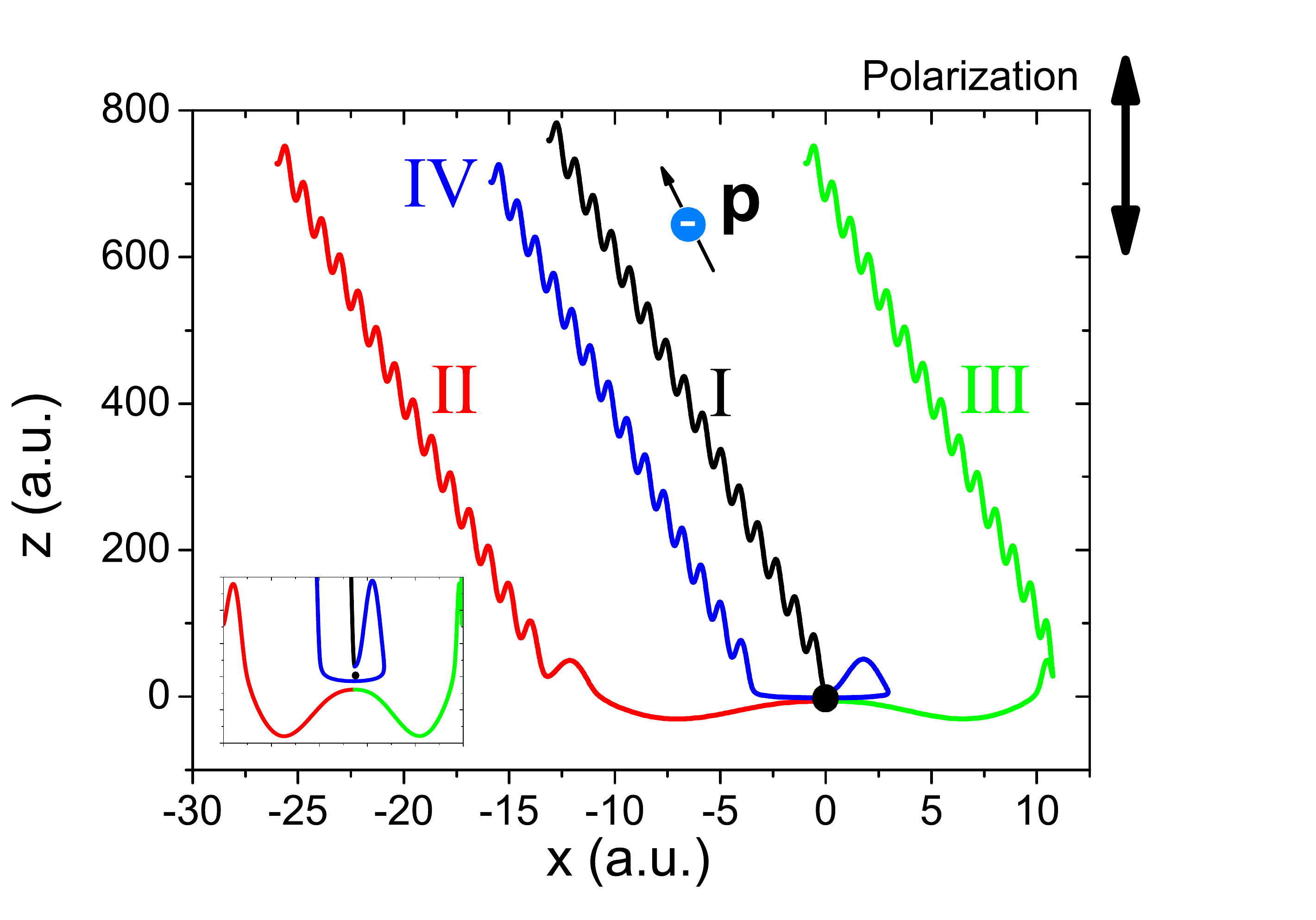}
	\caption{Illustration of the four types of CQSFA
		orbits in the laser-polarization plane for electrons with fixed final
		momentum $\textbf{p}$, computed for Hydrogen ($I_p=0.5$ a.u. and $C=1$ in Eq.~(\ref{eq:potential}) in a linearly polarized square
		pulse of intensity
		$I=2 \times 10^{14} \mathrm{W/cm}^2$ and frequency $\omega=0.057$
		a.u. The black dot
		denotes the position of the nucleus, and the inset shows the core region. From \cite{Lai2015a}. }
	\label{fig:orbit1} 
\end{figure}

The following classification has been first proposed in \cite{Yan2010} and has been successfully employed by us. It makes use of the initial and final momentum components, and of the direction of the tunnel exit with regard to the detector.  It may be summarized as follows:
\begin{itemize}
	\item \textit{Type 1 orbits:} the tunnel exit is located on the same side as the detector, i.e., $z_0p_{f\parallel}>0$, and the transverse momentum component does not change sign during the continuum propagation, i.e., $p_{0\perp}p_{f\perp}>0$. They can be related to the short SFA orbit in direct ATI. The residual binding potential decelerates an electron along this orbit.
	\item \textit{Type 2 orbits} are related to those events in which the electron is ionized from the opposite direction to the detector,  i.e., $z_0p_{f\parallel}<0$ but eventually changes its direction. It is deflected by the combined action of the field and the residual binding potential. Its Coulomb-free counterpart would be the long SFA orbit in direct ATI. Nonetheless, the presence of the Coulomb potential distorts this orbit and leads to field-dressed Kepler hyperbolae. The net effect of the Coulomb potential for an electron along this orbit is to accelerate it \cite{Lai2015a}.
	\item \textit{Type 3 orbits} are similar to type 2 orbits with regard to the direction of the tunnel exit ($z_0p_{f\parallel}<0$), and, in typical scenarios, its hyperbolic shape. The key difference is that, for type 3 orbits, the interaction with the core is stronger. This means that (i) the momentum component perpendicular to the laser-field polarization will change sign, i.e., $p_{0\perp}p_{f\perp}<0$, and (ii) this type of orbit has no counterpart in the direct SFA. Our recent work has in fact shown that rescattering is an important assumption with regard to type 3 orbits, if one is willing to recover the spider-like structures \cite{Maxwell2018}.
	\item \textit{Type 4 orbits} start in the same direction as the detector, but, unlike orbit 1, they go around the core. This means that $z_0p_{f\parallel}>0$, but  $p_{0\perp}p_{f\perp}<0$, and that these orbits are very strongly influenced by the Coulomb potential. These orbits can be loosely associated with backscattered trajectories in high-order ATI, although, strictly speaking, the rescattered SFA does not allow for events encircling the core.
\end{itemize}
For clarity, we provide an illustration of these orbits (Fig.~\ref{fig:orbit1}), together with a summary of their classification criteria (Table \ref{tab:orbits}). In the language employed in photoelectron holography, one may say that orbits type 2 and 3 are forward-scattered trajectories, and this scattering may in principle range from a light deflection to a hard collision. Type 4 orbits are backscattered trajectories. On a more rigorous note, the two laser-dressed hyperbolae in Fig.~\ref{fig:orbit1} actually form a torus in the three-dimensional space for ionization events parallel to the driving-field polarization, which becomes tilted for non-vanishing scattering angle. For the SFA, this torus decreases to a point, so that orbits type 2 and 3 become degenerate. For details see our previous publication \cite{Lai2015a}. 
\begin{table}
	\begin{center}
		\begin{tabular}{ c c c } 
			\hline
			\hline 
			Orbit & $z_0p_{f\parallel}$ & $p_{f\perp}p_{0\perp}$ \\ 
			\hline
			\hline
			1 & + & + \\ 
			2 & - & + \\ 
			3 & - & - \\ 
			4 & + & - \\ 
			\hline
		\end{tabular}
	\end{center}
	\caption{Summary of types of orbits identified in the CQSFA and the criteria employed in their classification. The + and - signs in the second and third columns indicate a positive or negative product, respectively. From \cite{Maxwell2018}.}
	\label{tab:orbits}
\end{table}

The orbit classification presented above invites the following questions: 
\begin{enumerate}
	\item[Q1:] \textit{Can one systematically relate the CQSFA orbits to those obtained with the SFA, be it direct or rescattered?} The main difficulty here lies on the fact that, while the SFA is constructed as a field-dressed Born series, the CQSFA is not. It is difficult to establish when and at what point in configuration space the electron ``returns'' or  ``rescatters'' if the binding potential and the driving field are considered on equal footing. In contrast, for the SFA the electron may only return to the origin, and its very structure ensures that either no, or a single act of rescattering occurs. 
	\item[Q2:] \textit{Does the classification provided suffice to describe the dynamics?} This is an important question as it could be that the conditions provided in Table \ref{tab:orbits} are satisfied for orbits of different topology.
\end{enumerate}

\begin{figure}
	\begin{center}
		\includegraphics[width=0.75\linewidth]{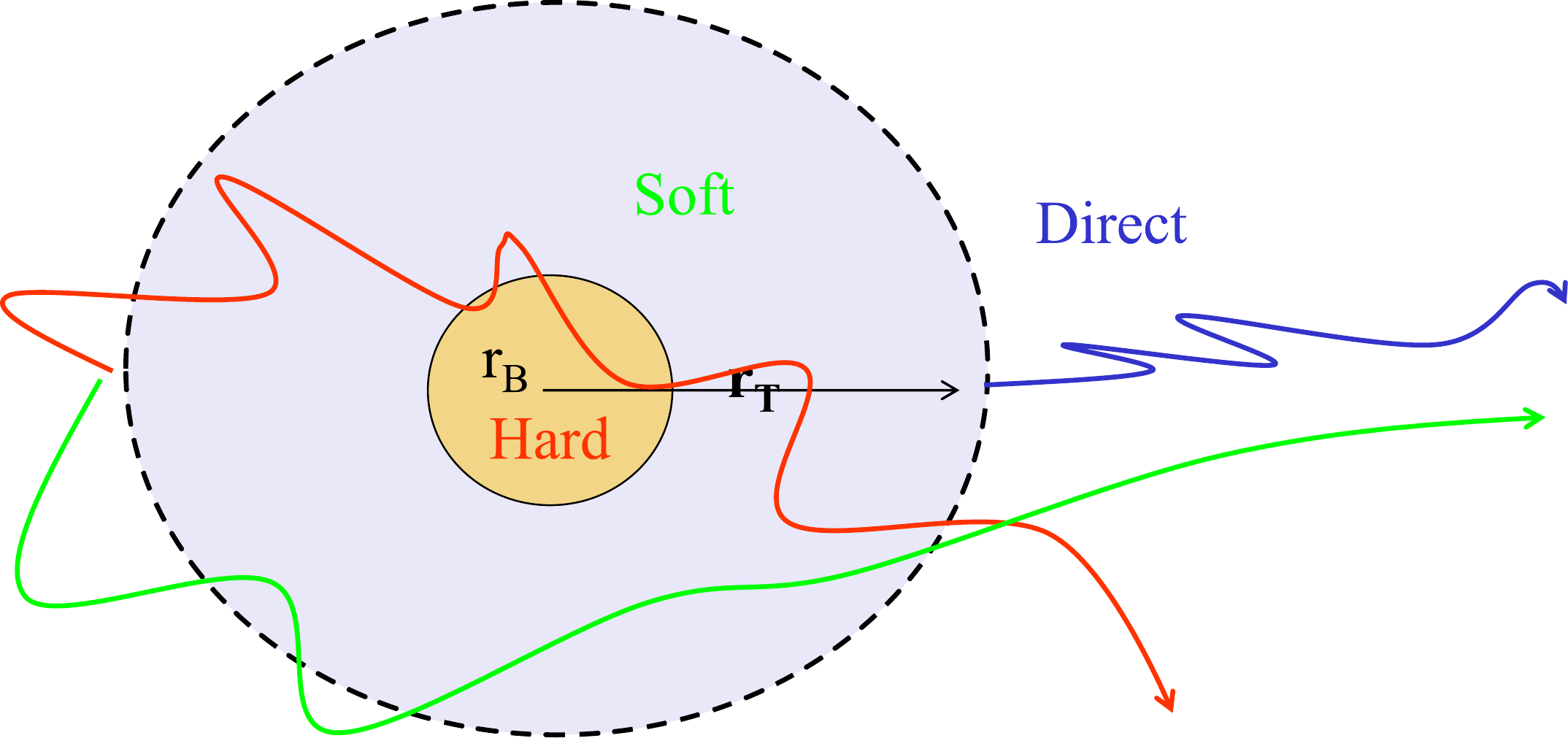}
	\end{center}
	\caption{Schematic representation of the different scattering regimes encountered in the CQSFA. In the figure, $\mathbf{r}_T$ refers to the tunnel exit, which is used to define a perimeter within  which soft collisions occur, and $\mathbf{r}_B$ denotes the Bohr radius. }
	\label{fig:regimes}.
\end{figure}

In order to address the question Q1, it was necessary to devise ad-hoc criteria for when the electron recollides and whether this collision is ``soft'' or "hard'', based on (i) the Bohr radius $r_B$; (ii) the perimeter $r_T$ defined by the (real) tunnel exit; (iii) the distance of closest approach $r_c$ of the electron subsequently to rescattering. We then assume that a hard collision will happen if an electron orbit enters the region defined by the Bohr radius. In this region, the Coulomb forces will be dominant. A soft collision will occur if $r_c$ is larger than the Bohr radius, but smaller than $r_T$. Finally, if the electron's orbit does not cross the region delimited by $r_T$, it may be related to the ATI direct orbits, as the laser field will be dominant. A schematic representation of this region is provided in Fig.~\ref{fig:regimes}. A boundary-based approach is also employed in the ARM to separate hard collisions from deflection.

\begin{figure}
	\centering
	\includegraphics[width=\linewidth]{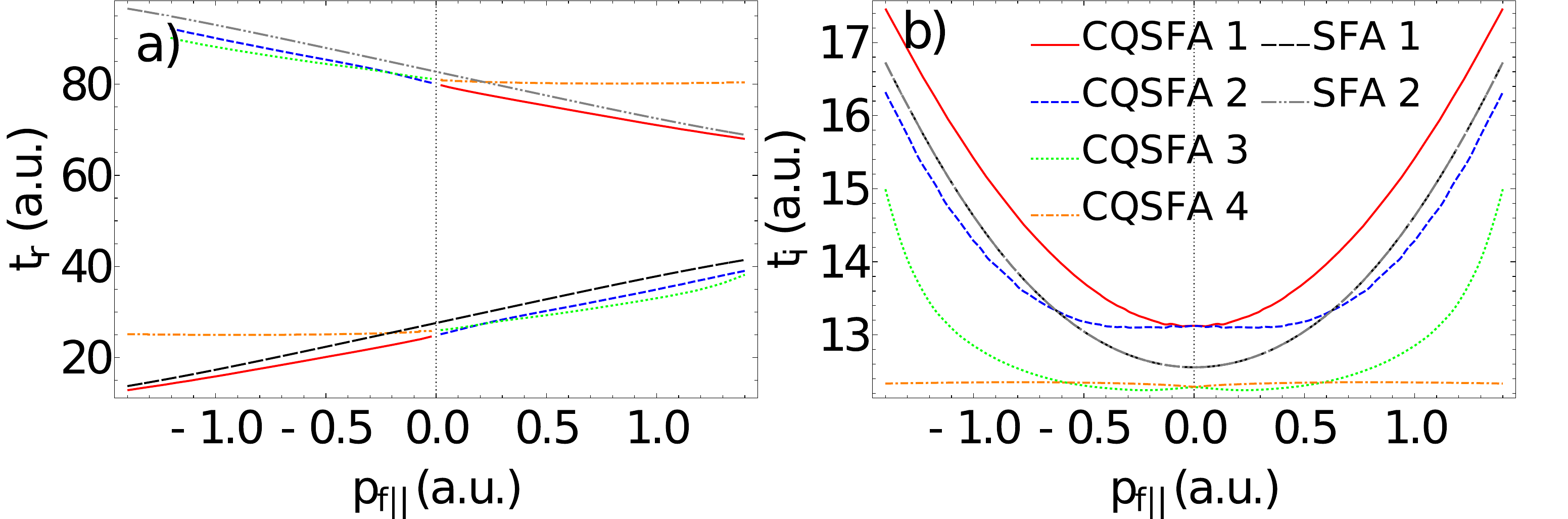}
	\caption{Real and imaginary parts of the ionization times $t'$ as functions of the final momentum component parallel to the driving field polarization (panels (a) and (b), respectively) computed using the CQSFA for final transverse momentum $p_{\perp}=0.25$ a.u. and the same parameters as in the previous figures.}
	\label{fig:realparts}
\end{figure}

\begin{figure*}
	\centering
	\includegraphics[width=\linewidth]{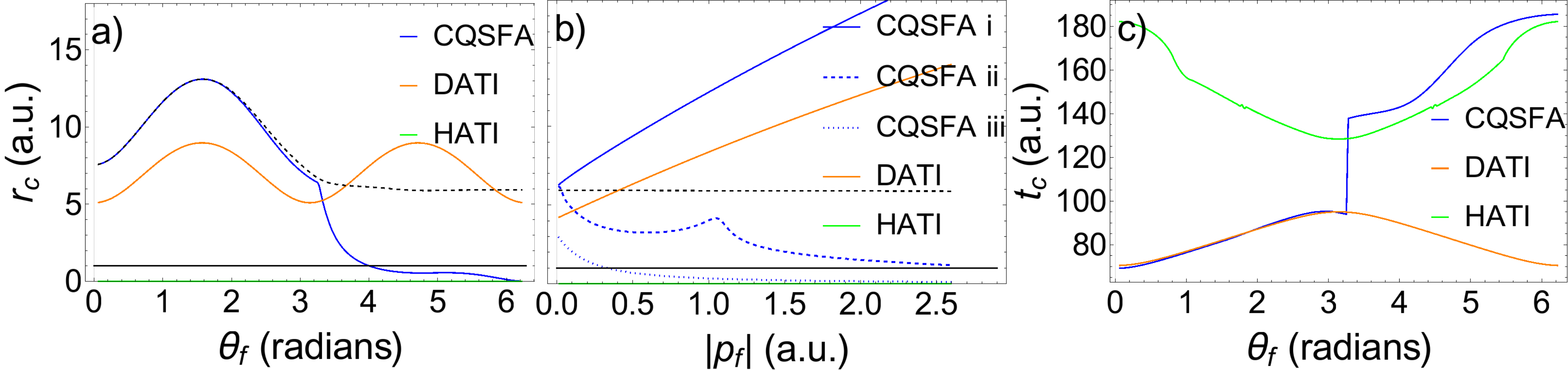}
	\caption{Distances $r_c$ and time $t_c$ of closest approach for the CQSFA, and the SFA direct and rescattered ATI solutions (labeled DATI and HATI, respectively), computed for Hydrogen ($I_p=0.5$ a.u.) in a field of wavelength $\lambda=800 $ nm, intensity $I_0=2\times 10^{14}$ W/cm$^2$, as functions of the scattering angle $\theta_f$ [panels a) and c)] and of the absolute value of the final parallel momentum component [panel (b)]. 
		In panels a) and c) a fixed energy of $E=0.26$ a.u was taken.  In panel b), $r_c$ is plotted for three fixed angles for the CQSFA. The indices i, ii and iii correspond to  $\theta_f=0.25\pi$, $\theta_f=1.10\pi$ and $\theta_f=1.75\pi$,  respectively.  For the  DATI and HATI, $r_c$ is plotted for a fixed angle of $\theta_f=1.75 \pi$. In panel c), the time of closest approach $t_c$ is plotted for a fixed energy of $E=0.26$ a.u. for the CQSFA, DATI and HATI. For HATI we compute $\theta_f$ as given in \cite{Becker2015}. In all cases, the tunnel exit for the CQSFA and the Bohr radius are marked with a dashed and solid black line, respectively.  From \cite{Maxwell2018}.} 
	\label{fig:ClosestApproach}
\end{figure*}

The real parts of the ionization times as functions of the final parallel momentum and scattering angle are good  indicators of how the CQSFA orbits behave. For instance, Fig.~\ref{fig:realparts}(a) shows that the real parts $t'_r$ of the CQSFA ionization times for orbits type 1, 2 and 3 follow the direct SFA solutions from below, but, depending on the orbit, behave differently in the high-energy limit, i.e., when $|p_{f\parallel}|$ is large. For orbit 1, $t'_r$ tends to the SFA solution, because increasing the photoelectron energy renders the influence of the Coulomb potential less and less relevant. In contrast, the real parts  $t'_r$ associated with the CQSFA orbits 2 and 3 never tend to the direct SFA solutions. For orbit 4, this behavior is particularly extreme and never follows that of the SFA. Instead, the ionization times are practically constant. The marked discrepancies for orbits 3 and 4 are a consequence of the fact that these types of orbit interact much more with the core. This means that to relate them to a direct ATI orbit is inaccurate for a broad parameter range. Fig.~\ref{fig:realparts}(b) also shows that the imaginary parts of such orbits do not approach those predicted by the SFA.  This issue will be discussed in more detail in Sec.~\ref{sec:subbarrrier}.

Other good indicators are the time and distance of closest approach, denoted $t_c$ and $r_c$ in our previous publications \cite{Maxwell2017a,Maxwell2018}. These quantities are plotted in Fig.~\ref{fig:ClosestApproach}, as functions of the final scattering angle [panels a) and c)] and of the absolute value of the final momentum $|\mathbf{p}_f|$ [panel (b)].  The figure shows that the distances of closest approach determined by the direct and rescattered SFA are limits to what is obtained from  the CQSFA. This is particularly clear if one looks at the behavior with regard to the final scattering angle, which, within the CQSFA framework, should encompass from direct orbits to hard scattering.  If one considers an orbit with small scattering angles, the distance of closest approach is the tunnel exit and follows what can be obtained if the Coulomb coupling is taken to zero from above, i.e., what one would expect from the direct SFA. Once the angle  $\theta_f=1.1\pi$ is reached, the distance of closest approach drops drastically, crossing the region delimited by the Bohr radius and tending towards the origin. That would be precisely the expected return coordinate for high-order ATI. An even clearer pattern can be observed for the CQSFA times of closest approach, which follow the solution of the saddle-point equation determined for direct and rescattered ATI very closely.
\begin{figure*}
	\centering
	\includegraphics[width=\textwidth]{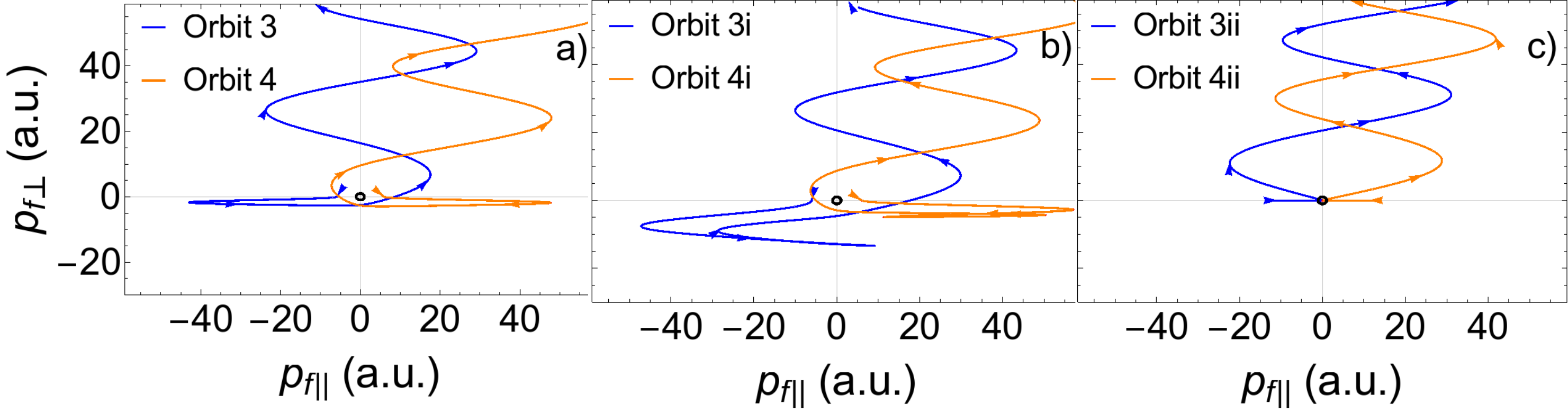}
	\caption{Different subtypes of orbit 3 and 4 that can occur for a final momentum $\mathbf{p}_f=(0.086,0.22)$ a.u., computed using the CQSFA. Panel a), b) and c) show the standard trajectories, the multi-pass orbits (denoted i) and the directly recolliding orbits (denoted ii), respectively.  The Bohr radius is marked by a black circle. From \cite{Maxwell2018}.}
	\label{fig:orbit3}
\end{figure*}

With regard to question Q2, i.e., of the classification  discussed in this work being sufficient, Fig.~\ref{fig:orbit3} provides a very good example of topologically very different trajectories, which however fall under the umbrella of type 3 and type 4 orbits, according to the classification in Table \ref{tab:orbits}. The standard case for such orbits, shown in Fig.~\ref{fig:orbit1} and described above, is only represented in panel a). Panel b) shows multi-pass orbits, which are driven past the core many times before recollision, and panel c) displays orbits that undergo hard scattering  before the laser field changes sign. These orbits, and in particular their influence on holographic patterns have not yet been investigated in detail.  There is however a strong indication that multipass orbits are responsible for near-threshold effects, and patterns such as the ``inner spider'', both from simplified methods \cite{Hickstein2012} and from the CQSFA \cite{Maxwell2017,Maxwell2017a,Maxwell2018}. A proper study of these orbits will however require novel asymptotic expansions, as there are radical changes in the number and nature of the saddles \cite{Maxwell2018}. 

It is important to stress that this is not the only classification employed in the literature. For instance, in \cite{Li2014}, orbits leading to holographic patterns have been classified as direct ($D_1$), and forward scattered ($R_1$ and $R_2$). Trajectories $R_1$ have large, initial positive transverse momenta, while for trajectories $R_2$ the initial transverse momenta are small and negative. Although we cannot directly state, as the direction of the final transverse momenta are not given, it seems that $R_1$ and $R_2$ may be associated with type 2 and type 3 orbits, respectively.  This impression is supported by the stronger influence of the core and higher ionization rate that exists for the $R_2$ trajectories in \cite{Li2014} (see also supplemental material therein).

\subsubsection{Complex trajectories and branch cuts} 
\label{sec:complextrajs}

\begin{figure}
	\centering
	\includegraphics[width=0.75\linewidth]{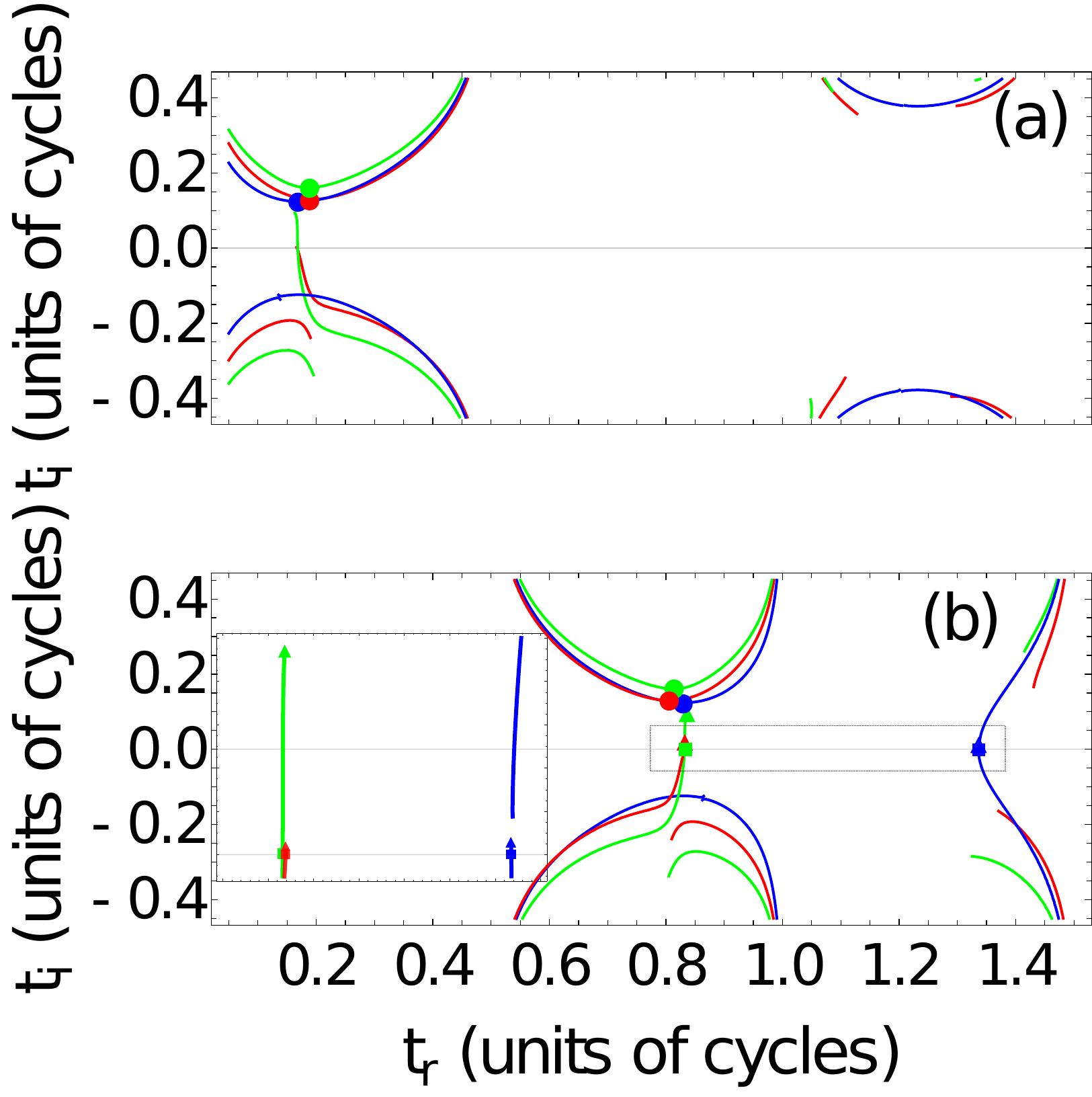}
	\caption{Branch cuts in the complex time plane calculated for the Coulomb-free orbits 1 [panel(a)] and 2 [panel (b)] using the same field and atomic parameters as in Figs.~\ref{fig:ClosestApproach} and  \ref{fig:orbit3}. The red (gray), green (light gray) and blue (dark gray) curves were computed using the final  the momentum components  $(p_{f\parallel},p_{f\perp})=(-0.63\mathrm{~a.u.}, 0.53\mathrm{~a.u.})$,     $(p_{f\parallel},p_{f\perp})=(-0.80\mathrm{~a.u.}, 1.05\mathrm{~a.u.})$   and  $(p_{f\parallel},p_{f\perp})=(-0.82\mathrm{~a.u.}, 0.01\mathrm{~a.u.})$. The start times $t'$ of each contour are indicated by the dots in the figure, while the times $t_b$ at which a branch cut crosses the real time axis under the condition that $t_b>\mathrm{Re}[t']$ are indicated by squares. The triangles mark the branching points $t_k$.  The left hand side of panel (b) provides a blow up of the region where the branch cuts meet the real axis for physically relevant parameters. From \cite{Maxwell2018b}. }
	\label{fig:Coulombfree}
\end{figure}

If complex trajectories are taken, two legitimate questions are: (i)\textit{ Will any of the trajectories discussed above, either Coulomb-free or Coulomb-distorted, meet a branch cut? }(ii) \textit{If so, what is the physical meaning of this encounter? }
In order to understand the problem and ultimately answer these questions, we will map the branch cuts in the complex momentum plane and match this behavior with that of qualitatively different trajectories.

In Fig.~\ref{fig:Coulombfree}, such a mapping is presented for Coulomb-free trajectories, i.e., the SFA orbits 1 and 2. The figure shows two sets of branch cuts separated by gaps that may get closer or distance themselves from the real axis depending on the parameters used. For orbit 1, the branch cuts do not cross the real time axis in the time range of interest, i.e, for $\mathrm{Re}[\tau]>t'_r$. However, they do for orbit 2. This implies that the contour must be altered for this latter orbit.  

\begin{figure}
	\begin{center}
		\includegraphics[width=0.75\linewidth]{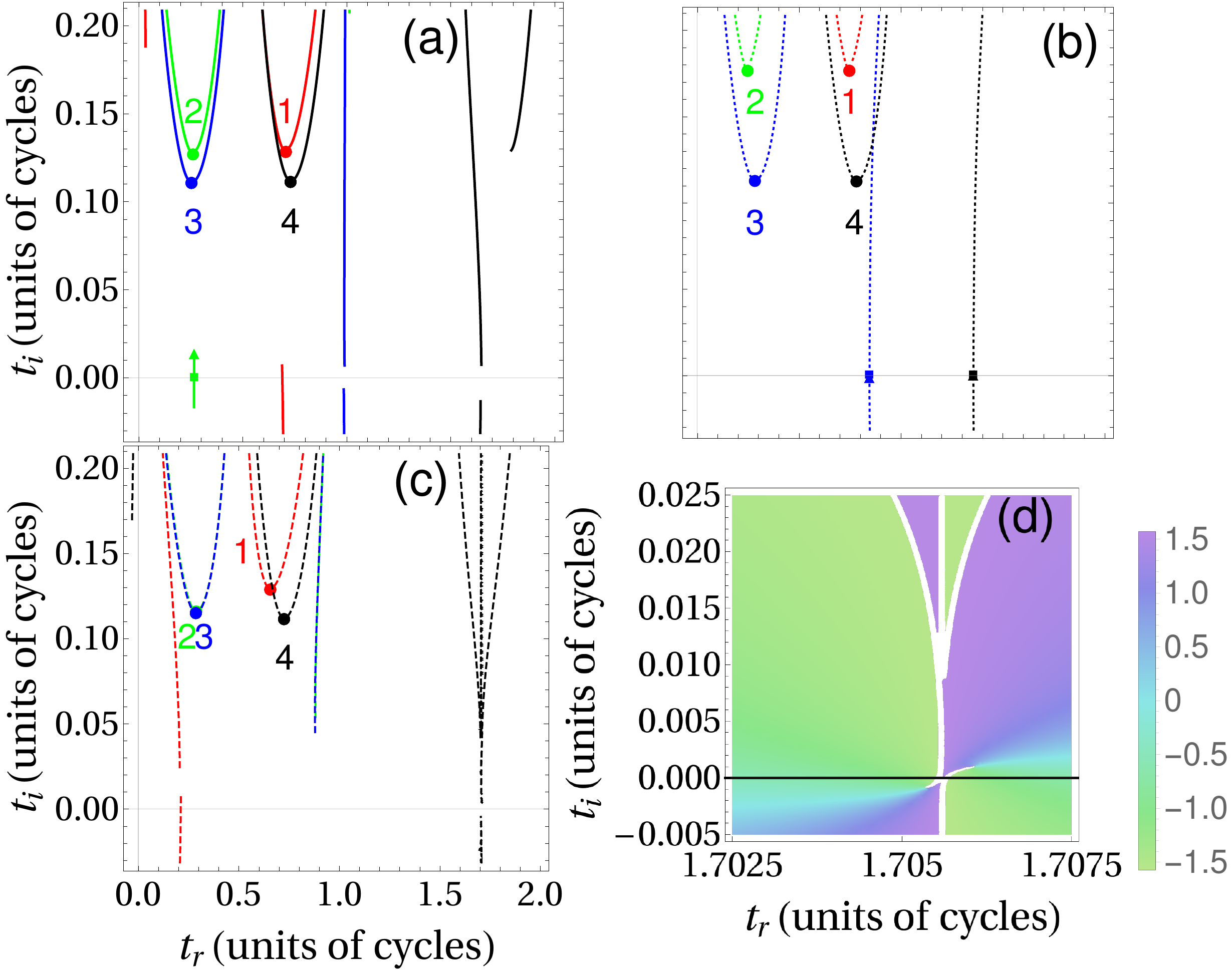}
	\end{center}
	\caption{Branch cuts in the complex time plane calculated using Coulomb-distorted trajectories for the same field and atomic parameters as in Fig.~\ref{fig:Coulombfree}. In panels (a), (b) and (c), respectively, the momentum components $(p_{f\parallel},p_{f\perp})=(-0.475\mathrm{~a.u.}, 0.400\mathrm{~a.u.}) $,   $(p_{f\parallel},p_{f\perp})=(-0.604\mathrm{~a.u.},0.980\mathrm{~a.u.}) $  and $(p_{f\parallel},p_{f\perp})=(-0.619\mathrm{~a.u.},0.0113\mathrm{~a.u.})$ were chosen. The branch cuts associated with orbits 1, 2, 3 and 4 are shown as the red (gray), green (light gray), blue (dark gray) and black lines. The corresponding ionization times, branching times and the intersection times of the branch cut with the real time axis are marked with dots, triangles and squares, respectively, using the same color convention. 	The number close to the ionization times indicate the type of orbit. In panel (d), we present $\arg(\sqrt{\mathbf{r}(\tau)\cdot\mathbf{r}(\tau)})$ for the parameters in panel (b), near the overlapping branch cuts. From \cite{Maxwell2018b}.}
	\label{fig:branchmapCQSFA}
\end{figure}

If the Coulomb potential is incorporated, as shown in Fig.~\ref{fig:branchmapCQSFA}, we also see that orbit 1 does not encounter a branch cut, but the other three types of orbits do depending on the choice of the electron momentum components. In Fig.~\ref{fig:branchmapCQSFA}(a), orbit 2 crosses the branch cut, while for the other types of orbits, branch cuts are avoided. In Fig.~\ref{fig:branchmapCQSFA}(b), the final momentum has been carefully chosen as to lead to a final momentum almost perpendicular to the driving-field polarization direction. In this case, the branch cuts intersect the real time axis for orbits 3 and 4. The situation is particularly complex in  Fig.~\ref{fig:branchmapCQSFA}(c), for which the branch cuts associated with orbit 4 cuts through the real time axis twice within a field cycle. A closer inspection [Fig.~\ref{fig:branchmapCQSFA}(d)] even shows that, in this case, the branch cuts overlap, which would rule out the applicability of the present procedure. 
Physically, a trajectory encountering a branch cut is strongly associated with an act of return or rescattering. For that reason, this may occur for all orbits except orbit 1.  The fact that orbit 4 may cross twice a branch cut within a field cycle reflects its behavior of going around the core.  This physical meaning has already been mentioned in \cite{Pisanty2016}, and in \cite{Keil2016}, which state that, every time a contour pass through the gates, soft recollisions do occur. Because the variables are complex, such recollisions influence not only the phases, but also the amplitudes associated with specific orbits. One should note that, due to the restrictions associated with the method in \cite{Pisanty2016,Keil2016}, in such studies only Coulomb-free trajectories were taken into consideration. In contrast, our method allow us to treat Coulomb-distorted trajectories. 

\subsection{Sub-barrier dynamics and single-orbit distributions}
\label{sec:subbarrrier}
In the standard formulation of the CQSFA, understanding the contributions of each orbit to the overall probability density means having a closer look at the sub-barrier dynamics and the imaginary parts of the ionization times. In the CQSFA framework, the ionization probability is proportional to $\exp[-2\mathrm{Im}[S(\mathbf{\tilde{p}}, \mathbf{r},t,t')]]$, where the variables upon which the action depends are determined using the saddle-point equations in the first part of the integration contour discussed in Sec.~\ref{subsubsec:contours}.
\begin{figure*}
	\begin{center}
		\includegraphics[width=0.75\linewidth]{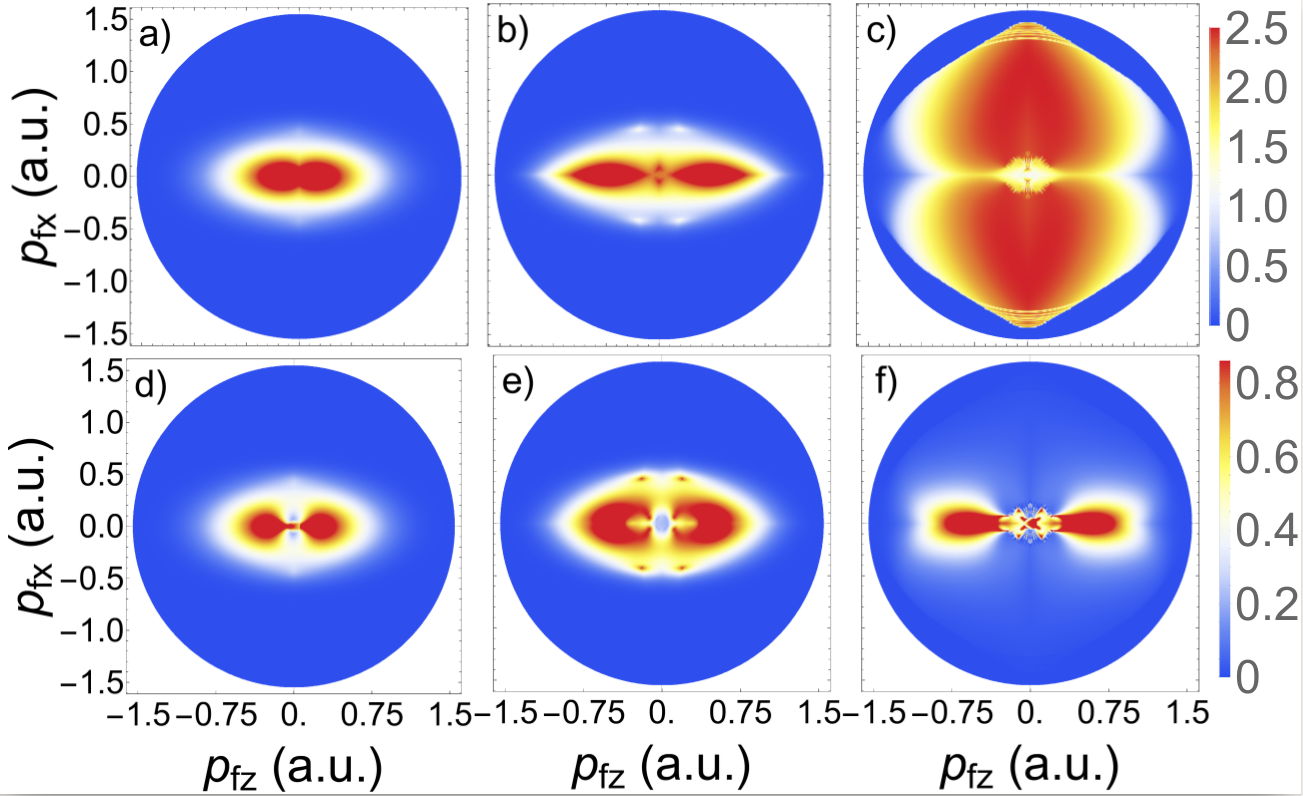}
	\end{center}
	\caption{CQSFA single-orbit electron momentum probability distributions plotted in arbitrary units and computed for the same field and atomic parameters as in the previous figures. The left, middle and right columns correspond to orbit 1, 2 and 3, respectively. The upper and lower panels have been computed without and with the prefactors, respectively. The upper panels have been multiplied by $10^3$. From \cite{Maxwell2017}.}
	\label{fig:SingleOrbit}
\end{figure*}

\begin{figure}
	\centering
	\includegraphics[width=0.75\linewidth]{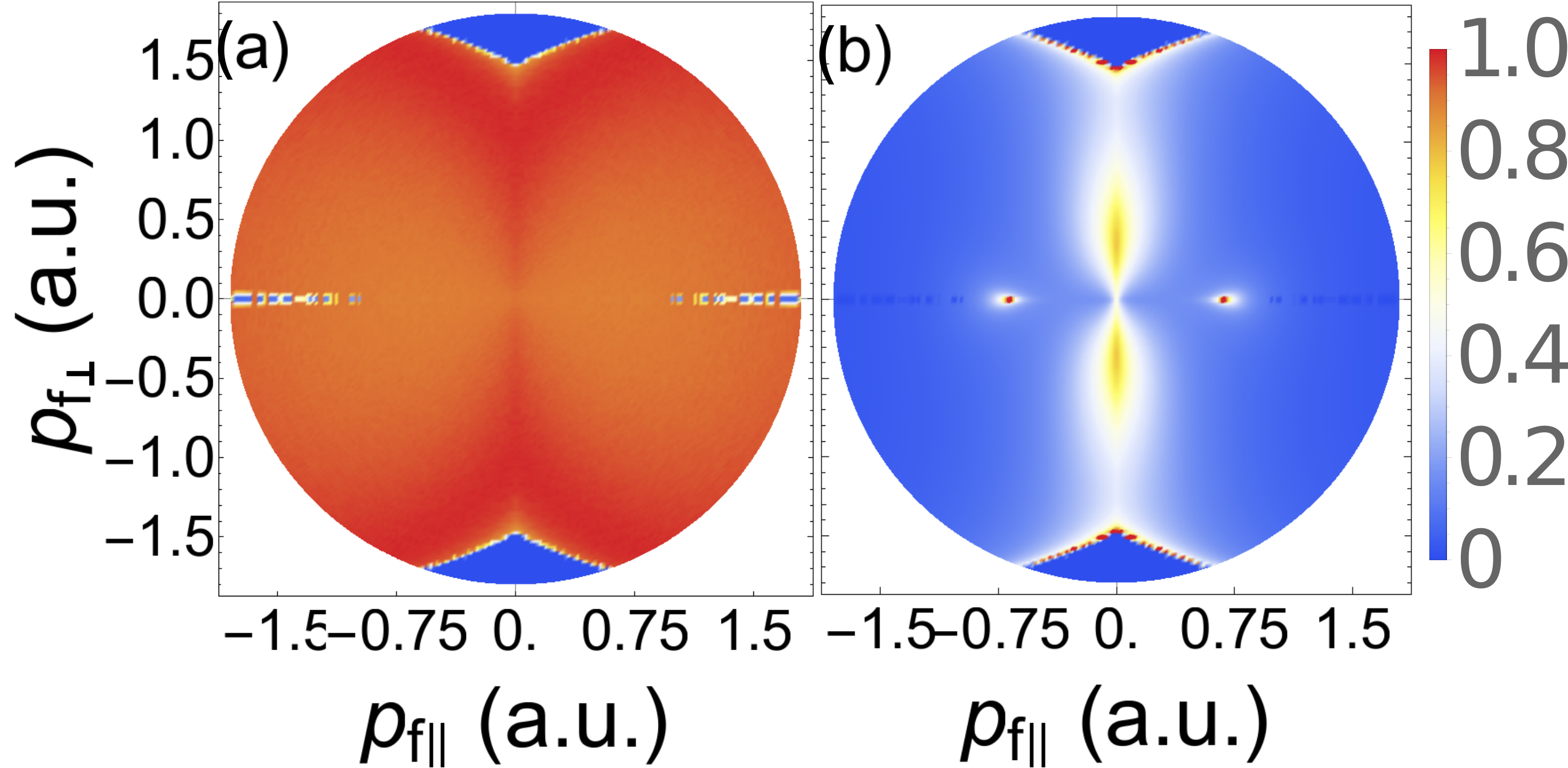}
	\caption{CQSFA single-orbit angle-resolved electron probability distributions in arbitrary units and computed using orbit 4 and the same field and atomic parameters as in the previous figures. Panels (a) and (b) have been computed without and with the prefactors, respectively.}
	\label{fig:orb4im}
\end{figure}

\begin{figure}
	\centering
	\includegraphics[width=0.75\linewidth]{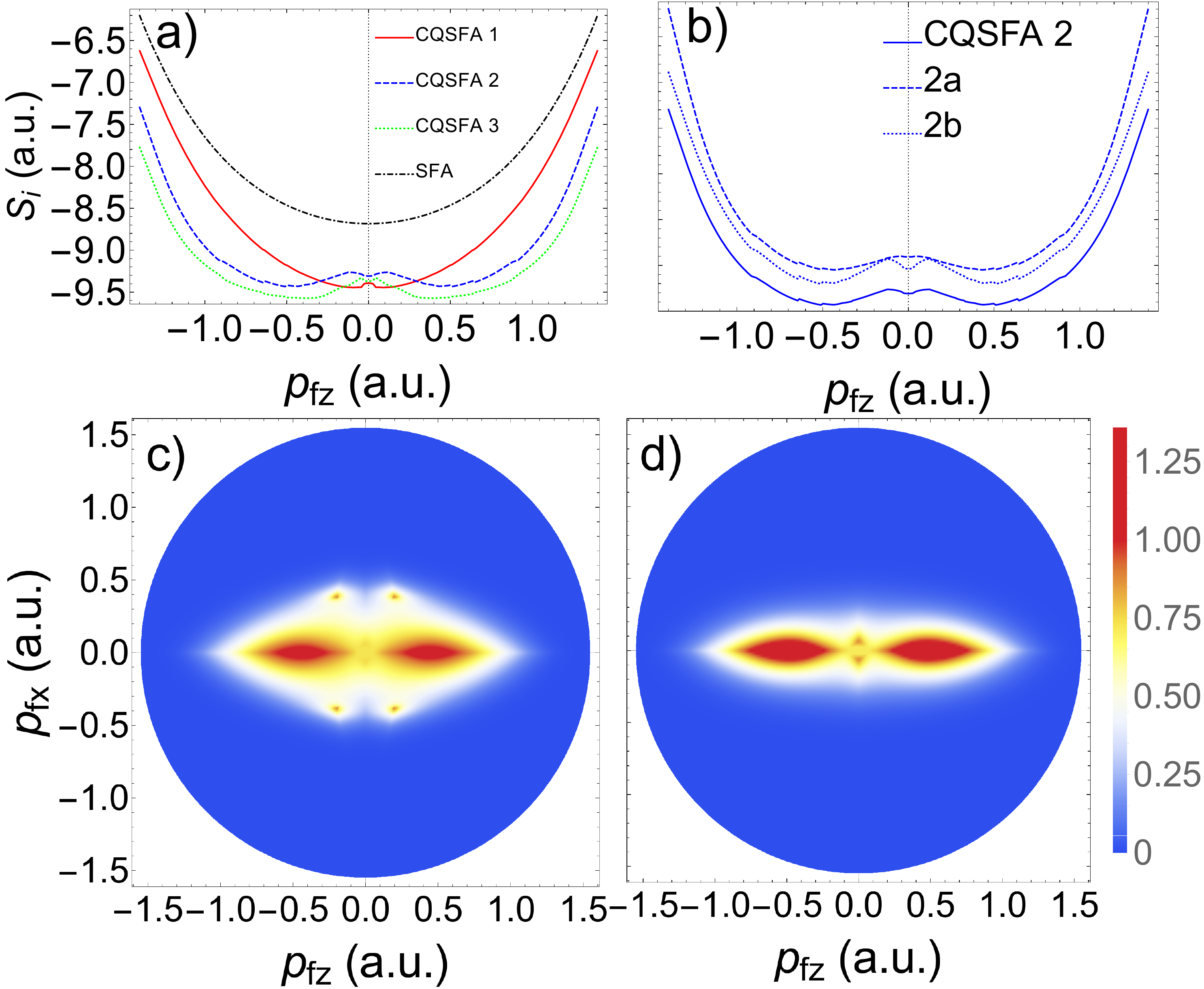}
	\caption{In panel (a), we present the imaginary parts of the actions $S_i$ (i=1,2,3) associated to the orbits 1, 2 and 3 of the CQSFA, as functions of the final parallel momentum for perpendicular momentum $p_{f\perp}=0.05$ a.u. For comparison, we also plot the SFA counterpart. In panel (b), we show the approximate analytic expressions obtained for orbit 2, with and without the integral over the binding potential (dashed and dotted lines, respectively), compared with the full CQSFA solution (solid line). In panels (c) and (d), we plot the single-orbit electron momentum distributions computed analytically with and without the sub-barrier Coulomb phase, respectively.  The atomic and field parameters are the same as in the previous figures. From \cite{Maxwell2017}.}
	\label{fig:ImAction}
\end{figure}

Fig.~\ref{fig:SingleOrbit} presents the single-orbit distributions computed for orbits type 1, 2 and 3, without and with prefactors. For the contributions of orbits 1 and 2 (left and middle panels), one may easily identify two peaks along the parallel polarization axis, for $p_{f\parallel} \neq 0$ and elongated shapes near the $p_{f\parallel}$ axis.  This behavior mirrors that of the imaginary parts of the CQSFA action, which are plotted in the upper panels of Fig.~\ref{fig:ImAction}, and that observed for $\mathrm{Im}[t']$, plotted in Fig.~\ref{fig:realparts}(b). The figure clearly shows two minima in both cases, whose positions correspond to the maxima in this distributions. 
Qualitatively, this doubly-peaked shape is very different from that observed for the SFA. In the SFA case, $\mathrm{Im}[S_d(\mathbf{p},t')]$  and $\mathrm{Im}[t']$ exhibit a single minimum for vanishing momentum, which leads to a single peak in the corresponding electron-momentum distribution. This may be understood as a result of the fact that the barrier to be overcome by the electron is narrowest at peak field. For peak-field times, the electron tunnels with vanishing momentum. Such events, however, are highly suppressed in the CQSFA due to the residual Coulomb attraction. This may be easily understood as an electron with vanishing momentum at the tunnel exit will be pulled back by the core. 
For orbit 3, one observes a much broader momentum range delimited by caustics. This is consistent with the fact that the imaginary part of the ionization time is very flat and much closer to zero for this orbit. Once the prefactor is included, however, the probability density associated with this orbit is restricted to a narrow range along the polarization axis. For orbit 4 the significant contributions are located perpendicular to the polarization axis, so that they only start to play a role for nearly right scattering angles. It is noteworthy that, for orbits 3 and 4, the shapes of the electron momentum distributions are strongly influenced by the prefactors, which also causes a drastic reduction of the yield. This is a strong indicator that these orbits are far less stable than orbits 1 and 2. 

Hence, we conclude that, for orbits 1 and 2, the key influence on these distributions comes from the Coulomb potential modifying the SFA-like part of the sub-barrier action $S^{\mathrm{tun}}(\mathbf{\tilde{p}},\mathbf{r},t'_r,t')$ via the ionization times, while for orbits 3 and 4 the prevalent contribution is less clear. Thus, it is legitimate to ask what role the sub-barrier Coulomb phase plays. Analytical estimates for each term can be used to disentangle both contributions. The imaginary parts of the action with and without the Coulomb integral are shown in Fig.~\ref{fig:ImAction}(b), while the corresponding single-orbits distributions for orbit 2 are shown in the lower panels of the figure. The main effect of the Coulomb integral is to broaden the distributions in the direction perpendicular to the polarization axis, and lead to secondary maxima. 

\begin{figure*}
	\centering
	\includegraphics[width=\textwidth]{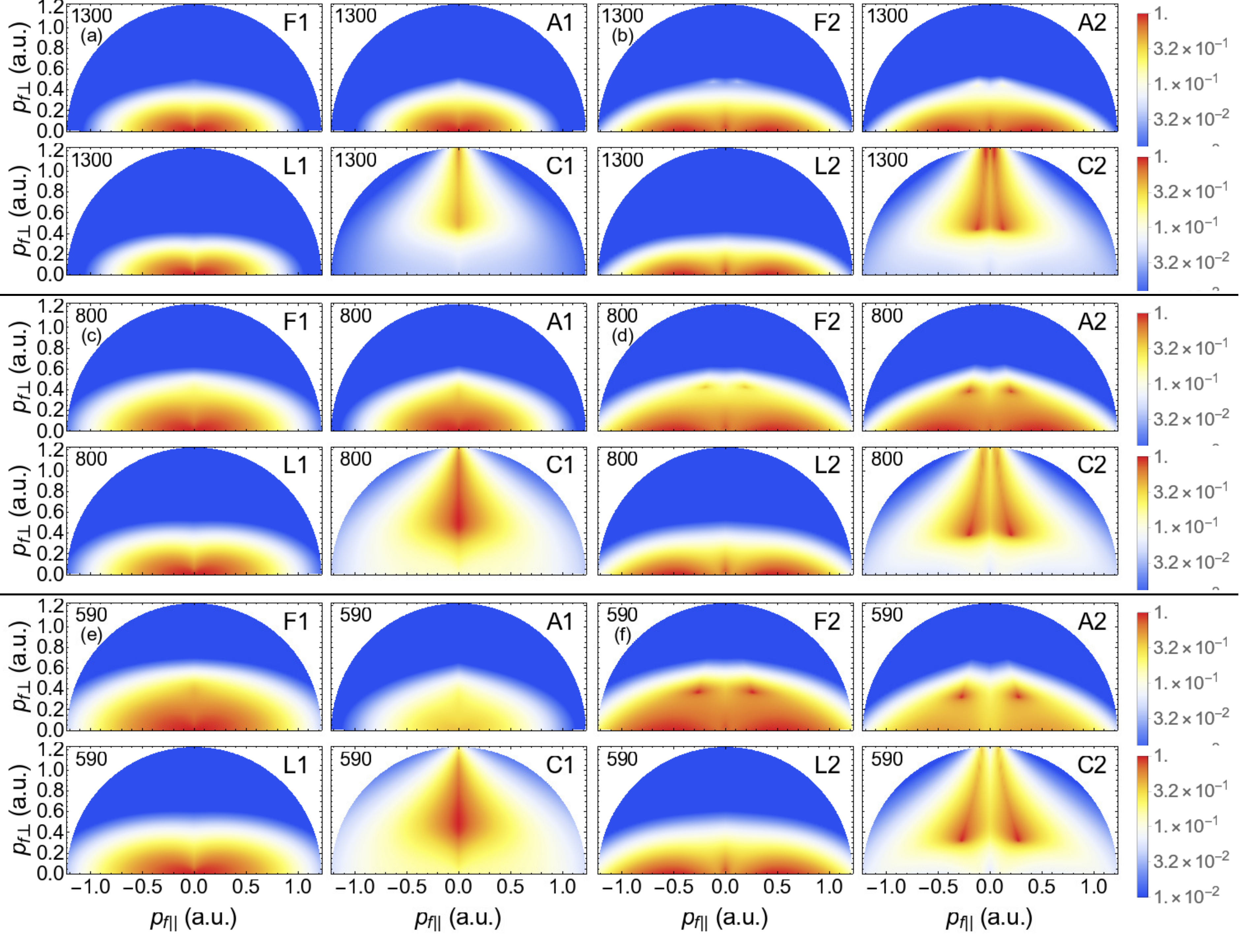}	
	\caption{Single-orbit CQSFA photoelectron momentum distributions computed without prefactors for hydrogen ($I_p=0.5$ a.u.) and orbits 1 and 2, for the field parameters $(I,\lambda)=(7.5\times 10^{13} \mathrm{W/cm}^2,1300 \hspace*{0.1cm}\mathrm{nm})$ [panels (a) and (b)], $(I,\lambda)=(2.0\times 10^{14} \mathrm{W/cm}^2,800\hspace*{0.1cm} \mathrm{nm})$ [panels (c) and (d)] and $(I,\lambda)=(3.75\times 10^{14} \mathrm{W/cm}^2,590\hspace*{0.1cm} \mathrm{nm})$ [panels (e) and (f)], where $I$ and $\lambda$ give the field intensity and wavelength, respectively. The acronyms $\mathrm{F}n$ $(n=1,2)$, $\mathrm{A}n$ $(n=1,2)$ on the right top corner indicate the full and analytic CQSFA solution for orbits 1 or 2, respectively, while $\mathrm{L}n$ $(n=1,2)$ and $\mathrm{C}n$ $(n=1,2)$ give the laser and Coulomb terms, respectively, in the analytic expressions. Each panel has  been plotted in the logarithmic scale and normalized to its highest yield. The thick horizontal lines separate panels with different field parameters. From \cite{Maxwell2017a}.}
	\label{fig:SingleOrbNoPref}
\end{figure*}

More detailed information is given  in Fig.~\ref{fig:SingleOrbNoPref}, which shows the analytic estimates together with the full CQSFA electron momentum distributions for a wide parameter range, computed using orbits 1 and 2. Throughout, the SFA-like terms lead to elongated shapes along the polarization axis, while the contributions from the sub-barrier Coulomb phase have non-trivial behavior. In most cases, the latter are located at or around the $p_{f\perp}$ axis. This is consistent with the momentum regions occupied by the probability distributions associated with orbits 1 and 2, along the polarization axis, and for orbit 4, in the direction perpendicular to it (see Fig.~\ref{fig:orb4im}). For orbits 1 and 2, the coupling with the laser field dominates, while for orbit 4 the binding potential prevails. 
For very long wavelengths, the influence of $S^{\mathrm{tun}}_{C}(\mathbf{r},t'_r,t')$ is minimal. This may be attributed to the long electron excursion amplitudes for the electron in this regime, which reduce the influence of the core.  As the wavelength decreases, the sub-barrier Coulomb phase becomes more important by modifying the shape of the electron momentum distributions. 
Our studies also show that sidelobes are already present for single-orbit distributions, so that they are primarily caused by sub-barrier contributions. In the seminal work \cite{HuismansScience2011}, they were attributed to quantum-interference effects. Quantum interference does enhance the sidelobes, but it is only a contributory cause. This assertion was only made possible because the CQSFA allows one to separate and visualize the contributions of specific types of orbits in the presence of the binding potential and assess the influence of sub-barrier dynamics in each. Interestingly, features that could be associated with sidelobes have also been reproduced by a classical model and associated with rescattering in ATI \cite{Paulus1994}.

Finally, it is noteworthy that many semi-classical approaches that include the residual Coulomb potential weight the electron's release in the continuum with the quasi-static ADK tunnelling rate, such as the QTMC \cite{Li2014,Xie2016}, and the SCTS  \cite{Shvetsov2016,Shilovski2018} methods. It is a well-known fact, however, that these rates are maximal at peak field. However, the results presented in this section do not seem to support this picture. 
Still, both the QTMC and the SCTS exhibit a good agreement with \textit{ab-initio}  methods and reproduce many of the holographic features encountered in experiments. This apparent contradiction may be understood if one considers that an electron freed with vanishing momentum would be recaptured by the core into a high-lying excited state. This mechanism is known as frustrated tunnel ionization (FTI) \cite{Nubbemeyer2008}, and has been recently incorporated in the SFA in \cite{Popruzhenko2017}. For early studies of transient recapture of an electron in a high-lying state see \cite{Yudin2001}.  This mechanism is explicitly mentioned in the context of the QTMC \cite{Li2014}, and of an in-depth study using classical-trajectory methods \cite{Shvetsov-Shilovski2009}. It is important to note the resemblance between our single-orbit distributions for orbits 1 and 2 and those in \cite{Shvetsov-Shilovski2009}, with two  peaks stretched along the direction of the laser-field polarization. Frustrated tunneling ionization has also been studied in conjunction with the attoclock \cite{Landsman2013}, with the very low energy structure \cite{Wolter2014a} and with the fan-shaped holographic fringes \cite{Liu2012}. Nonetheless, the issue is not fully settled, as recent studies indicate that non-adiabatic ionization conditions play an important role in enhancements that occur for the ATI signal near photoelectron energy $2U_p$ \cite{He2018b}. For a thorough discussion of several tunneling criteria see \cite{Ni2018}, a phase-space analysis of tunneling using initial value representations see \cite{Zagoya2014}, and non-adiabatic effects in circularly polarized fields see \cite{Barth2011,Ivanov2014,Hofmann2014}.

\subsection{Holographic patterns: well-known and overlooked}
\label{sec:resholographic}
The objective of this section is to discuss the key types of interfering trajectories contributing to specific patterns encountered in ATI photoelectron momentum distributions.  This will include not only the fan and the spider-like structures, but also patterns not yet reported in experiments. 
\subsubsection{Preamble: several types of temporal double slits.}
\label{sec:SFAinterf}
\begin{figure*}
	\includegraphics[width=15cm]{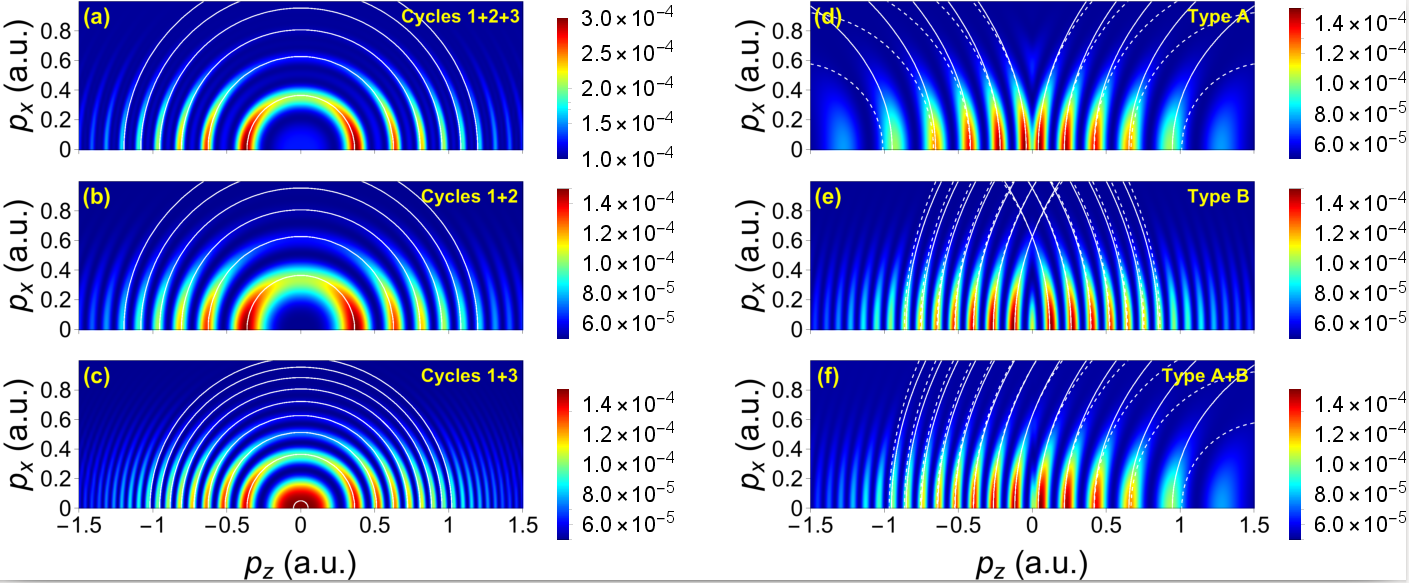}
	\caption{Electron momentum distributions computed with the direct SFA for Hydrogen $I_p=0.5$ a.u. in a linearly polarized monochromatic field of intensity $I=2 \times 10^{14} \mathrm{W/cm}^2$ and frequency $\omega=0.057$ a.u.. In panels (a) to (c), we display inter-cycle interference patterns obtained using orbit 1 using different cycles.  In  panels (d) to (f), we present intracycle interference patterns computed using the times $t'_{1}$ and $t'_{2}$ within the first field cycle. Panels (d) and (e) exhibit type A and B intracycle interference, for which $\mathrm{Re}[t'_{2}-t'_{1}]$ and  is less than or greater than half a cycle, respectively. Panel (f) was computed following the solutions $t'_{1}$ and $t'_{2}$ from negative to positive parallel momenta without imposing any restriction upon the time difference. The white dashed and full lines give exact and approximate analytic interference conditions derived in \cite{Maxwell2017}.  From \cite{Maxwell2017}.} \label{fig:interfSFA1}
\end{figure*}
Thereby, a good place to start is the direct SFA. An example of several possible types of inter- and intracycle interference  within the SFA is provided in Fig.~\ref{fig:interfSFA1} (left and right panels, respectively). Intercycle interference leads to the ATI rings, which  become finer if more and more cycles are added, while intracycle interference leads to double-slit patterns.  A striking feature is that, even if only two solutions per cycle exist, \textit{several} intra-cycle patterns can be constructed. This follows from the variety of \textit{phase differences} that may be obtained depending on the time difference between both solutions is chosen. One may for instance choose the interfering orbits so that  $\mathrm{Re}[t'_{2c}-t'_{1c}]$ is smaller (panel (d)) or greater (panel (e)) than half a cycle, respectively, or if no restriction is imposed (panel (f)). We denote the interference stemming from the conditions  $\mathrm{Re}[t'_{2c}-t'_{1c}]<\pi/\omega$ as type A interference, while that coming from  $\pi/\omega\le\mathrm{Re}[t'_{2c}-t'_{1c}]\le2\pi/\omega$ is called type B interference. The former restriction has been widely used in the literature, while the latter case has only been addressed systematically in our previous publication \cite{Maxwell2017}. A methodical  study of the type A double-slit interference is presented in \cite{Arbo2012}. The present  classification is not related to that used in the seminal paper \cite{Bian2011}. 
Type A interference, when symmetrized with regard to $p_{\parallel}=0,$ leads to the double slit pattern reported in  \cite{Arbo2012}. Fig.~\ref{fig:interfSFA1}(d) shows such fringes, which are nearly vertical and slightly divergent at the origin. Type B interference, shown in Fig.~\ref{fig:interfSFA1}(e), leads to much finer, slightly convergent fringes. The differences in the widths may be understood by inspecting the differences in the real parts of the ionization times and actions of the two contributing orbits (see Fig.~\ref{fig:MomReTimes}). The further apart they are, the finer the fringes they will be. If no restriction is imposed upon $\mathrm{Re}[t'_{2c}-t'_{1c}]$, the fringes move from finer to thicker in the example presented. In Fig.~\ref{fig:MomReTimes}, types A and B intracycle interference are marked by circles and rectangles, respectively. Analytic interference conditions for all double-slit cases were derived in our previous publication \cite{Maxwell2017}.

\subsubsection{The fan and the spider.}

\begin{figure*} [tb]
	\centering
	\includegraphics[width=\linewidth]{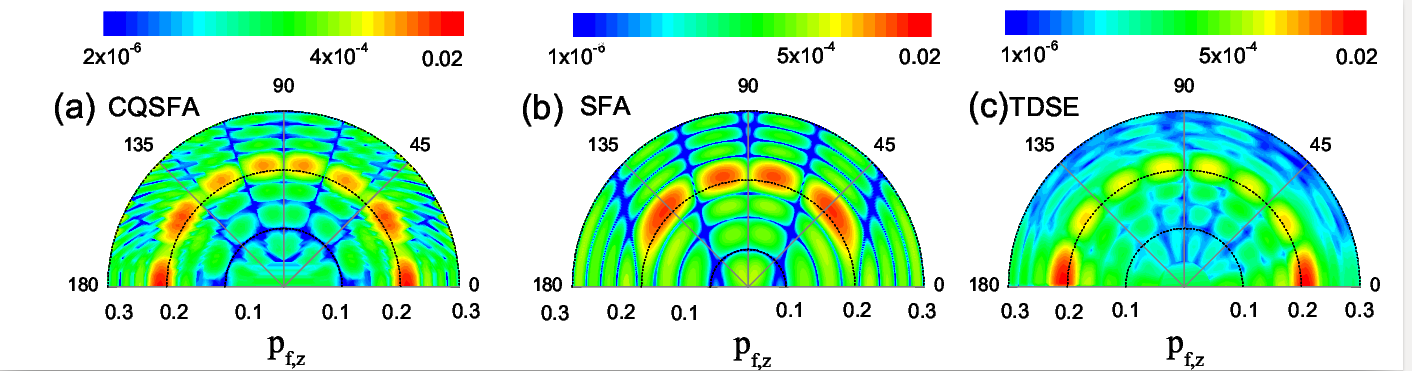}
	\caption{Angle-resolved photoelectron momentum  distributions for hydrogen  ($I_p=0.5$ a.u.) in a linearly polarized laser field of intensity
		$I=2\times 10^{14}$ W/cm$^2$ and wavelength $\lambda=800$ nm, for
		momenta $p_f<0.3$ a.u as functions of the momentum component $p_{f,z}$ along the laser
		polarization direction. Panels (a), (b) and (c) refer to the CQSFA,
		SFA, and TDSE, respectively. In panels (a) and
		(b), we have employed a monochromatic linearly polarized field over five cycles, while in panel (c) we have used a trapezoidal pulse (up and down-ramped over 2
		cycles, and constant over 8 cycles). the TDSE results have been computed with Qprop \cite{qprop}. All panels have been normalized to the same range. From \cite{Lai2017}. } \label{fig:fan1}
\end{figure*}
\begin{figure}
	\centering
	\includegraphics[width=0.75\linewidth]{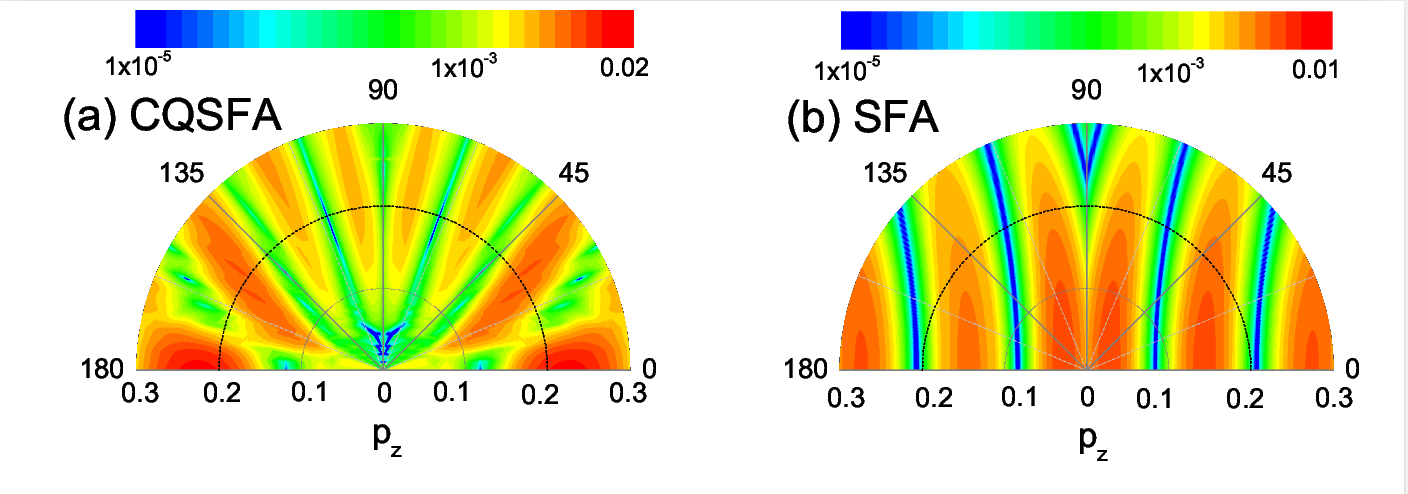}
	\caption{(Color online) Same as Fig.~\ref{fig:fan1}(a) and
		Fig.~\ref{fig:fan1}(b), but calculated over a single cycle restricting the differences between the real parts of ionization times $t'_{1c}$ and $t'_{2c}$ to at most half a cycle.  Both panels have been normalized to the same range  to facilitate a direct comparison.  From \cite{Lai2017}.} \label{fig:fan2}
\end{figure}

We will now view the fan-shaped pattern discussed in Sec.~\ref{sec:fan} in the light of the CQSFA.  In Fig.~\ref{fig:fan1}, we plot photoelectron momentum spectra computed in the polarization plane, using only orbits 1 and 2 over five cycles. Apart from the ATI rings caused by inter-cycle interference, the figure shows the fan-shaped structure, both in the CQSFA and \textit{ab-initio} results. This is in striking disagreement with the SFA results, which exhibit the nearly vertical fringes that are characteristic of the type A interference. One should note as well that there is excellent agreement between the number of fringes in the fan for the CQSFA and the TDSE. This is not the case in many Coulomb corrected approaches, such as the CCSFA and CVA (see Sec.~\ref{sec:beyondsfa}).
If only type A intra-cycle interference is allowed, the fan becomes even clearer (see Fig.~\ref{fig:fan2}).  A  comparison with the SFA results, displayed on the right panels of Figs.~\ref{fig:fan1} and \ref{fig:fan2}, supports the statement that the fan is due to unequal angular distortions in the temporal double-slit pattern caused by the Coulomb potential. The influence of the binding potential decreases with photoelectron energy and scattering angle, being very pronounced near the polarization axis, but practically cancelling out close to $\theta=90^{\circ}$. Thus, resonances with excited bound states are not necessary to reproduce the fan.

\begin{figure}
	\centering
	\includegraphics[width=\linewidth]{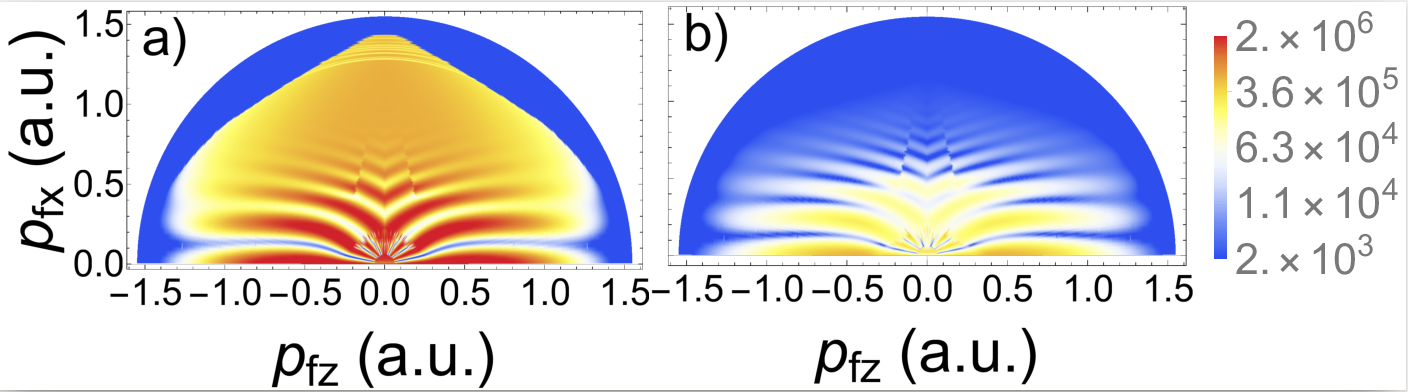}
	\caption{Photoelectron momentum distributions calculated with the CQSFA using orbits 2 and 3 for the same parameters as in Fig.~\ref{fig:fan1},  without and with prefactor [panels (a) and (b),
		respectively] and plotted in a logarithmic
		scale. From \cite{Maxwell2017}. } \label{fig:spiderCQSFA}
\end{figure}

\begin{figure*}
	\centering
	\includegraphics[width=\linewidth]{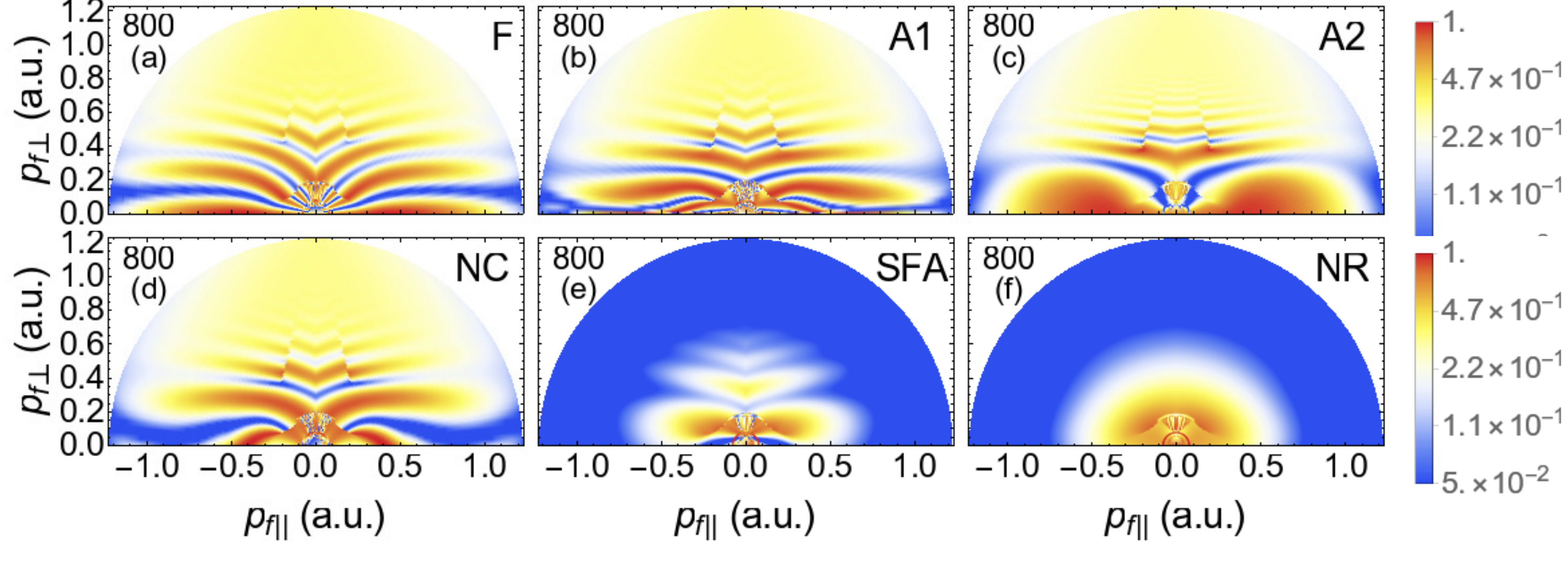}
	\caption{Photoelectron momentum distributions computed with the full CQSFA [panel (a), indicated by F] and its analytical counterparts [remaining panels] without prefactors, for the same parameters as in Figs.~\ref{fig:fan1} and \ref{fig:fan2}. Panel (b), denoted A1, depicts the analytic solution with all phases. Panel (c), denoted A2, shows the analytical approximation without rescattering. In panels (d), (e) and (f) we have omitted the Coulomb phase $S^{\mathrm{prop}}_{C}(\mathbf{r},t,t'_r)$ (NC), the phase  $S_{\mathcal{P}}(\mathbf{\mathcal{P}},t,t'_r)$ associated with the acceleration caused by the residual potential in the continuum (SFA), and both phases together with rescattering (NR). From \cite{Maxwell2017a}. } \label{fig:spideranalytic}
\end{figure*}
In Fig.~\ref{fig:spiderCQSFA}, we plot the spider-like pattern calculated with the CQSFA.  In agreement with previous studies in the literature \cite{HuismansScience2011}, our results show that its main contributors are trajectories 2 and 3, which are forward-scattered orbits starting in the same half cycle.  However, the CQSFA allows a much deeper analysis as we can use it to probe directly how the spider forms. An interesting feature is that there is a caustic around the spider, which is very visible if the prefactor is not included. If the prefactor is inserted, the fringes become much more localized near the $p_{f\parallel}$ axis. One should note that there are no type A or B interferences for the spider. 

One may also use the analytic approximations mentioned in Sec.~\ref{sec:analyticCQSFA} to dissect what each of the phases  $S_{\mathcal{P}}(\mathbf{\mathcal{P}},t,t'_r)$ and $S^{\mathrm{prop}}_{C}(\mathbf{r},t,t'_r)$ given by Eqs. (\ref{eq:accphase}) and (\ref{eq:Coulombphase}), as well as soft recollision, does to the spider-like fringes. This can be seen in Fig.~\ref{fig:spideranalytic}, in which we also include the outcome of the full CQSFA and that of its analytic approximation including all phases [Fig.~\ref{fig:spideranalytic}(a) and (b), respectively]. The sub-barrier Coulomb phase in taken into consideration in all cases.
If soft recollision is neglected [Fig.~\ref{fig:spideranalytic}(c)], the central part of the spider and its fringes are poorly reproduced. This shows that it is an essential part of the physics leading to the spider. The role of the Coulomb phase is to modify the slope of the spider-like fringes. If it is removed, they are ``bent down'' beyond the polarization axis [Fig.~\ref{fig:spideranalytic}(d)], but are still present as they are caused by rescattering and the phase $S_{\mathcal{P}}(\mathbf{\mathcal{P}},t,t'_r)$ related to the acceleration caused by the Coulomb potential. A remarkable example of this is provided in Fig.~\ref{fig:spideranalytic}(e), for which the acceleration phase has been removed. This leads to the spider-like fringes being truncated at the direct-ATI cutoff energy $2U_p$ prescribed by the SFA. If rescattering is removed [Fig.~\ref{fig:spideranalytic}(f)], the electron momentum distribution bears a striking resemblance to those obtained for direct ATI using the SFA if only the long orbits are taken. This is expected as the ionization times for orbits 2 and 3 are relatively close. 

It is important to mention that there exist  interpretations for the spider that are alternatives to the assumption that it is caused by the interference of two types of trajectories. For instance, in \cite{Hickstein2012}, spider-like fringes are obtained by considering the interference of a spherical wave with an incoming plane wave, and, in \cite{Xia2018}, the spider is attributed to an axial caustic singularity and to glory effects in rescattering. This is justified by computing the transverse width of the electron momentum distributions and the positions of the fringes. Nonetheless, the issue is not yet settled as the glory trajectories in \cite{Xia2018} resemble orbits 2 and 3, which form a torus in the three-dimensional space (see discussion in Sec.~\ref{sec:realtrajsCQSFA} and \cite{Lai2015a}). Furthermore, the trajectory-based methods used in the comparisons therein were less sophisticated than many of those reported here as it used either the plain SFA or trajectories in which the binding potential has been added perturbatively. Still, it raises important questions with regard to the role of caustic singularities along the polarization direction.

\subsubsection{Overlooked holographic patterns.}  
\begin{figure*}
	\centering
	\includegraphics[width=\linewidth]{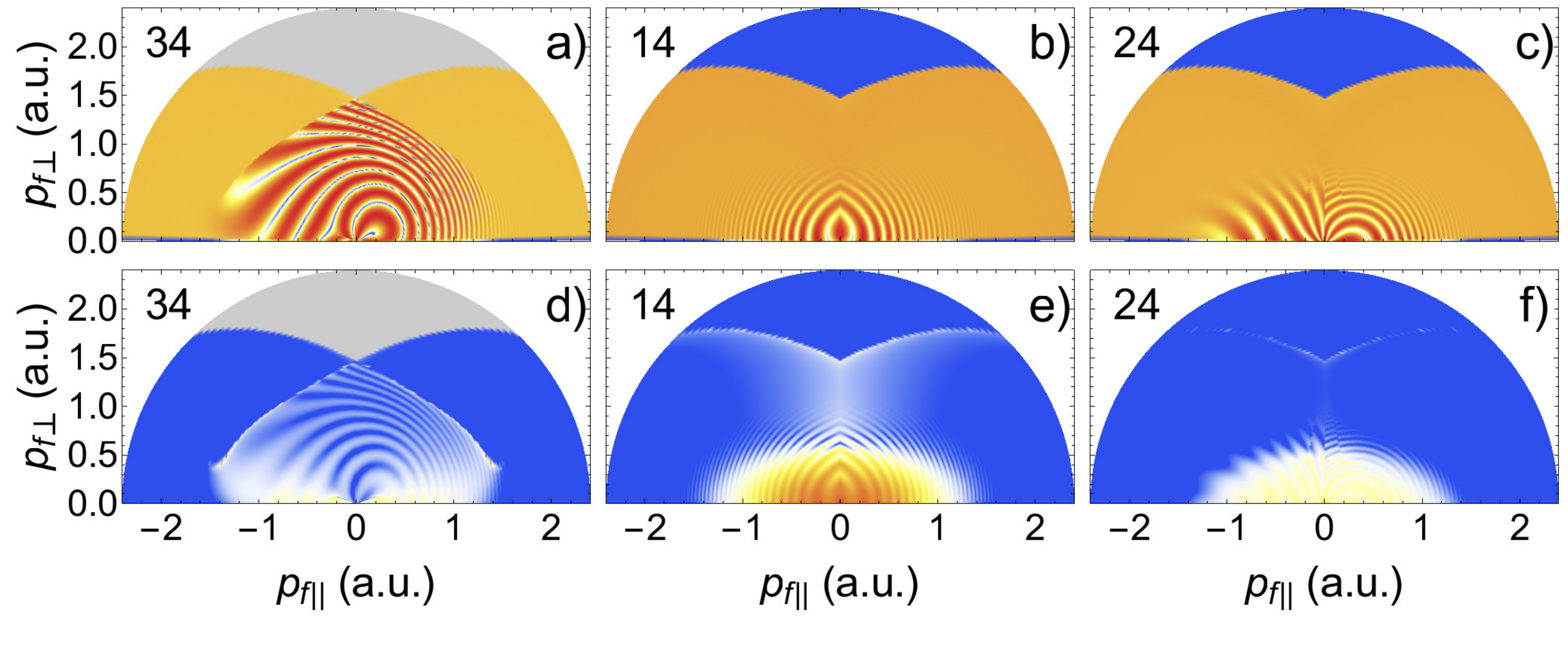}
	\caption{Photoelectron momentum distributions calculated with the CQSFA without and with prefactors (upper and lower panels, respectively) for the same field and atomic parameters as in Figs.~\ref{fig:fan1} and \ref{fig:fan2} using pairwise combinations of orbits 1, 2 and 3 with orbit 4. The orbits used to compute the distributions are indicated on the top left corner of the figure.  All distributions have been normalised by their peak intensity and are plotted in a logarithmic scale. No restriction has been imposed upon the real parts of the ionization times. From \cite{Maxwell2018}.} \label{fig:orb4pairs}
\end{figure*}

Apart from the fan and the spider, there are many overlooked patterns that may be obtained if (i) type 4 orbits are taken into consideration; (ii) if one considers ionization times displaced by more than one half cycle and different types of orbits. Below we provide a few examples, taken from \cite{Maxwell2017,Maxwell2018}. In Fig.~\ref{fig:orb4pairs}, we show spiral-like patterns that are obtained if orbit 4 is incorporated. In experiments and \textit{ab-initio} computations, these patterns can be easily obfuscated and/or mistaken for ATI rings.  They are most visible close to the perpendicular momentum axis and relatively high photoelectron energies, because in this region the spider-like fringes and the fan-shaped structure are strongly suppressed. Caustics associated with this orbit and with orbit 3 are also visible in the figure.
\begin{figure}
	\centering
	\includegraphics[width=0.75\linewidth]{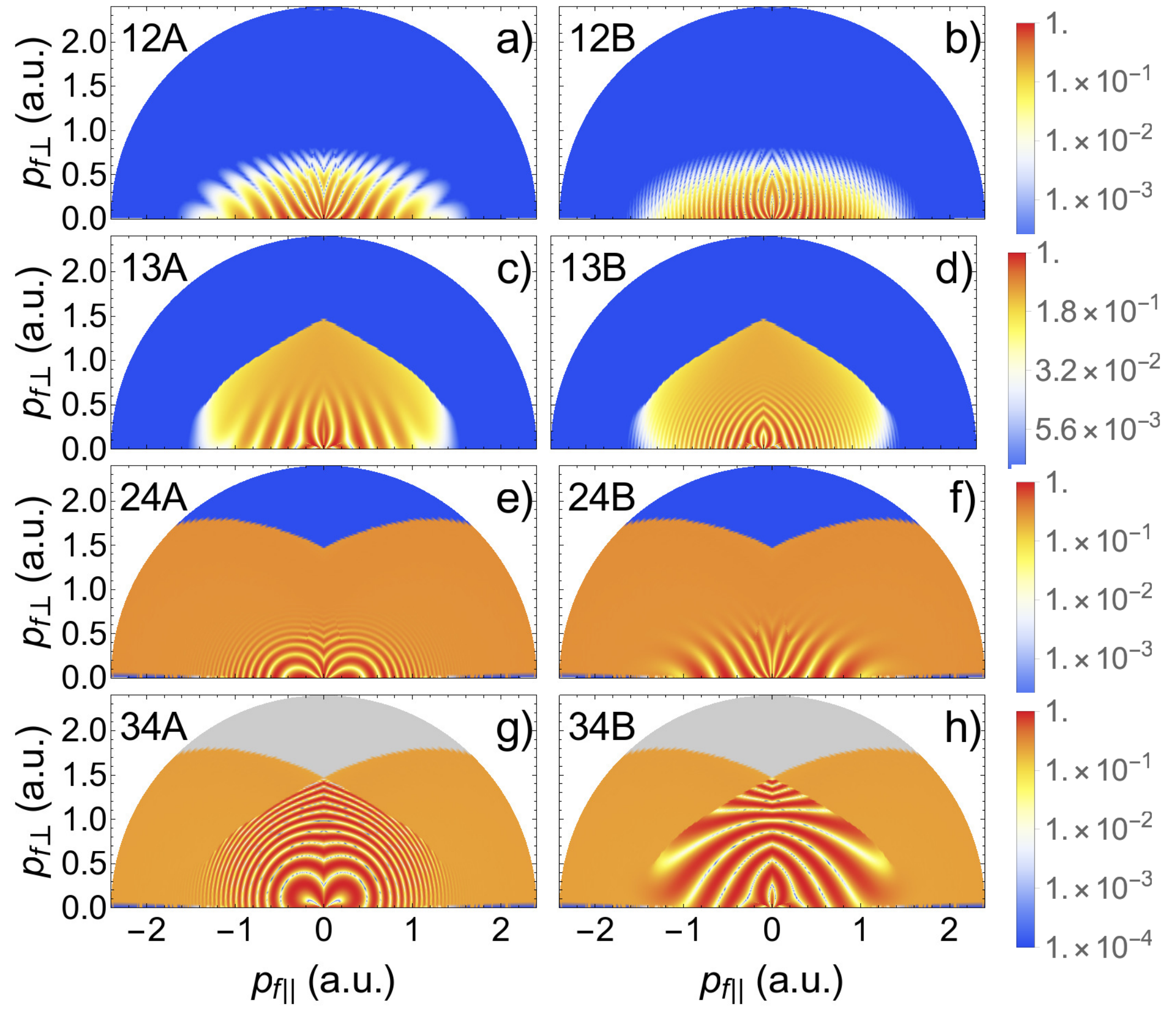}
	\caption{CQSFA photoelectron momentum distributions computed for the same parameters as in the previous figure using pairwise combinations symmetrized with regard to $p_{\parallel}=0$ and the time restrictions A and B as explained in Sec.~\ref{sec:SFAinterf}. The two interfering orbits and the type of interference are indicated on the top left corner of each panel.  All distributions have been normalised by their peak intensity and are plotted in a logarithmic scale. No prefactors have been employed. From \cite{Maxwell2018}.} \label{fig:allpairsAB}
\end{figure}

By restricting the real parts of the ionization times to specific intervals, one may obtain a myriad of patterns. One such pattern is the fan-shaped structure given in Fig.~\ref{fig:fan1} and \ref{fig:fan2}, which results from differences smaller than half a cycle and symmetrization with regard to $p_{f\parallel}=0$ [see also Fig.~\ref{fig:allpairsAB}(a)]. Another possibility is to take this difference to lie between half a cycle and a whole cycle (type B interference according to the notation in \cite{Maxwell2017}). Examples of these patterns are provided in Fig.~\ref{fig:allpairsAB}.  Depending on the choice of interfering orbits, the interference can be divergent [Figs.~\ref{fig:allpairsAB}(a) and (f)], convergent [Figs.~\ref{fig:allpairsAB}(b), (c) and (d)], spiral-like [Figs.~\ref{fig:allpairsAB}(e) and (g)] or borderline cases. Clearly, if no symmetrization or restriction is employed, we will find quite different fringes on the left and right sides of  $p_{f\parallel}=0$. 
\begin{figure*}
	\centering
	\includegraphics[width=\linewidth]{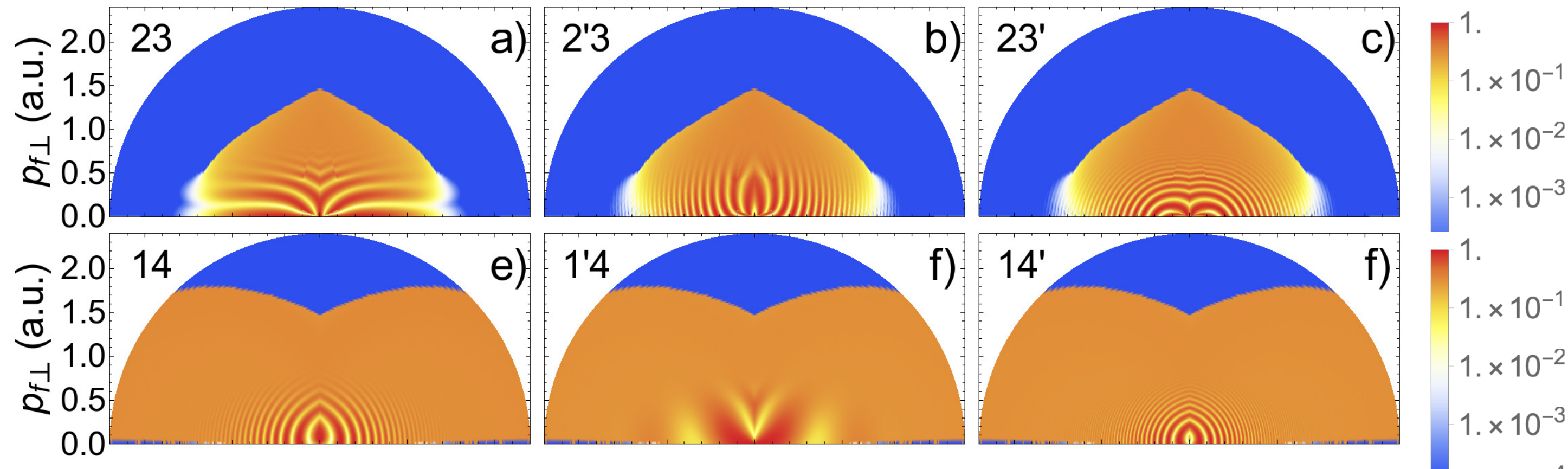}
	\caption{CQSFA photoelectron momentum distributions computed using forward scattered or backscattered trajectories (upper and lower panels, respectively), no prefactors and the same field and atomic parameters as in the previous figures. The prime indicates orbits starting one cycle later, with regard to their counterparts. The orbit combinations are provided on the top left corner of each panel. The yield in each panel has been normalized to its peak value and is plotted in a logarithmic scale. From \cite{Maxwell2018}. } \label{fig:spiderpairs}
\end{figure*}

One may also shift $\Re[t']$ in a whole cycle for one of the interfering orbits in a pair, and this will lead to strikingly different patterns. However, because the types of orbits in the interfering pairs are different, the patterns are very distinct from ATI rings. 
In Fig.~\ref{fig:spiderpairs}, we display results obtained using orbits starting from the same side of the core, i.e., 2 and 3 (upper panels), and 1 and 4 (lower panels). For forward scattered trajectories, apart from the spider [\ref{fig:spiderpairs}(a)], if one of the orbits are shifted of a full cycle, the fringes will become much finer, and either convergent [Fig.~\ref{fig:spiderpairs}(b)] or spiral-like [Fig.~\ref{fig:spiderpairs}(c)]. Finer fringes indicate that the phase shift between orbits 2 and 3 has been increased. 
If one considers the interference of backscattered and direct trajectories (lower panels of Fig.~\ref{fig:spiderpairs}), it is clear that, if orbit 1 starts one cycle later [Fig.~\ref{fig:spiderpairs}(e)], the fringes become much broader and there is a fan-shaped structure. This is due to the fact that it will be more ``in phase'' with a backscattered orbit that started much earlier but went around the core. Clearly, if orbit 4 starts later, or within the same cycle as orbit 1, the fringes will be much finer [Fig.~\ref{fig:spiderpairs}(d) and (f)]. 

The message to be taken out of this section is that the interplay between the Coulomb potential and the driving field leads to a far more complex scenario than anticipated in the seminal paper \cite{Bian2011}.  Those pictures have been drawn by assuming that the tunnel exit remained fixed, the amplitudes associated with each orbit were equal and that rescattering took place at one single point: the origin. The CQSFA made it possible for us to blur many of these distinctions. First, both the weight of the orbit and the tunnel exit are dictated by the sub-barrier dynamics and the first part of our contour, as discussed in Sec.~\ref{sec:subbarrrier}. This means that the tunnel exits will vary with regard to the field and that the probability amplitudes associated with each interfering orbits will be unequal, which, in some instances, will lead to the blurring of the fringes. Second, the phase shifts acquired in the continuum will be determined by the interplay between the laser field and the long-range binding potential, which means that they will differ a lot from those predicted in \cite{Bian2011}. Finally, it is not trivial to define whether an electron has returned to the core, or suffered rescattering, and what type of rescattering it underwent. Indeed, to make this kind of statement we followed two strategies: (i) to employ analytic approximations in which individual contributions to the overall action were switched on and off at will; (ii) to compare the CQSFA orbits with those in the direct or rescattered SFA. Thus, some of the patterns predicted in \cite{Bian2011} that have been sought in experiments are an over simplification and may need to be reassessed.

\subsection{Comparison with \textit{ab-initio} methods} 
\label{sec:abinitio}
\begin{figure*}
	\centering
	\includegraphics[width=\linewidth]{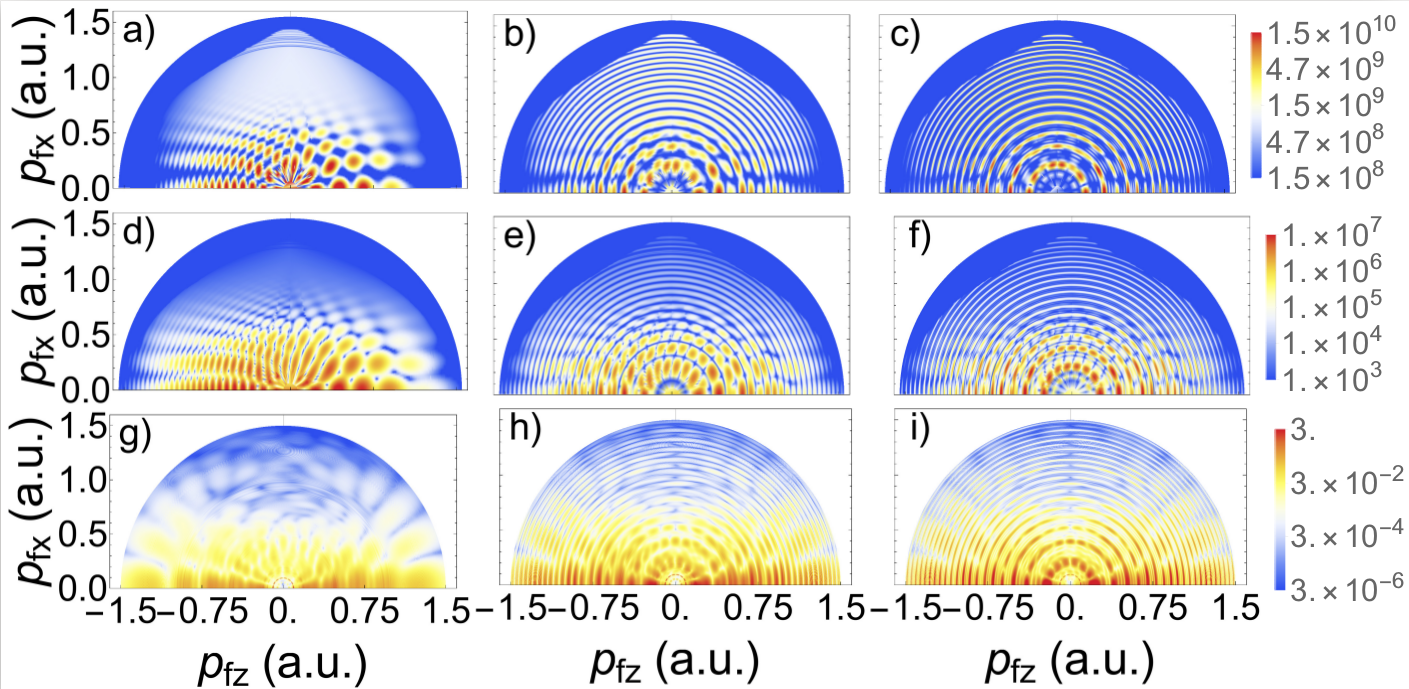}
	\caption{Photoelectron momentum distributions computed with the CQSFA using the orbits 1, 2 and 3 (upper and middle panels) compared with the \textit{ab-initio} solution of the time-dependent Schr\"odinger equation (lower panels), for Hydrogen in a field of intensity $I= 2 \times 10^{14} \mathrm{W/cm}^2$ and wavelength $\lambda=800 \mathrm{nm}$. The left, middle and right panels have been computed using one, two and four cycles of constant amplitude. For the TDSE, we have employed the freely available software Qprop \cite{qprop}, using and additional half a cycle linear turn on and off. All distributions are given in arbitrary units. The upper and middle panels exclude and include the prefactors, respectively. From \cite{Maxwell2017}.} \label{fig:comparisontdse1}
\end{figure*}
\begin{figure*}
	\centering
	\includegraphics[width=\linewidth]{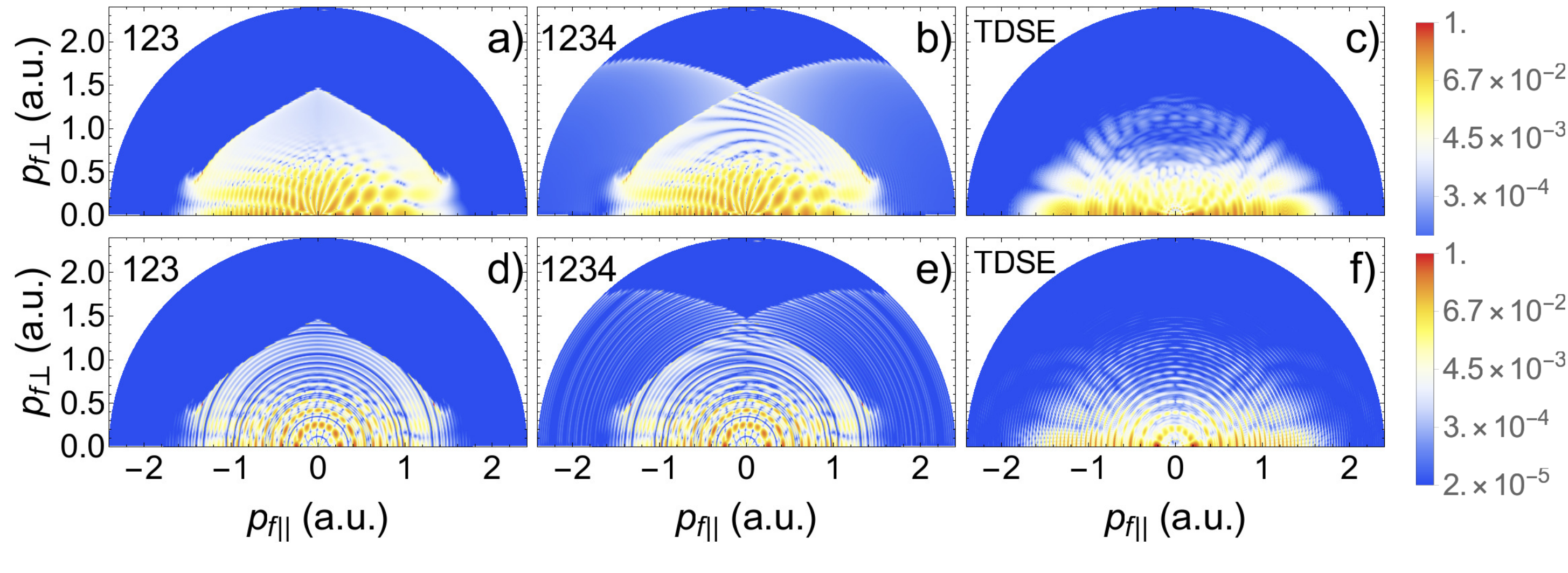}
	\caption{Photoelectron momentum distributions computed with the full CQSFA for the same parameters as in Fig.~\ref{fig:comparisontdse1} excluding and including the contributions of orbit 4 (left and middle panels, respectively). The far right panels provide the outcome of the TDSE, obtained using Qprop \cite{qprop}. The upper and lower panels have been computed over a single and four laser cycles, respectively. From \cite{Maxwell2018}.  } \label{fig:comparisontdse2}
\end{figure*}
\begin{figure*}
	\centering
	\includegraphics[width=\linewidth]{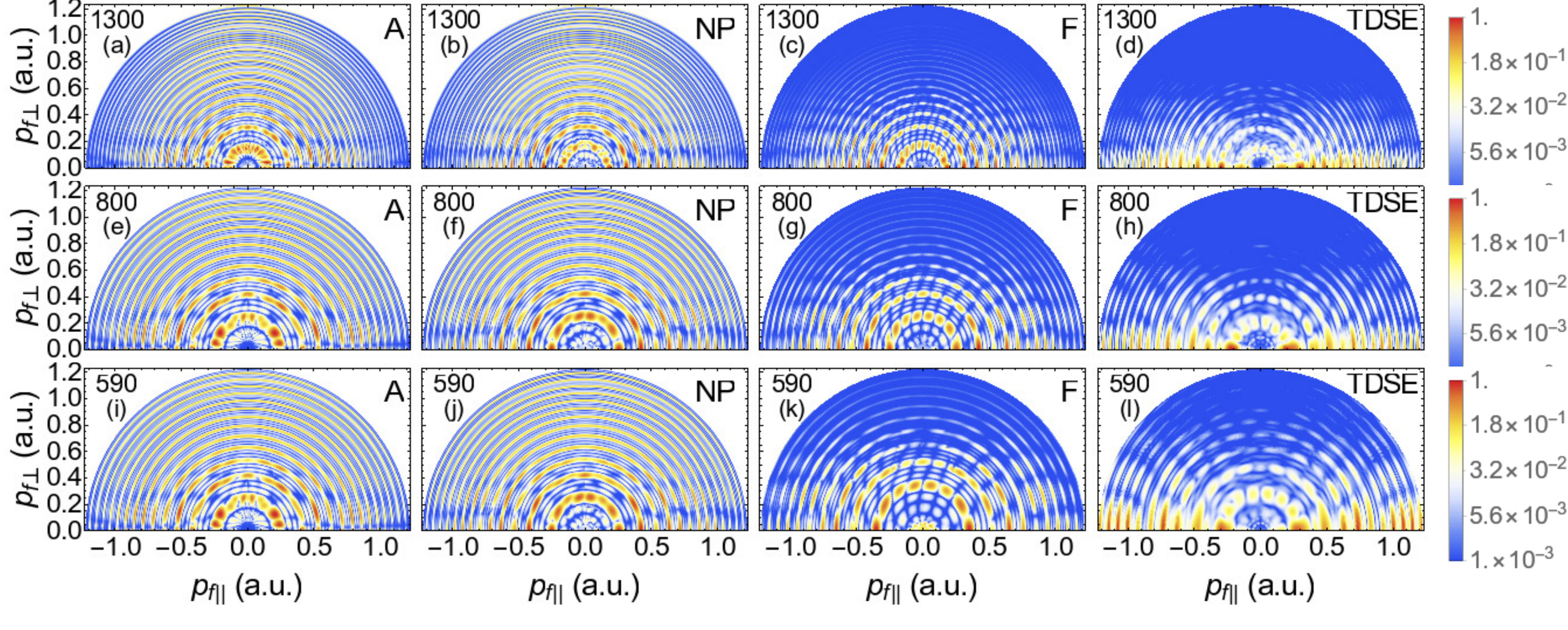}
	\caption{Photoelectron momentum distributions calculated for hydrogen  ($I_p=0.5$ a.u.) using orbits 1 to 3 over four driving-field cycles, using the field parameters $(I,\lambda)=(7.5\times 10^{13} \mathrm{W/cm}^2,1300 \hspace*{0.1cm}\mathrm{nm})$ [panels (a) to (d)], $(I,\lambda)=(2.0\times 10^{14} \mathrm{W/cm}^2,800\hspace*{0.1cm} \mathrm{nm})$ [panels (e) to (h)] and $(I,\lambda)=(3.75\times 10^{14} \mathrm{W/cm}^2,590\hspace*{0.1cm} \mathrm{nm})$ [panels (i) to (l)], where $I$ and $\lambda$ give the field intensity and wavelength, respectively. The far left (denoted by A), left (denoted by NP), right (denoted by F) and far right (denoted by TDSE) columns have been computed using the analytic approximations for the CQSFA, the CQSFA without prefactors, the full CQSFA solution  and the TDSE, respectively. The density plots have been plotted in a logarithmic scale and normalized to the highest yield in each panel. The numbers on the top left corner of each panel give the driving-field wavelength. From \cite{Maxwell2017a}.  } \label{fig:comparisontdse3}
\end{figure*}
Apart from allowing the in-depth analysis presented above, the CQSFA exhibits a good agreement with the \textit{ab-initio} solution of the time-dependent Schr\"odinger equation over a wide momentum range.  This is shown explicitly in the following three figures, which also provide information about what types of interference would be prominent in a realistic setting, and how this is influenced by the field parameters. The CQSFA electron momentum distributions were computed without any restriction upon the real parts of the ionization times and using no symmetrization. 

Fig.~\ref{fig:comparisontdse1} displays the results from the CQSFA taking into consideration the contributions of orbits 1, 2 and 3. Throughout, both the near-threshold fan and the spider-like pattern are very visible, and become more and more symmetric as the pulse length increases. For the shortest pulses, one may see traces of the converging fringes that result from the interference of orbits 1 and 3. Once the number of cycles increases, these fringes become washed out and there is the presence of very clear ATI rings.
The effect of the prefactor is mainly to suppress contributions of high-energy photoelectrons for large scattering angles, improving the resemblance of the CQSFA with the TDSE computation. 

It is noteworthy that the number and approximate location of the spider-like and fan-shaped fringes agree both in the CQSFA and TDSE. This agreement is particularly good for moderate and near-threshold photoelectron energy. The slope of the spider-like structure is however slightly steeper for the \textit{ab-initio} computation. In the CQSFA, this slope is caused by the interplay of the phases $S_{\mathcal{P}}(\mathbf{\mathcal{P}},t,t'_r)$ and $S^{\mathrm{prop}}_{C}(\mathbf{r},t,t'_r)$, which undergo some approximations as the continuum trajectories are taken to be real. Furthermore, all CQSFA results reveal a caustic for high-energy photoelectrons, which marks a boundary within which the present asymptotic expansions are valid for orbit 3. This caustic is also exacerbated by the continuum trajectories being real. Finally, one may see that, close to the perpendicular momentum axis, the ATI rings appear to be ``broken'' for the TDSE distributions, while they remain intact for the CQSFA. This caustic has also been reported in \cite{Morishita2017}.

Fig.~\ref{fig:comparisontdse2} shows that this effect is caused by the contributions of orbit 4. If this orbit is included in the computations, the spiral-like structures resulting from its interference with orbit 3 will cause the ATI rings to break close to the $p_{f\perp}$ axis. In \cite{Korneev2012a}, the broken rings have been attributed to the constructive and destructive interference of the direct SFA short and long orbits. However, in the energy range for which they occur, one expects rescattering to play an important role. Therein, discrepancies with regard to the experimental results have also been reported. This means that the problem is still open to a great extent.  Orbit 4 will also lead to an additional caustic in the distribution [see Figs.~\ref{fig:comparisontdse2}(b) and (e)]. This caustic has been identified in \cite{Morishita2017} using the adiabatic theory, and has been rightly attributed to backscattered trajectories. In \cite{Cerkic2009a}, this shape has been obtained from backscattered SFA orbits with maximum photoelectron momentum, and was explained analytically using the mapping (\ref{eq:causticmap}).

Finally, in Fig.~\ref{fig:comparisontdse3}, we show that, depending on the driving field parameters, the agreement between the CQSFA and the TDSE may improve or worsen. For instance, in the limit of long wavelengths (see upper panels in the figure),  the slope of the spider obtained in the TDSE computation decreases, so that it becomes more similar to the outcome of the CQSFA. Overall, in this wavelength regime the similarity between the result of both approaches is striking. This is possibly due to the long electron excursion amplitudes involved, which reduce the influence of the core.  For the very same reason, the agreement worsens with decreasing wavelength (see middle and lower panels in the figure). The figure also shows that the analytic approximations employed in order to analyze the patterns in greater depth work well together and bear a strong resemblance to the results  obtained with the full CQSFA (see far left and left panels in the figure). 

\subsection{Branch-cut corrections}
\label{sec:branchcutPADs}
\begin{figure}
	\begin{center}
		\includegraphics[width=\linewidth]{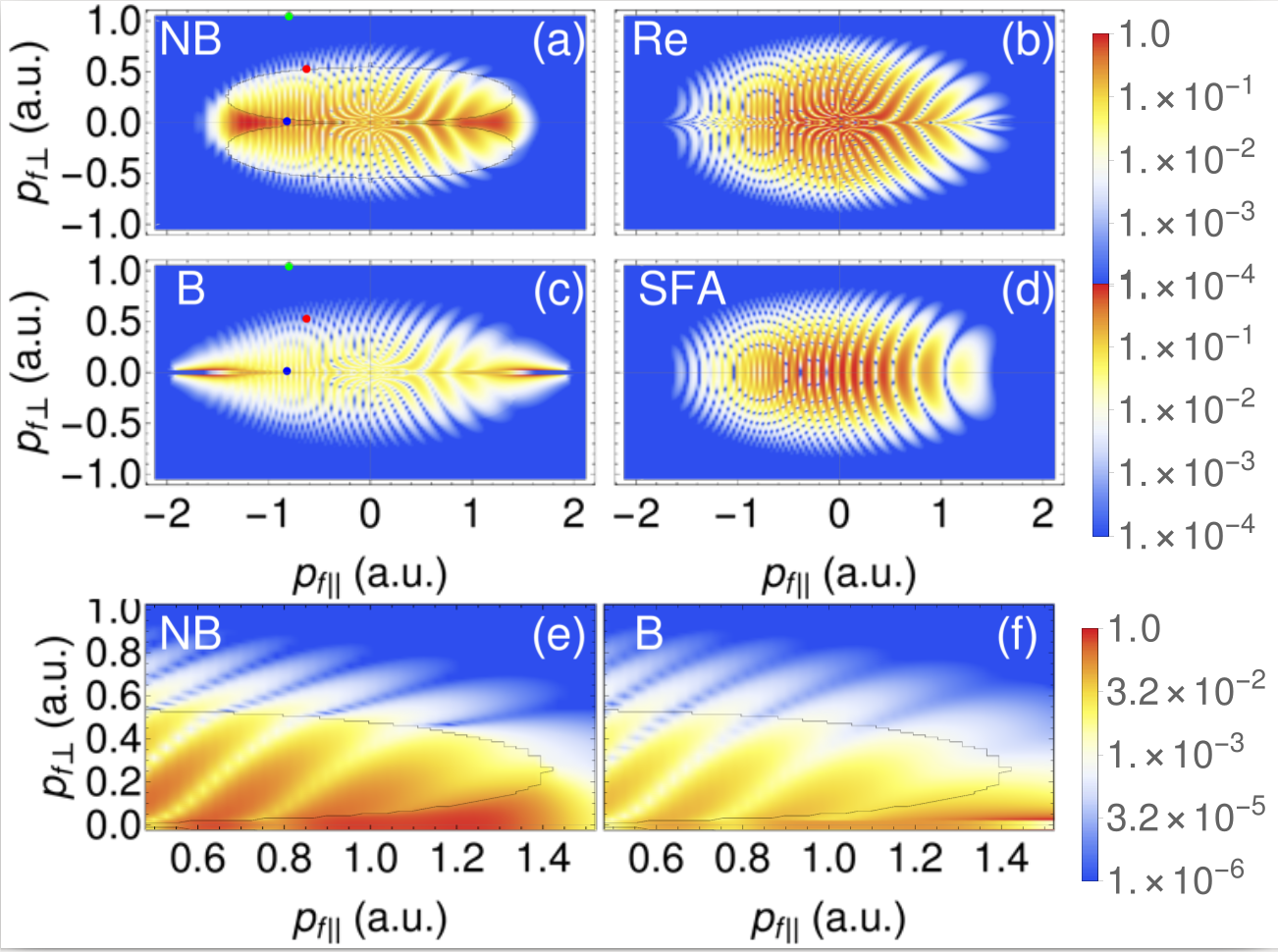}
	\end{center}
	\caption{Photoelectron momentum distributions computed using Coulomb-free trajectories for the same atomic and field parameters as in Figs.~\ref{fig:comparisontdse1} and \ref{fig:comparisontdse2} including the Coulomb phase [panels (a) to (c)], together with the outcome of the standard SFA [panel (d); acronym SFA]. In panels (a) and (c) we have employed complex trajectories without (acronym NB) and with branch-cut corrections (acronym B), while in panel (b) we have used only their real parts in the continuum part of the contour (acronym Re).  Panels (e) and (f) exhibit blow ups of the distributions in panels (a) and (c) in the momentum region for which the branch cut corrections are applied. The red (gray), green (light gray) and black dots in panels (a) and (c) mark the  momentum components  $(p_{f\parallel},p_{f\perp})=(-0.63 \mathrm{~a.u.}, 0.53 \mathrm{~a.u.})$    $(p_{f\parallel},p_{f\perp})=(-0.80\mathrm{~a.u.}, 1.05\mathrm{~a.u.})$   and  $(p_{f\parallel},p_{f\perp})=(-0.82\mathrm{~a.u.}, 0.01\mathrm{~a.u.})$ used in the branch-cut mapping in Fig.~\ref{fig:Coulombfree}. For clarity, the same colors have been employed.  The thin black lines in panels (a), (e) and (f) separate the regions in momentum space for which branch cuts cross (outside the oval) and do not cross (inside the oval) the real axis for orbit 2. All panels have been displayed in a logarithmic scale and plotted in arbitrary units. From \cite{Maxwell2018b}. }
	\label{fig:padcoulfree}
\end{figure}
In the following, we incorporate complex trajectories in the CQSFA and analyze the effect of branch cuts in the photoelectron momentum distributions. The simplest scenario is given in Fig.~\ref{fig:padcoulfree}, in which the Coulomb phase has been incorporated in the electron propagation but Coulomb-free, SFA orbits were used. The ovals in Fig.~\ref{fig:padcoulfree} and in the lower panels of the figure delimit the region for which orbit 2 does not cross a branch cut. In the absence of branch-cut corrections, fringe discontinuities and a trumpet-shaped structure near the $p_{f\parallel}$ axis closely following the oval boundary are very visible [Figs.~\ref{fig:padcoulfree}(a) and (e)]. These features are eliminated once the branch-cut corrections have been applied [Figs.~\ref{fig:padcoulfree}(c) and (f)]. Overall, we also see that the distributions computed with the strong-field approximation [Fig.~\ref{fig:padcoulfree}(d)], or with real trajectories [Fig.~\ref{fig:padcoulfree}(e)] are much broader and decay more slowly. This supports the statement that complex trajectories may lead to an effective deceleration of the electronic wave packet \cite{Torlina2013}. One should bear in mind, however, that their imaginary parts may also lead to enhancements in the photoelectron yield, as reported in \cite{Keil2016}. Throughout, the Coulomb phase considerably modifies the shape of the quantum-interference fringes, although the holographic structures discussed in the previous section are absent. The figure also makes it clear that our procedure does not work well on the $p_{f\parallel}$ axis, as previously stated [see Fig.~\ref{fig:padcoulfree}(c)].
\begin{figure*}
	\begin{center}
		\includegraphics[width=0.75\linewidth]{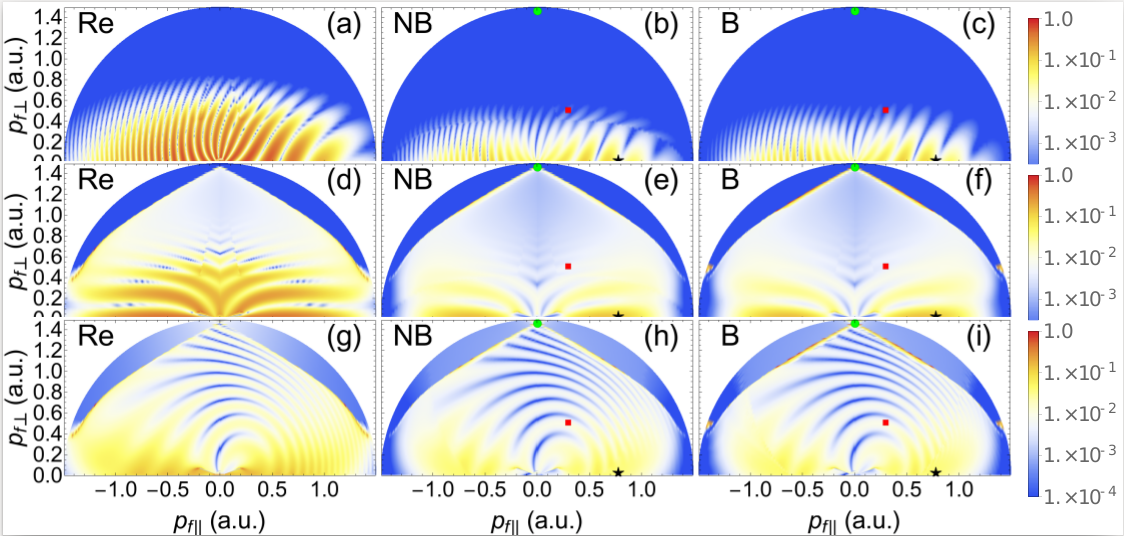}
	\end{center}
	\caption{Contributions from specific pairs of orbits to the ATI photoelectron momentum distributions computed using different versions of the CQSFA for the same field and atomic parameters as in the previous figure. In the left, middle and right panels, we have employed real trajectories (acronym Re), complex trajectories and no branch cut corrections (acronym NB), and complex trajectories with branch-cut corrections (acronym B), respectively. The first row [panels (a) to (c)] considers orbits 1 and 2, the middle row [panels (d) to (f)] orbits 2 and 3, and the lower row [panels (g) to (h)] to orbits 3 and 4.  The red (gray), green (light gray) and black dots corresponds to the momenta $(p_{f\parallel},p_{f\perp})=(-0.475\mathrm{~a.u.}, 0.400\mathrm{~a.u.}) $,   $(p_{f\parallel},p_{f\perp})=(-0.604\mathrm{~a.u.}, 0.980\mathrm{~a.u.}) $  and $(p_{f\parallel},p_{f\perp})=(-0.619\mathrm{~a.u.}, 0.0113\mathrm{~a.u.})$ that have been used to compute the branch cuts in Fig.~\ref{fig:branchmapCQSFA}. All plots have been displayed in a logarithmic scale. From \cite{Maxwell2018b}. }
	\label{fig:PADCQSFA2}
\end{figure*} 

If Coulomb-distorted orbits are used, similar features are observed. An example is provided in Fig.~\ref{fig:PADCQSFA2}, in which specific holographic patterns are shown. For the interference between orbit 1 and 2, one clearly sees that the distributions computed with complex orbits [Fig.~\ref{fig:PADCQSFA2}(b) and (c)] decay much faster than that in which real orbits are taken into the continuum propagation [Fig.~\ref{fig:PADCQSFA2}(a)]. Furthermore, orbit 2 crossing a branch cut also leads to discontinuities, which our method eliminates. Similar features can also be observed for the spider, and the spiral (middle and lower rows in the figure, respectively). Furthermore, employing complex trajectories alters the slope of the spider and changes the spacing and the contrast of the spiral. Finally, the fringes near the $p_{f\perp}$ axis that exist for the spider if real trajectories are taken are softened in the complex-trajectory case. This is due to a faster decay in the contributions of orbit 2.
\begin{figure}
	\begin{center}
		\includegraphics[width=0.75\linewidth]{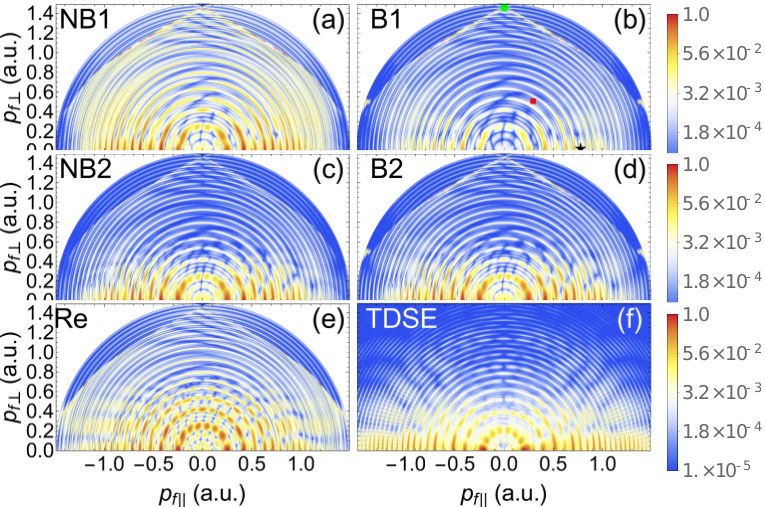}
	\end{center}
	\caption{Photoelectron momentum distributions computed using several versions of the CQSFA [panels (a) to (e)] over four cycles of the same driving field as in the previous figures, compared with the outcome of an \textit{ab-initio} computation performed with the freely available software Qprop \cite{qprop,Mosert2016} [panel (f)]. In panels (a) to (d) we have employed complex trajectories without (left panels; acronyms NB) and with (right panels; acronyms B) branch-cut corrections, while  panel (e)(acronym Re) was calculated with real trajectories.  In panels (c) and (d) (acronyms NB2 and B2, respectively) the contributions of orbits 3 and 4 to the overall transition amplitude have been reduced multiplying by a factor 0.2, while in panels (a) and (b) (acronyms NB1 and B1, respectively) no ad-hoc adjustment was made. The red (gray), green (light gray) and black dots mark the momentum components  $(p_{f\parallel},p_{f\perp})=(-0.475\mathrm{~a.u.}, 0.400\mathrm{~a.u.}) $,   $(p_{f\parallel},p_{f\perp})=(-0.604\mathrm{~a.u.},0.980\mathrm{~a.u.}) $  and $(p_{f\parallel},p_{f\perp})=(-0.619\mathrm{~a.u.},0.0113\mathrm{~a.u.})$ that have been used to compute the branch cuts in Fig.~\ref{fig:branchmapCQSFA}. All plots have been displayed in a logarithmic scale. From \cite{Maxwell2018b}.
	}
	\label{fig:PADCQSFA1}
\end{figure}
In Fig.~\ref{fig:PADCQSFA1}, we show a direct comparison of all the versions of the CQSFA developed and employed by us with the \textit{ab-initio} solution of the time-dependent Schr\"odinger equation. If branch-cut corrections are taken into consideration, complex trajectories make the electron momentum distribution decay faster for increasing transverse momentum components $p_{f\perp}$. This means that they become more localized near the polarization axis, improving the agreement with the \textit{ab-initio} solution. Furthermore, the slope of the spider-like fringes is nearly horizontal if real trajectories are taken, but becomes closer to the TDSE outcome in the complex-trajectory case.
There is however a drawback in the present procedure for correcting branch cuts, which tends to worsen the overall agreement between the CQSFA and \textit{ab-initio} computations. The approximations employed in the Coulomb-distorted case to render the problem tractable (see Sec.~\ref{subsubsec:branchcuts1}) work well for orbit 1, reasonably for orbit 2, but less so for orbits 3 and 4. This is expected, as those two latter orbits interact very strongly with the core, and are therefore quite different from their direct SFA counterparts (see upper panels of Fig.~\ref{fig:PADCQSFA1}). An immediate consequence is that the contributions of orbits 3 and 4 to the overall probability densities are over-enhanced. This is caused by the imaginary parts of orbits 3 and 4 being  overestimated. In order to counteract this problem, we reduce the contributions of orbits 3 and 4 by employing an empirical factor. This considerably improves the agreement, as  a direct comparison of Figs.~\ref{fig:PADCQSFA1}(d) and (f) shows.

\section{Trends}
\label{sec:trends}

\begin{figure}[t]
	\begin{center}
		\includegraphics[width=0.65\linewidth]{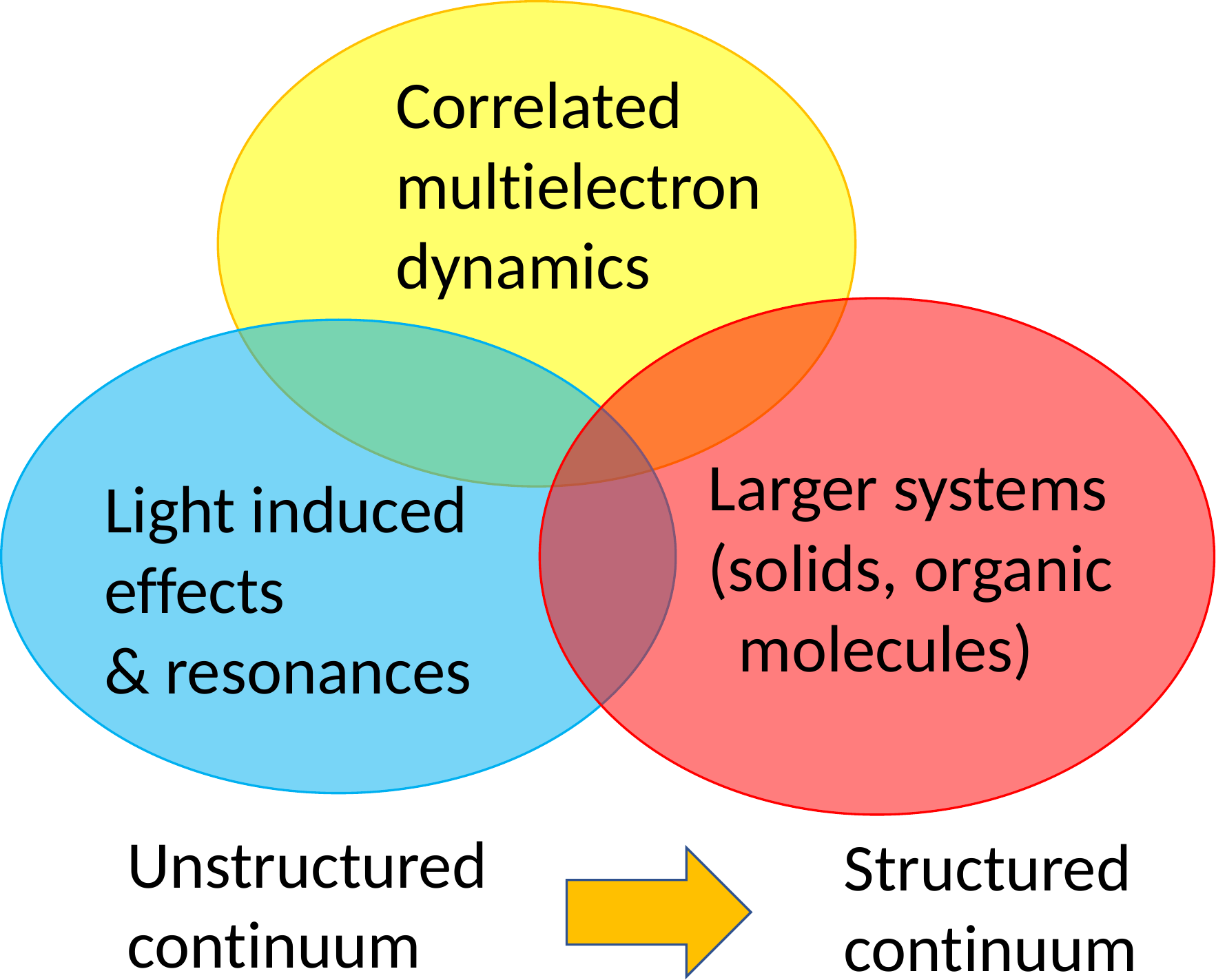}
	\end{center}
	\caption{Schematic representation of the current trends for subfemtosecond imaging in general and ultrafast photoelectron holography in particular.}
	\label{fig:trends}
\end{figure}

Overall, the existing trends for photoelectron holography reflect those one may observe for ultrafast imaging as a whole. There is a shift from merely structural questions towards the aim of steering electron and core dynamics in real time, which requires more sophisticated modeling, more complex targets and tailored fields, which are chosen in such a way as to probe specific properties of these targets. Thereby, an important issue is to move from an unstructured continuum, which can be approximated by field-dressed plane waves, towards a structured continuum.  This, together with an appropriate treatment of multielectron dynamics, is a key requirement for photoelectron holography in polyatomic molecules, solids and nanostructures. A schematic representation is presented in Fig.~\ref{fig:trends}.

The simplest possible example of modeling a structured continuum, which is to include the residual Coulomb potential and compute photoelectron spectra for a one-electron atom, has been extensively discussed in this review, with particular emphasis on the CQSFA. Our results show that the interplay between the residual potential and the laser field plays a huge role. The very fact that the Coulomb potential alters the ionization dynamics and the number and topology of the interfering orbits means that the resulting holographic patterns are strikingly different from those predicted with the strong-field approximation and in the seminal papers \cite{Spanner2004,Bian2011}. Clearly, the present studies have only paved the way for the understanding and control of this interplay in more complex systems. 
Below we elaborate on some of the  present trends:
\begin{itemize}
	\item \textit{Molecules and multielectron atoms.}  These are the quintessential ultrafast imaging targets, and have called a great deal of attention since the mid 2000s. Apart from the geometry of the bound states involved, that may not be spherically symmetric, removing one electron will alter the dynamics of the the core. This removal may for instance create a hole that will migrate and/or induce further rearrangement processes in the remaining charge, such as relaxation, excitation and electron migration. The external field may also polarize the core electrons, or even induce double and multiple ionization. In case of molecules, there may be also the coupling of electronic and nuclear degrees of freedom, such as vibration and rotation, or even dissociation and fragmentation. The latter effects will be particularly important for large systems, or highly excited states. To be able to probe such systems using photoelectrons, one must ensure that a strong interaction with the core takes place. This can be achieved by focusing on the near-threshold energy regime, for which the electron orbits are substantially distorted by the core, or on patterns involving orbits that undergo appreciable acceleration by the residual ion.
	Well-known examples of the former and the latter situation are the fan and the spider, respectively. Enhancements in the fan have for instance been related to resonances with Rydberg states in dimers \cite{Veltheim2013}, nuclear-electronic coupling of degrees of freedom \cite{Mi2017}, and suppression of the fan-shaped fringes has been attributed to frustrated tunnel ionization \cite{Liu2012}.
	Phase shifts related to molecular alignment \cite{Meckel2014,Liu2016,He2018}, as well as imprints of nodal planes  and evidence of coupling with nuclear degrees of freedom  have been identified in the spider \cite{Walt2017,Fernandez2009}. Recent computations indicate that polarization effects influence both the fan and the spider \cite{Shvetsov2018}.  Backscattered trajectories and the resulting quantum interference may also be employed. For instance, in \cite{Kreicinic2018} backscattered trajectories have been used to retrieve information about multiple orbitals in LIED of an aligned organic molecule, and required modelling beyond the single-active orbital and plane rescattering wave approximation. However, special care must be taken that they are not obfuscated by more prominent patterns. An example is the fishbone structure in  \cite{Haertelt2016}, which could only be identified by using differential holograms in which the spider was removed. This fishbone pattern is highly dependent on the internuclear separation, and thus could be used for tracking nuclear dynamics. 
	
	\item \textit{Tailored fields,} such as linearly \cite{Xie2013,Arbo2014,Skruszewicz2015,Xie2016b,Luo2017,Porat2018} or orthogonally polarized \cite{Gong2017,Han2017,Han2018} two-color fields, few-cycle pulses \cite{Shvetsov2014,Oi2017}, elliptically or circularly polarized fields \cite{Pfeiffer2011,Pfeiffer2012,Li2013,Ivanov2014,Han2017b,Landsman2013}, or bicircular fields \cite{Milos2016,Hoang2017,Eckart2018,Milos2018}, allow a greater deal of control in the electron ionization, propagation and deflection/recollision by the potential. Hence, they may be used to probe a myriad of properties such as tunneling times \cite{Eckle2008,Pfeiffer2011,Pfeiffer2012,Li2013,Ivanov2014,Han2017b,Landsman2013,Torlina2015,Klaiber2018}, the influence of the Coulomb potential \cite{Zhang2014} and holographic patterns that are not easily resolved \cite{Li2016}. In principle, obfuscated holographic patterns could be revealed by (a) moving more prominent holographic patterns to other momentum regions or suppressing them altogether; (b) changing the curvature and shapes of specific holographic fringes. 
	For instance, in \cite{Li2016} it has been proposed, within the SFA framework, that orthogonal two-color (OTC) fields may be employed to distinguish the fishbone holographic patterns related to backscattered orbits from those stemming from the temporal double slit. The SFA predictions, however, are very distinct from the outcome of \textit{ab-initio} computations in which the residual potential is taken into consideration. Still, OTC fields may move overlapping holographic patterns into distinct momentum regions.
	Furthermore, orthogonally polarized fields  provide a good testing ground for the influence of the Coulomb potential on photoelectron momentum distributions. In \cite{Zhang2014}, asymmetries in photoelectron momentum distributions for neon in orthogonally polarized two-color fields could be traced back to the influence of the Coulomb potential using classical-trajectory Monte Carlo (CMTC) methods. Thereby, a strong deviation from the mapping $\mathbf{p}_0=-\mathbf{A}(t')$	has been encountered, which has been related to the Coulomb potential holding back departing electrons, i.e., to frustrated tunnel ionization.	
	Bicircular fields are of paramount interest as they may be used for probing chiral molecules, and also allow for rescattering. Above-threshold ionization in such fields is expected to reveal a myriad of LES and interference patterns \cite{Milos2016,Eckart2018,Milos2018}, some of which exist only due to nonvanishing initial momenta and the presence of the Coulomb potential \cite{Eckart2018}. 
	\item \textit{Plasmonically enhanced fields.} It has been suggested that plasmon-induced resonant enhancements may amplify modest fields in several orders of magnitudes \cite{Park2011}. This amplification does however lead to driving fields with an inhomogeneous spatial profile. Effective models in which this inhomogeneity is approximated within a one-electron framework show that this causes significant distortions in electron-momentum distributions \cite{Ciappina2013}. A quantum-orbit analysis using the SFA also shows that the photoelectron orbits significantly change in this case \cite{Shaaran2013}. It is noteworthy that even very simplified models as those used in both papers may already reveal a very rich, non-trivial dynamics. For instance, in \cite{Zagoya2016}, in the context of HHG, we have found that even very small inhomogeneities lead to spatial confinement, bifurcations, symmetry breaking, two very distinct time scales and different stability regimes. This all plays a huge role in the continuum electron dynamics, and is expected to influence above-threshold ionization as well.
	\item \textit{Longer wavelengths}. Typically, the complex systems mentioned above have lower ionization potentials. Furthermore, in specific cases, such as in the seminal work of \cite{HuismansScience2011}, it is advantageous to prepare the target in an excited metastable state. 
	This means that, to probe such systems, it is desirable (i) to reduce the driving-field intensity and keep the ponderomotive energy fixed; (ii) to ensure tunneling is still the dominant ionization mechanism. Condition (i) implies that high-energy photoelectrons may be created, as the ATI cutoffs  both for direct and rescattered electrons are functions of $U_p$, and that one is still operating below the saturation intensity. Condition (ii) means that all the formalism and the physical interpretations developed for the quasi-static regime are valid, and that ionization is very much dependent on the instantaneous field strength. Both conditions can be achieved by using fields of longer wavelengths, which have become increasingly popular since the beginning of decade \cite{Colosimo2008,Wolter2015}. For that reason, many of the studies reported in this review have been performed in the mid-IR regime. Thereby, another difficulty arises: if the
	magnetic field associated with the laser field causes the electron motion perpendicular to the
	driving-field polarization to become greater than a Bohr radius, then the dipole approximation breaks down \cite{Reiss2008}. This will alter several features described in this work, such as Coulomb focusing and defocusing \cite{Ludwig2014}, and the topology of the electron orbits \cite{Keil2017}. For studies of radiation pressure in the context of LIED see, e.g.,  \cite{Chelkowski2015,Brennecke2018}, and for non-dipole effects with elliptically polarized fields see \cite{Maurer2018}.
\end{itemize}
The above-stated list is obviously non-exhaustive, and ultrafast photoelectron holography will bring with itself many challenges and questions, both theoretical and experimental. With the present review, we hope to have provided a thorough and multifaceted contribution to the understanding of photoelectron holography's past, present and future.

\section*{Funding}
This research was is part funded by the UK Engineering and Physical Sciences Research Council (EPSRC) (grants EP/J019143/1 and EP/P510270/1). The latter grant is within the remit of the  InQuBATE Skills Hub for Quantum Systems Engineering.

\section*{Acknowledgements}
We would like to thank  X. Liu, X. Y. Lai, T. Das, A. Al-Jawahiry, and  S. Popruzhenko for their collaboration, which led to some of the publications discussed in this review. We are also grateful to D. B. Milo\v{s}evi\'c, D. Bauer, M. Yu. Ivanov, W. Becker, M. Lewenstein, E. Pisanty, L. Torlina, J. Z. Kaminski, N. I. Shvetsov-Shilovski, W. Koch and D. J. Tannor for useful discussions.

\section*{Abbreviations}
The following abbreviations are used in this manuscript:

\begin{tabular}{@{}ll}
	ADK & Ammosov-Delone Krainov\\
	ATI & Above-threshold ionization\\
	ARM & Analytical R-Matrix theory\\
	CCSFA& Coulomb-corrected strong-field approximation\\
	CMTC& Classical-trajectory Monte Carlo method\\
	CQSFA & Coulomb Quantum-orbit Strong-Field\\& approximation\\
	CVA & Coulomb-Volkov approximation\\
	DATI & Direct above-threshold ionization\\
	EVA& Eikonal Volkov approximation\\
	HATI & High-order above-threshold ionization\\
	HHG & High-order harmonic generation\\
	HOMO & Highest-occupied molecular orbital \\
	ISFA & Improved strong field approximation\\
	KFR & Keldysh-Faisal-Reiss\\
	LIED& Laser-induced electron diffraction\\
	LES & Low-energy structure\\
	LFA & Low-frequency approximation\\
	NSDI & Nonsequential double ionization\\
	OTC & Orthogonally polarized two-colour\\
	QRS & Quantitative Rescattering Theory\\
	QMTC & Quantum-trajectory Monte Carlo method\\
	SCTS & Semi-classical two-step model\\
	SFA & Strong-field approximation\\
	RABITT & Attosecond Burst By Interferences of\\& Two-photons Transitions\\
	TCSFA & Trajectory-based Coulomb strong-field\\& approximation\\
	TDSE & Time-dependent Schr\"odinger equation\\
	VLES & Very low energy structure\\
	WKB & Wentzel-Krammers-Brillouin\\
	ZES & Zero energy structure\\
\end{tabular}

\section*{References}
\bibliographystyle{iopart-num}
\bibliography{Review2}

\end{document}